\documentclass[11pt,reqno]{amsart}

\usepackage{packages}
\usepackage{notations}

\date{\today}
\title{A Modular Engine for Quantum Monte Carlo Integration}
\makeatletter
\let\inserttitle\@title
\makeatother

\author{Ismail Yunus Akhalwaya~$^{1,3}$, Adam Connolly~$^1$, Roland Guichard~$^{1,\dagger}$, Steven Herbert~$^{1,4}$, Cahit Kargi~$^1$, Alexandre Krajenbrink~$^1$, Michael Lubasch~$^2$, Conor Mc Keever~$^2$, Julien Sorci~$^1$, Michael Spranger~$^1$, Ifan Williams~$^1$}

\address[Ismail Yunus Akhalwaya, Adam Connolly, Roland Guichard, Steven Herbert, Cahit Kargi, Alexandre Krajenbrink, Julien Sorci, Michael Spranger, Ifan Williams]{$~^1$Quantinuum, Terrington House, 13–15 Hills Road, Cambridge CB2 1NL, United Kingdom}

\address[Michael Lubasch, Conor Mc Keever]{$~^2$Quantinuum, Partnership House, Carlisle Place, London SW1P 1BX, United Kingdom}

\address[Ismail Yunus Akhalwaya]{$~^3$School of Computer Science and Applied Mathematics, University of the Witwatersrand, Johannesburg, South Africa}

\address[Steven Herbert]{$~^4$Department of Computer Science and Technology, University of Cambridge, United Kingdom}

\email{\href{mailto:steven.herbert@quantinuum.com}{steven.herbert@quantinuum.com} \& \href{mailto:alexandre.krajenbrink@quantinuum.com}{alexandre.krajenbrink@quantinuum.com}}

\thanks{$\dagger$ Current address: PASQAL, 7 Rue L\'eonard de Vinci, 91300 Massy, France}

\begin{document}

\begin{abstract}
We present the Quantum Monte Carlo Integration (QMCI) engine developed by Quantinuum. It is a quantum computational tool for evaluating multi-dimensional integrals that arise in various fields of science and engineering such as finance. This white paper presents a detailed description of the  architecture of the QMCI engine, including a variety of distribution-loading methods, a novel quantum amplitude estimation method that improves the statistical robustness of QMCI calculations, and a library of statistical quantities that can be estimated. The QMCI engine is designed with modularity in mind, allowing for the continuous development of new quantum algorithms tailored in particular to financial applications. Additionally, the engine features a resource mode, which provides a precise resource quantification for the quantum circuits generated. The paper also includes extensive benchmarks that showcase the engine's performance, with a focus on the evaluation of various financial instruments.\\
{ } \\
\end{abstract}

\maketitle

{\hypersetup{linkcolor=black}
\setcounter{tocdepth}{2}
\makeatletter
\def\l@subsection{\@tocline{2}{0pt}{2.5pc}{2.5pc}{}}
\makeatother
\tableofcontents
}


{\hypersetup{linkcolor=black}
\listoftables
}

\newpage
\part{Introduction and background}
\label{part:intro-background}

\section{Introduction}
\label{sec:introduction}

For high-dimensional integrals with no analytic solution \textit{Monte Carlo integration} (MCI) is invariably the most efficient method of (classical) numerical integration, and for this there is a proven quadratic quantum advantage. In particular, \textit{quantum} Monte Carlo integration (QMCI) maps the integral value to the amplitude of a qubit, and then uses quantum amplitude estimation (QAE) to estimate this value. QAE is such that the estimator's root mean-squared error (RMSE) is proportional to the reciprocal of the number of samples, whereas for (classical) MCI the RMSE is proportional to the reciprocal of the square root of the number of samples. Over the past few years, a number of breakthroughs have fuelled the hope that QMCI will be an application of quantum computing that enjoys quantum advantage at a relatively early stage. Various works have shown how QAE can be performed without requiring the computationally costly subroutine of quantum phase estimation (QPE) \cite{Suzuki_2020, grinko2019iterative, Aaronson_2020, nakaji2020faster} (hereafter these are generally referred to as QPE-free QAE algorithms). Furthermore it has been shown that the Monte Carlo integral can be decomposed as a Fourier series, such that each harmonic can be estimated using only a shallow quantum circuit, whilst maintaining the full quantum advantage \cite{herbert1}.\\

 In this paper we present Quantinuum's new end-to-end QMCI engine, which is principally tailored to applications in mathematical (quantitative) finance. Whilst MCI touches on almost every area of science, technology, operations research and business, there has been specific enthusiasm for quantum solutions within the financial sector \cite{Rebentrost2018, Stamatopoulos_2020, Woerner2019, bouland2020prospects, QCfinance, egger2019credit, kaneko2020quantum, chakrabarti2020threshold, rebentrost2018quantum, financeppr, an2020quantumaccelerated, jpmc, goldmangradients}. Many of the problems within mathematical finance are naturally posed in a way that makes them amenable to quantisation, and thus able to benefit from the theoretical quantum advantage. In particular, financial applications often have an appropriate blend of simplicity and complexity. For example, when considering derivative pricing, standard random processes such as geometric Brownian motion may typically be used to model some underlying asset price\footnote{At least in textbook models -- in time we will supplement these with more advanced models for financial time-series.}. We can therefore design efficient ways to load these relatively \textit{simple} random processes as quantum states. However, the actual Monte Carlo integral will itself be \textit{complex} (or, more precisely, high-dimensional) when the derivative is path-dependent, and hence in principle can benefit from QMCI. Furthermore, when it comes to path-dependent derivative instruments, there is potential for almost infinite variation. In particular, various thresholds (e.g., in barrier options), sums (e.g., to calculate the average as required in Asian options) and maxima (e.g., in look-back options) of random variables pertaining to time-slices of the time-series may be taken\footnote{See Table~\ref{tab:all-calcs1} for more details on these instruments.}. We are thus motivated to provide as key components of the QMCI engine: (i) a library of quantum circuits encoding models of financial time-series; and (ii) a generic way to construct computations that are commonplace in mathematical finance. In the former case, we include some simple explicit circuits that are required for our benchmark results, and we also comprehensively detail a number of generic methods for constructing quantum circuits that efficiently encode probability distributions. In the latter case, we introduce the \textit{enhanced} $P$-\textit{builder}\footnote{``$P$'' has become the standard notation for the quantum circuit that loads a probability distribution -- and this is how we use it here.} -- a module that allows efficient construction of quantum circuits encoding a wide variety of financial derivative payoff and portfolio risk calculations.\\

The quantum circuit encoding the (financial) quantity of interest is then inputted into the Monte Carlo integrator, and for this we use the aforementioned Fourier series decomposition of the Monte Carlo integral. The Fourier series decomposition is possible when taking expectations of finitely supported probability distributions (as is any probability distribution represented on a digital computer -- quantum or classical -- even if it actually approximates an infinitely supported distribution) owing to the fact that a periodic piecewise function can be constructed to match the desired function applied to the random variable over the support, and joined up to be smooth elsewhere. Of particular importance are periodic piecewise functions that are (i) linear; and (ii) exponentially increasing in the piece corresponding to the support of the probability distribution, as these enable computation of the mean of the random variable and the mean of its exponential, as required when working in price and return space, respectively. For these statistical quantities, along with the second moment (which we include as another statistical quantity which is often of interest) the QMCI engine includes optimised periodic piecewise functions and we give an explicit analytic upper-bound on the RMSE convergence, which is demonstrated with a simple numerical example.\\

This upper-bound on QMCI convergence itself relies on a complete characterisation of the underlying QAE subroutine and, whilst the existing literature is reasonably comprehensive, we have identified a significant lacuna that requires attention before such an end-to-end QMCI engine could be proposed. Specifically, the quadratic quantum advantage in RMSE convergence of QAE algorithms masks the fact that the estimators are not statistically robust -- in that they obtain the claimed accuracy whilst being skewed and heavy-tailed. This is a potentially significant deficiency of QMCI as, even though the convergence is slower, classical MCI estimators are distributed as unbiased Gaussian random variables. To resolve the issue of statistical robustness, we propose \textit{linear combination of unitaries quantum amplitude estimation} (LCU QAE) which achieves comparable speed of convergence to existing QAE algorithms, whilst outputting an estimator which is (to an acceptably-close approximation) an unbiased Gaussian random variable.\\

\begin{figure}[t!]
\centering
\includegraphics[width=\textwidth]{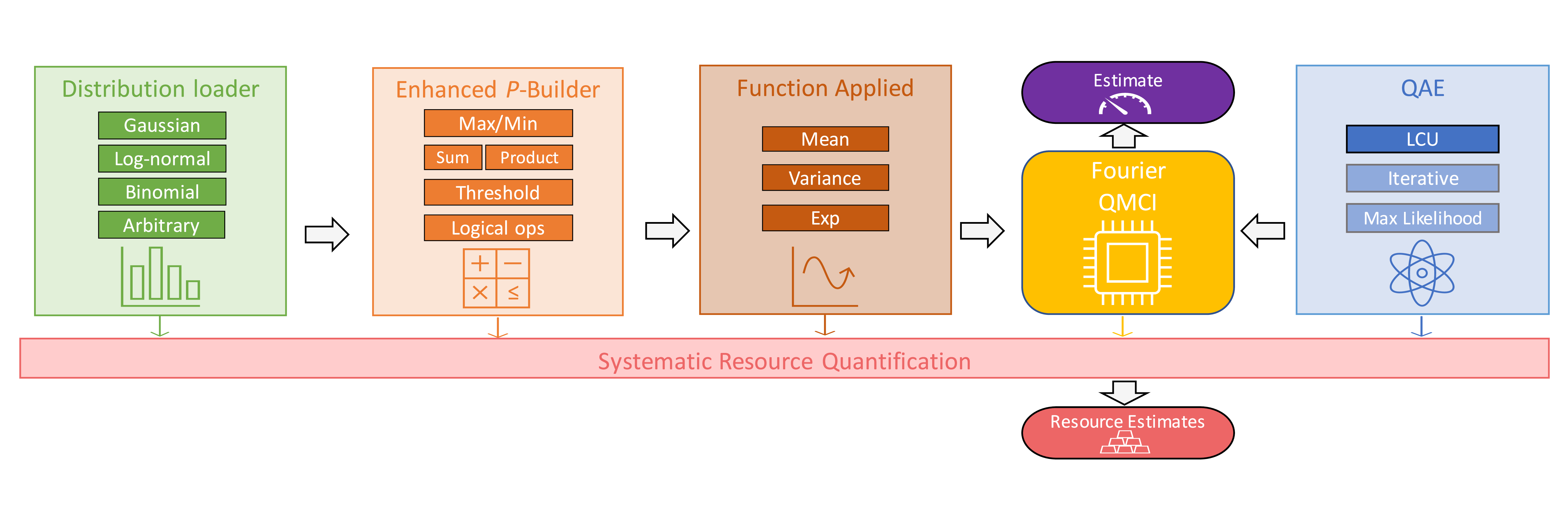}
\caption{A graphical illustration of the end-to-end QMCI engine.}
\label{fig:graphical-abstract}
\end{figure}

As well as executing quantum circuits to compute small-scale QMCI problems, the QMCI engine also includes a \textit{resource mode}, which precisely quantifies the quantum resources (numbers of qubits and gates) required to run any QMCI problem (i.e., including full-scale problems, whose size exceeds what can be executed or simulated on present-day quantum hardware and simulators). Using resource mode, we present a set of benchmarks, corresponding to simple, but not trivial, financial derivative instruments. Thus, the inclusion of resource mode enables the QMCI engine to bridge between the small-scale problems that can be executed on today's quantum hardware and simulators, and the full-scale QMCI problems that will eventually fulfil the promised useful quantum advantage. On this subject, we expect applications to enjoy useful quantum advantage if they are such that (i) MCI is the best approach classically; (ii) precision is crucial -- and it's hard to achieve adequate precision by classical means; and (iii) there is a realistic prospect of complementary quantum advantage in distribution loading. (The third of these is actually optional, and refers to the fact that, if the random process modelling the financial time-series in question can be sampled more efficiently quantumly than classically -- which we refer to as ``complementary quantum advantage in distribution loading'' -- then the overall quantum advantage may arise even sooner.) In the context of mathematical finance, we expect calculating the Greeks for financial derivatives, and portfolio risk / optimisation to be the principal beneficiaries. Fig.~\ref{fig:graphical-abstract} illustrates the core components and operation of the QMCI engine.\\

\section{Preliminaries}
\label{sec:preliminaries}

\subsection{Probability and statistics}
\label{subsect:probability-prelims}

\subsubsection{Random variables}

A random variable, $X$, is realised as some actual value according to its probability distribution. The probability distribution is typically described by its probability density function (PDF) in the case of a continuous random variable, or a probability mass function (PMF) in the case of a discrete random variable. Mathematically, given a $d$-dimensional domain $\Omega$, we define 
a PDF $f(x)$ as

\begin{equation}
    \begin{split}
        f_X: \; \Omega \subset \R^d &\to \R_+\\
        x &\mapsto f_X(x)\\
        s.t. \; \int_\Omega  f_X(x) \, & \rmd x=1
    \end{split}
\end{equation}

For our purposes, the random variables always take real numerical values, and the support of the random variable is a single interval of the real line outside of which the PDF or PMF is always zero which means that if the domain $\Omega$ is not bounded, the support of the distribution $f(x)$ will be truncated to a bounded space.\\

The probability that the random variable, $X$, takes value, $x$ is typically denoted $p(X=x)$ and, slightly overloading notation for simplicity, in the case of discrete random variables we also use $p(x)$ to denote the PMF. In the case of continuously distributed random variables, such a conflation of PDF and probability would be an abuse too far, and so instead we denote the PDF $f_X(x)$, such that in one dimension:
\begin{equation}
    p(a \leq X \leq b) = \int_a^b f_X(x) \mathrm{d}x
\end{equation}

We present in Example~\ref{ex:standard-distrib} two standard distributions commonly encountered in mathematical finance.
\begin{example}[Standard distributions]
\label{ex:standard-distrib}
Two common univariate probability distributions are:
\begin{enumerate}
    \item The Gaussian (normal) distribution $X \sim \mathcal{N}(\mu,\sigma^2)$ where
    \begin{equation}
    \label{eq:distrib-normal}
        f_X(x)=\frac{1}{\sigma \sqrt{2\pi}}e^{-\frac{1}{2}(\frac{x-\mu}{\sigma})^2}
    \end{equation}
    \item The log-normal distribution $X \sim \mathrm{Lognormal}(\mu,\sigma^2)$ where
    \begin{equation}
    \label{eq:def-lognormal-distrib}
        f_X(x)=\frac{1}{\sigma x \sqrt{2\pi}}e^{-\frac{1}{2}(\frac{\log x-\mu}{\sigma})^2}
    \end{equation}
    Equivalently, we have $\log X \sim \mathcal{N}(\mu,\sigma^2)$.\\
\end{enumerate}
\end{example}

\subsubsection{Expectation}

The expectation of a continuous random variable is defined as
\begin{equation}
    \E (X) = \int_{-\infty}^{\infty} x f_X(x)  \mathrm{d} x
\end{equation}
and the expectation of a discrete random variable is defined:
\begin{equation}
    \E (X) = \sum_{x\in \Omega} x p(x)
\end{equation}
where the sum is taken over the entire support of $x \in \Omega$. It is worth noting that expectation is a linear function, and therefore is such that should an affine transform be applied to the random variable, say $Y = aX+b$, then we have:
\begin{equation}
    \E(Y) = a \E(X) + b
\end{equation}
It turns out that this basic property is very useful when it comes to QMCI.\\

Some function, $g(\cdot)$, may be applied to the random variable prior to the expectation, in which case we have for continuous random variables:
\begin{equation}
    \E_{g(x)}(X) = \E(g(X)) = \int_{-\infty}^{\infty} g(x) f_X(x)  \mathrm{d} x,
\end{equation}
and for discrete random variables:
\begin{equation}
    \E_{g(x)}(X) = \E(g(X)) = \sum_{x\in \Omega} g(x) p(x).
\end{equation}
For instance, the (non-centralised) second moment is such that $g(X) = X^2$.

\subsubsection{Indicator functions and ``conditional expectation''}

Indicator functions are defined
\begin{equation}
    \mathds{1}_A(x) = \begin{cases} 1 & \text{ if } x \in A \\ 0 & \text{ otherwise}
    \end{cases}
\end{equation}
where $A$ is a set. For our purposes, $A$ is always the subset of the support of a random variable. We use indicator functions to ``filter out'' certain points of probability mass when computing expectations. Such calculations have the general form:
\begin{equation}
    \E_{g(x)}(X|A) = \sum_{x \in \Omega} g(x) p(x) \mathbb{1}_A(x)
\end{equation}
where $\E_{g(x)}(X|A)$ is a quantity that we (somewhat informally) refer to as the ``conditional expectation'' (given the subset of the support, $A$).

\subsubsection{Multivariate distributions and marginalisation}

Let $X$ and $Y$ be two random variables. If they are independent, then their distributions factorise, i.e.
\begin{equation}
\forall (x,y): p(X=x \, , \, Y=y) = p(X=x)p(Y=y)    
\end{equation}
 However, in general this equivalence need not hold, in which case they are dependent random variables that are realised as actual values according to a joint probability distribution. The joint probability distribution is typically described by a multivariate PDF ($f_{X,Y}(x,y)$) in the continuous case; or a multivariate PMF ($p(x,y)$) in the discrete case. The multivariate setting extends naturally to any number of dimensions, and from a joint distribution it is possible obtain the \textit{marginal} distribution of any particular random variable. For instance, if some $d$-dimensional multivariate PDF is $f_{X_1, \dots X_d}(x_1,\dots x_d)$, then the random variable pertaining to the first dimension is distributed according to the PDF obtained by \textit{marginalisation}:
\begin{equation}
    f_{X_1}(x) = \int_{x_2 \dots x_d} f_{X_1, X_2, \dots, X_d}(x, \dots , x_d) \mathrm{d}x_2 \dots \mathrm{d}x_d 
\end{equation}
and similarly with sums in place of integrals for the discrete case (note this applies to any dimension, the first has simply been selected to keep the notation light). As the marginal distribution \textit{is} the distribution of the first random variable (obtaining by averaging over all of the other random variables), we can also define the expectation:
\begin{equation}
    \E_{g(x)}(X_1) = \int g(x) f_{X_1}(x) \mathrm{d}x = \int_{x_1 \dots x_d} g(x_1) f_{X_1, \dots ,X_d}(x_1,x_2, \dots ,x_d) \mathrm{d}x_1 \dots \mathrm{d}x_d 
\end{equation}
once again, similarly with sums in place of integrals for the discrete case.

\subsubsection{Bayes' law}

PDFs and PMFs are typically \textit{parametric} in nature; as well as taking $x$ as an argument, they contain parameters that dictate the instance of some family of functions. For example, the PDF of the Gaussian distribution, given in Eq.~\eqref{eq:distrib-normal} and denoted $\mathcal{N}(\mu, \sigma^2)$, is described by two parameters:
$\mu$, the mean, and $\sigma^2$, the variance.\\

Collectively, the parameters of a PDF / PMF are often denoted $\Theta$, and one common task in applied statistics is to infer the parameter values from some data, $\mathcal{D}$. This data is generally sampled from the random process (i.e., as specified by the PDF / PMF in question) or some variant thereof (for instance, the sample with some function applied). Thus the aforementioned task requires evaluating the \textit{conditional} probability of the parameters \textit{given} the observed data denoted $p(\Theta | \mathcal{D})$. Conditional probabilities are related to joint probabilities by
\begin{equation}
    p(\Theta|\mathcal{D}) p(\mathcal{D}) = p(\Theta,\mathcal{D}) = p(\mathcal{D}|\Theta)p(\Theta)
\end{equation}
From this, \textit{Bayes' law} follows as an easy corollary
\begin{equation}
    p(\Theta | \mathcal{D}) = \frac{p(\mathcal{D}|\Theta) p(\Theta)}{p(\mathcal{D})}
\end{equation}
which is useful as the probability of the \textit{data given the parameters}, $p(\mathcal{D}|\Theta)$, is simple to ascertain (this is either the PDF / PMF itself, or some simple function thereof). In Bayesian statistics $p(\mathcal{D}|\Theta)$ is known as the likelihood (of the data), $p(\Theta)$ is the \textit{prior} belief about the parameter values, $p(\Theta | \mathcal{D})$ is known as the \textit{posterior} (which is generally what is sought), and $p(\mathcal{D})$ is treated as a normalising constant.

\subsubsection{Estimators}

The final step in the ``common task'' introduced above is to extract some estimate, $\hat{\Theta}$ of the parameters from the posterior distribution. Suppose that some random process is parameterised by a single parameter, $\theta$ (as it indeed is in QAE): one important figure of merit is the mean-squared error (MSE) of the estimator $\hat{\theta}$, (itself a random variable whose probability distribution, $p(\hat{\theta})$, depends on data and the estimator type), with respect to the true $\theta$ 
\begin{align}
\label{eq:MSE-def}
    \text{MSE} & =  \E ((\hat{\theta} - \theta)^2) = \int (\hat{\theta} - \theta)^2 p(\hat{\theta} ) \mathrm{d} \hat{\theta} \\
    \text{RMSE} & =  \sqrt{\E ((\hat{\theta} - \theta)^2)} = \sqrt{\int (\hat{\theta} - \theta)^2 p(\hat{\theta} ) \mathrm{d} \hat{\theta} } 
\end{align}
where we also formally define the RMSE.\\

By treating the MSE as the key figure of merit, one important estimator immediately follows, namely the minimum mean-squared error (MMSE) estimator. However, even though the aim is to choose a value of $\hat{\theta}$ that suppresses the MSE, this cannot be done directly from Eq.~\eqref{eq:MSE-def} as the true parameter value, $\theta$, is in general unknown (that is after all what is being estimated). Instead, the probability of $\theta$ given the empirical data is used:
\begin{equation}
    \hat{\theta}_{\text{MMSE}} = \argmin_{\hat{\theta}} \left(( \E ((\hat{\theta} - \theta)^2)\right) = \argmin_{\hat{\theta}} \left( \int (\hat{\theta} - \theta)^2 p(\theta | \mathcal{D}) \mathrm{d} \theta \right)
\end{equation}
and similarly with a sum in place of the integral for the discrete case.\\ 

Another important estimator is the \textit{maximum a-posteriori} (MAP) estimator:
\begin{equation}
    \hat{\theta}_{\text{MAP}} = \argmax_{\hat{\theta}}(p(\hat{\theta} | \mathcal{D}))
\end{equation}
In the case of a uniform prior, the MAP estimator is equal to the \textit{maximum likelihood estimator}:
\begin{equation}
    \hat{\theta}_{\text{MLE}} = \argmax_{\hat{\theta}}(p(\mathcal{D} |\hat{\theta} ))
\end{equation}
In each of these cases (MAP estimator and MLE), the PDF / PMF is supported on a range of $\theta$.

\subsubsection{Monte Carlo integration}

Suppose that there is some random process which we can sample (termed ''\textit{sample access}''), but for which we cannot necessarily query the probability mass / probability density at any point in the support (termed ``\textit{query access}'') -- then one way to estimate the mean of the distribution is by MCI: 
\begin{equation}
    \E(X) \approx \frac{1}{q} \sum_{i=1}^q X_i
\end{equation}
where $X_i$ are independent, identically distributed (\iid) samples from the random process (of which there are $q$ in total). Furthermore, this also applies to the case where a function is applied to the random variables:
\begin{equation}
    \E_{g(x)}(X) \approx \frac{1}{q} \sum_{i=1}^q g(X_i)
\end{equation}
If we let $Y$ be the random variable $Y = g(X)$ (with special case $g(X) = X$ such that $Y=X$), then we can express the convergence of the estimator for the mean, $\hat{\mu}_Y$, as approximated by MCI, in terms of RMSE:\footnote{Note that in the case that the samples are not \iid, a similar expression with a different exponent of $q$ may hold. Note also that for some (heavy-tailed) distributions the standard deviation could be infinite, in which case this is not a meaningful estimator.}
\begin{equation}
    \sqrt{\E ( (\hat{\mu}_Y - \E(Y))^2 )} = \frac{\sigma_Y}{\sqrt{q}}
\end{equation}
where $\sigma_Y$ is the standard deviation of $Y$. If we further let $\max(Y)$ and $\min(Y)$ be the maximum and minimum points of non-zero probability density / mass for the random variable $Y$ (i.e., the extrema of its support), then we further get:
\begin{equation}
\label{eqn:classicalMCI-convergence}
    \sqrt{\E ( (\hat{\mu}_Y - \E(Y))^2 )} \leq \frac{\max(Y) - \min(Y)}{2\sqrt{q}}
\end{equation}
with the bound saturated when half of the probability mass is concentrated at each of the extrema. 

\subsubsection{Implicitly-defined random variables and marginal Monte Carlo integrals}
\label{subsubsect:impl-defined}

Suppose some random variable, $Y$, is defined as a function of other random variables, $\{X_i \}$, and it is not possible to analytically express the PDF / PMF thereof. In this case, we have a joint distribution over $Y \cap \{X_i \}$, and even if the desired statistical quantity is some expectation only of $Y$, in general (for sufficiently large number of random variables, i.e., large $|\{X_i \}|$) the most efficient method will still be to sample from the joint distribution and estimate the expectation of $Y$ as a marginal variable accordingly. This is a manifestation of the essential property that MCI is the most efficient means of estimating expectations of high-dimensional random processes, and for applications of QMCI it is of paramount importance to appreciate that this essential principle holds even when only the marginal expectation of a single, implicitly-defined random variable, is desired.

\subsection{Quantum states as probability distributions}
\label{subsect:qstates-as-prob}

In this section we assume that readers are familiar with the essentials of quantum computation\footnote{We recommend Ref.~\cite{nielsenchuang2010} if this is not the case.}, and focus in on the specifics of how quantum circuits are used to represent statistical quantities.

\subsubsection{Qubit bundles as registers}

Suppose we have a computational basis state over some $n$ qubits, then by treating the first qubit as the most significant bit, we can read the computational basis state, $\ket{b_1 b_2 \dots b_n}$, as the bitstring $b_1 \dots b_n$. We refer to blocks of qubits that we treat as representing some such binary values as \textit{registers} and, in the absence of any additional information, treat $x$ as a binary value. 

\subsubsection{Quantum states as probability distributions over binary values}

If we have some superposition of $n$ qubits (for arbitrary relative phases, $\varphi_i$)
\begin{equation}
\label{eq:def-prob-dist-state}
    \ket{p} = \sum_{ x_i \in \{0,1\}^n} e^{2 \pi \I  \varphi_i} \sqrt{p_{i}} \ket{x_i}
\end{equation}
then measurement in the computational basis will have the effect of sampling a random variable distributed such that the binary value $x_i$ occurs with probability $p_i$. The nomenclature ``$\ket{p}$'' is deliberately suggestive of probability, and we denote a circuit that prepares this state from the all-0 state $P$, that is: $\ket{p} = P \ket{0^n}$. Building a circuit $P$ to efficiently encode financial random processes is one of the central topics we address in this paper, in particular we propose our enhanced $P$-builder in Section~\ref{sec:enhanced-data-loader}.

\subsubsection{Quantum states as multivariate probability distributions over real numbers}

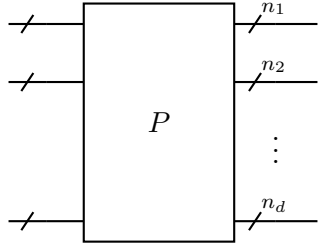
\begin{figure}[t!]
\centering
\begin{quantikz}
   & \qwbundle{} & \gate[wires=4, nwires={3}][2cm]{P} & \qwbundle{n_{1}} & \qw \\
    & \qwbundle{} & & \qwbundle{n_{2}} & \qw \\
    & & & \vdots & \\
    & \qwbundle{} & & \qwbundle{n_d} & \qw
\end{quantikz}
    \caption{An illustration of how a general circuit, $P$, can be interpreted as loading a multivariate probability distribution as a quantum state.}
    \label{fig:simple-P}
\end{figure}

In general, we do not wish simply to sample binary values, but real numbers corresponding to (regularly spaced) discrete points on the real line. Furthermore, as MCI is only the best solution (classically) for high dimensional integrals, it is necessary to use quantum states to encode multivariate distributions. Both of these can easily be achieved, taking a lead from the notation in Ref.~\cite{herbert1}, any quantum circuit with some $n$ qubits may be interpreted as preparing a $d$-dimensional multivariate distribution where each dimension is represented by $n_i$ qubits, also called wires hereafter, where $i$ indexes the dimension, and is such that $n = \sum_{i=1}^d n_i$. Furthermore, for each dimension, the binary numbers supported are interpreted as real numbers starting at some position $x_{\ell}^{(i)}$ and with equal spacing $\Delta^{(i)}$. This is sketched in Fig.~\ref{fig:simple-P}.

\subsubsection{Implicit marginalisation}
\label{subsubsect:implicit-marg}

The Born rule for evaluating the measurement outcome probabilities when a subset of the qubits are measured in the computational basis is such that marginalisation occurs implicitly. That is, if the qubits corresponding to the $i^{th}$ dimension are measured in the computational basis, and all of the other qubits are left alone, then the measurement outcomes will sample from the marginal distribution of the $i^{th}$ random variable. This is critical, as the settings we consider for QMCI are those covered in Section~\ref{subsubsect:impl-defined}, where only the marginal expectation is of interest.

\subsection{Quantum Monte Carlo integration}
\label{subsect:QMCI-intro}

Suppose we wish to perform (marginal) MCI for some implicitly-defined random variable, $X$, where sample access to the corresponding joint distribution has been encoded in a circuit, $P$. We let $d$ be the number of dimensions in the joint distribution, and to simplify the following notation and diagrams, we let the dimension in question be the final, $d^{th}$, dimension. We now perform an affine transformation\footnote{Here we assume that the desired expectation is $\E(X)$ and any applied functions (other than $g(X) = X$) have been included in the circuit $P$.} on the support of the $d^{(th)}$ dimension,
\begin{equation}
    \tilde{X}^{(d)} = a X^{(d)} + b
\end{equation}
such that the support of $\tilde{X}^{(d)}$ is $[0,1]$, which simply amounts to changing the description of how samples from the $d^{th}$ dimension are interpreted as real numbers (we later perform the inverse affine transform in classical post-processing to recover the mean of the actual random variable with this shifting and re-scaling). \\

\begin{figure}[t!]
    \centering
    \begin{quantikz}
   & \qwbundle{} & \gate[wires=4, nwires={2}][1cm]{P} & \qwbundle{} & \qw \rstick[wires=3]{$x^{(1)} \ldots  x^{(d-1)}$} & & & & &\\
   & & & \vdots & & & & & & &\\
    & \qwbundle{} & & \qwbundle{} & \qw & & & & & \\
    & \qwbundle{} & & \qwbundle{} & \gate[wires=5, nwires={4}][1cm]{R} & \qwbundle{} \rstick[wires=1]{$\tilde{x}^{(d)}$} & & & &\\
     & & \lstick[wires=4]{0} & \qw & \qw & \qw &  \qw & \qw & \qw & \ctrl{4} & \qw \rstick[wires=4]{$\text{arcsin}\sqrt{\tilde{x}^{(d)}}$} \\
    & & & \qw & \qw & \qw & \qw & \qw & \ctrl{3} & \qw & \qw \\
    & & & \vdots  & &  & \iddots & & & &\\
    & & & \qw & \qw & \ctrl{1} & \qw & \qw & \qw & \qw & \qw \\
    & & & \qw & \qw & \gate{R_y(\phi_1)} & \qw \hspace{0.5cm} \dots &  & \gate{R_y(\phi_{n_{d-1}})} & \gate{R_y(\phi_{n_d})} & \qw \rstick[wires=1]{$\sqrt{\tilde{x}^{(d)}}$} \\
\end{quantikz}
    \caption{An illustration of how a probability distribution loading circuit, $P$, can be supplemented with a quantum arithmetic circuit, $R$, such that an expectation value of interest is encoded in the amplitude of a qubit. In order for the bank of controlled rotations to implement the sine, angles $\{ \phi_1, \dots ,\phi_{n_d}\}$ will be some constant multiplied by a 2 raised to incrementally increasing (integer) powers. Note that for illustrative purposes we have used a mixed convention in the qubit labelling in this figure: $x^{(1)} \dots x^{(d-1)}$, $\tilde{x}^{(d)}$, the label $0$ to the left of $R$ and $\arcsin \sqrt{\tilde{x}^{(d)}}$ all represent values encoded in binary registers, whereas $\sqrt{\tilde{x}^{(d)}}$ on the bottom-most qubit represent the amplitude of this qubit (such that the outcome one is measured with probability $\tilde{x}^{(d)}$.
    }
    \label{fig:QMCI-noFourier}
\end{figure}
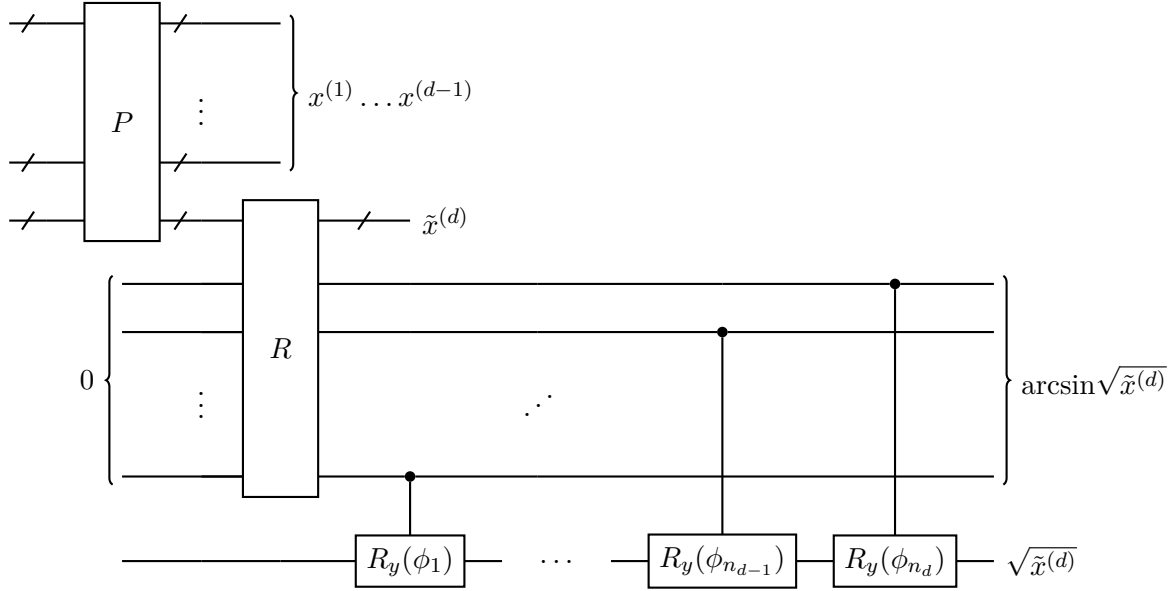

A circuit, denoted $R$ and termed the ``quantum arithmetic circuit'', is then executed to (approximately) reversibly apply the function $\arcsin (\sqrt{\tilde{x}^{(d)}})$, such that the result is placed in a new qubit register (of appropriate size). This register is then used to control a bank of $R_y$ rotation gates, such that the state\footnote{This bank of controlled rotation gates is of the same form as that shown in Ref.~\cite[Fig. 1]{herbert1} -- that is, such that the rotation angle is equal to the value in the register.}
\begin{equation}
    \cos (\arcsin \sqrt{\tilde{x}^{(d)}}) \ket{\Phi_0}\ket{0} + \sin (\arcsin \sqrt{\tilde{x}^{(d)}}) \ket{\Phi_1} \ket{1} = \sqrt{1- \tilde{x}^{(d)}} \ket{\Phi_0}\ket{0} + \sqrt{\tilde{x}^{(d)}}\ket{\Phi_1}\ket{1}
\end{equation}
is prepared. Here $\ket{\Phi_0}$ and $\ket{\Phi_1}$ are quantum states whose exact expression is not of concern (but, slightly departing from the usual convention, they are each considered to have an explicit global phase -- otherwise this expression is not sufficiently general). In general, this will actually happen for a superposition of computational basis states, as dictated by the circuit, $P$, which thus gives:
\begin{equation}
    \ket{\psi} = \sum_i \sqrt{p_i} \left( \sqrt{1- \tilde{x}_i^{(d)}} \ket{\Phi_0}\ket{0} + \sqrt{\tilde{x}_i^{(d)}}\ket{\Phi_1}\ket{1} \right)
\end{equation}
and thus, by the Born rule, the probability of measuring 1 on the final qubit is
\begin{equation}
   \sum_i \left(\sqrt{p_i \tilde{x}_i^{(d)}}\right)^2 = \E(\tilde{x}^{(d)})
\end{equation}
as desired. QAE can then be used to estimate this quantity, with RMSE convergence proportional to inverse of the number of samples (number of uses of the circuit $P$). Defining $\hat{\mu}$ as the estimate,  $\cqae$ as the constant of proportionality, and applying the inverse affine transform, we get overall convergence: 
\begin{equation}
\label{eq:QMCIconvergence-prelim}
    \text{RMSE} \leq \cqae  \frac{x^{(d)}_u - x^{(d)}_{\ell}}{q}
\end{equation}
where $x^{(d)}_u$ is the largest value in the support of the $d^{th}$ random variable (this represents the scaling of the error, when the original affine transform to map the support to $[0,1]$ is reversed). Equation~\eqref{eq:QMCIconvergence-prelim} therefore represents a quadratic advantage over classical MCI (i.e., in Eq.~\eqref{eqn:classicalMCI-convergence}). A sketch of how the circuits $P$ and $R$ are composed is given in Fig.~\ref{fig:QMCI-noFourier}.

\subsection{Basics of mathematical finance}

Mathematical finance is a rich and wide-ranging field in its own right, and so it is therefore only possible to address a relatively simple abstraction -- even in a paper such as this, which primarily focuses on applications of QMCI to mathematical finance. For instance, we largely gloss over the important general distinction between the ``P'' and ``Q'' worlds. The latter includes the pricing of derivative instruments which covers most of the examples we include -- although we do mention how the same computational methods may be applied to risk and portfolio management calculations (which belong to the P world). This conflation of the two worlds is perhaps made a little more palatable (and mathematically sound) by the fact that we set the risk-free rate to zero (for simplicity) in the examples and benchmarks we give\footnote{Setting the risk-free rate to zero is a simplification that we make for convenience, and does not limit the applicability -- or quantum advantage -- of the techniques we use in this paper.}. Even within the Q world, our examples are somewhat specialised, and we essentially focus on equities -- even though the methods apply equally to other assets such as FX instruments. For the reader who wishes to delve deeper into mathematical finance, we recommend Ref.~\cite{Wilmott07}.

\subsubsection{Assets}

An asset is a resource with economic value, such as a stock in a listed company. For our purposes, we will assume that all assets have a well-defined price (given by some market) which varies with time.

\subsubsection{Portfolios}

A portfolio is a set of one or more assets, which in general may be correlated. For theoretical purposes a portfolio is any set of assets, whether or not it represents a sensible real-world example thereof.

\subsubsection{Financial instruments: derivatives  \& options}

Rather than simply buying an asset, one may purchase a \textit{derivative instrument} (or simply ``derivative'') such as an option, where one has the option but not obligation to buy (termed a ``call option'') or sell (termed a ``put option'')  a certain asset (termed the ``underlying'') on a certain date at a pre-defined price (termed the ``strike price''). Such an option is clearly advantageous for the holder as it need only be exercised if the market conditions are favourable -- and so it follows that options are themselves assets, and indeed the ``underlying'' in an option contract need not be a stock, but could itself be some derivative instrument\footnote{Derivatives of FX and interest rate instruments are also traded -- but this distinction is not required in what follows.}.

\subsubsection{Pricing derivatives}

When two counterparties enter into a derivative they need to agree on a price. According to basic theory, the price they should (optimally) agree on is known as the ``fair value'' which depends on the expected trajectory of the price of the underlying and the strike price (if applicable); for instance a call option with a low strike price on an asset whose value is expected to increase should be relatively expensive to purchase. Option prices can be calculated from the expected payoff. As an example, consider a European call option with strike price, $S$; in this case the payoff is
\begin{equation}
   \text{Payoff} \, = \max(X - S,0)
\end{equation}
where $X$ is a random variable modelling the price of the underlying at the expiry time of the contract. Typically the \textit{expected} payoff -- averaging over $X$ -- is what needs to be calculated (or estimated) to price a derivative. Here the ``max'' condition captures the fact that the option will only be exercised if the underlying has reached a price exceeding the strike price, otherwise the option will not be exercised and the payoff will be zero.

\subsubsection{Path dependency}

There are myriad types of financial derivatives, however an important distinction is whether a derivative is \textit{path independent}, where the expected payoff only relies on the underlying's price at the derivative's expiry time; or \textit{path dependent}, where the expected payoff depends on the price of the underlying throughout the lifetime of the derivative (and thus depending on the ``path'' that the asset price has taken).\\

Discretised versions of path-dependent options are high-dimensional random processes (represented by multivariate probability distributions with a dimension for each time-slice), and it follows that MCI (and variants thereof) is usually the most effective means of computing the expected payoff classically, and hence there is an opportunity for quantum advantage by using QMCI.\\

Conversely, path-independent options are generally low-dimensional random processes, and hence no such opportunity exists -- MCI is unlikely to be the best approach classically, unless high-dimensionality is introduced in another way, such as in basket options where the underlying is a group of (non-trivially correlated) assets.

\subsubsection{Price and return space}

In mathematical finance, estimates are made given statistical models for how the price of an asset or a portfolio will evolve in the future. If this is modelled directly, this is termed working in ``price space''; however, in practise it is often beneficial to work with the (natural) logarithm of the price; this is termed working in ``return space''.

\subsubsection{Volatility}

How much an asset's price is expected to vary is measured by its volatility. For the purposes of this paper, we need only consider a simple textbook model where the price varies according to geometric Brownian motion, and hence in return space the price is modelled by a Gaussian random variable. The \textit{total volatility} is then the standard deviation of this Gaussian random variable at the final time in the time series.

\subsubsection{Use of Monte Carlo integration in mathematical finance}

MCI is generally the most efficient means of numerically evaluating \textit{high-dimensional} integrals, other methods such as \textit{Quasi-Monte Carlo} \cite{caflisch1998} suffer the ``curse of dimensionality'' meaning that the computational cost grows exponentially with the number of dimensions. Alternative approaches such as solving stochastic differential equations (and so bypassing numerical integration altogether) also only work well in the low-dimensional setting.\\

As already mentioned, within mathematical finance, path-dependent derivatives are typically modelled as high-dimensional random processes, as are portfolios of correlated assets (whose risk may need to be MCI in this paper -- as it is general purpose -- there exist a host of techniques that have been tailored to various settings, and may represent (possibly marginal) classical improvements for some of the examples we give, see e.g. Refs~\cite{MCI-1, MCI-2, MCI-3}.\\

\section{Summary of our previous works}
\label{sec:state-of-the-art}

\subsection{The problem with Grover-Rudolph state preparation}
\label{subsec:Grover-Rodolph}

In the overview of QMCI in Section~\ref{subsect:QMCI-intro}, the number of classical samples is equated with the number of uses of the quantum circuit encoding the sample access, $P$. This equivalence assumes that whenever it is possible to classically sample, then a corresponding circuit $P$ can be constructed. The \textit{Grover-Rudolph} method of state preparation has frequently been cited as evidence of this equivalence \cite{grover2002creating}, however in an earlier work we showed that, owing to the reliance on (classical) MCI in the Grover-Rudolph method, this represents a false economy \cite{herbert2021problem}. Specifically, as the desired accuracy of the estimator increases, so does the circuit depth of the Grover-Rudolph method, meaning that the overall convergence is no better than classical MCI. For this reason, we do not rely on Grover-Rudolph as a means to deliver quantum advantage in our QMCI engine.

\subsection{Sampling with reversible circuits}
\label{subsec:SamplingReversible}

Whilst the Grover-Rudolph method was shown to be inadequate as a state preparation method for QMCI, we showed that simply encoding the classical sampling circuit itself as a reversible circuit \textit{is} sufficient to prove the equivalence between classical samples and quantum uses of $P$ \cite{herbert2} To do so, we noted that classical samples are obtained by mapping uniform (pseudo) randomness to samples from the required distribution, and this ``classical sampling circuit'' can be used directly (in its reversible form) as the circuit $P$. As this construction leaves any intermediate results as further dimensions of a joint distribution, its utility relies directly on the implicit marginalisation, discussed in Section~\ref{subsubsect:implicit-marg}. Encoding probability distributions into quantum states using reversible circuits constitutes one of the methods mentioned in Section~\ref{sec:data-loaders}.

\subsection{Fourier quantum Monte Carlo integration}
\label{subsec:FourierQMCI}

The seminal observation amongst our earlier works is that the Monte Carlo integral can always be decomposed as a Fourier series which means that the computationally-expensive quantum sub-circuit, $R$ (as in Section~\ref{subsect:QMCI-intro}), which performs a quantum arithmetic operation using quantum operations can be omitted and furthermore, that the full quadratic quantum advantage can still be achieved in this way\footnote{The subject invention described in the article is US patent pending and titled "Quantum Computing System and Method" with Publication Number US2023/0036827.}.\\

The central idea is that, any function, $g(.)$ (including the simple case $g(x) = x$) applied to a random variable can be altered outside of the support of the random variable, without affecting the expectation value. So it follows that a periodic piecewise function, $\mathrm{g}(x)$, can be constructed such that $\mathrm{g}(x) = g(x)$ within the support of $x$ and elsewhere adheres to certain smoothness conditions\footnote{Note that $f(.)$ and $\mathrm{f}(.)$ were used in place of $g(.)$ and $\mathrm{g}(.)$ in Ref.~\cite{herbert1}.}. As periodic functions always have a Fourier series, it follows that the Monte Carlo integral can be evaluated by performing a weighted sum of terms of the form $\E_{\sin(m \omega x)} (x)$ and $\E_{\cos(m \omega x)} (x)$ and that such quantities can be calculated ``naturally'' in a quantum circuit by simply applying the bank of controlled rotation gates to the random variable in question. To see how this works in more detail, we begin by formally and generally defining a QAE algorithm:
\begin{definition}[Quantum amplitude estimation algorithm]
\label{def1}
A QAE algorithm takes as an input an $n+1$ qubit circuit $A$ where $A \ket{0^{n+1}} \! = \! \cos \theta \ket{\Phi_0} \! \ket{0} + \sin \theta \! \ket{\Phi_1} \ket{1}$ (where $\ket{\Phi_0}$ and $\ket{\Phi_1}$ are arbitrary $n$-qubit states), and returns an estimate, $\hat{a}$, of $  a= \sin^2 \theta$. The allowed number of uses, denoted ``$q$'', of the circuit $A$ is given as an input to the QAE algorithm. 
\end{definition}
\noindent From this definition, we can write any QAE algorithm as a function in pseudocode form:
\begin{equation}
\hat{a} = \texttt{QAE}(A, q)
\end{equation}
such that convergence of any QAE algorithm can be expressed as:
\begin{equation}
\label{qae-converge}
\text{RMSE} \leq \cqae q^{-\lambda/2}
\end{equation}
for some constant $\cqae$, and where $1 \leq \lambda \leq 2$.\\

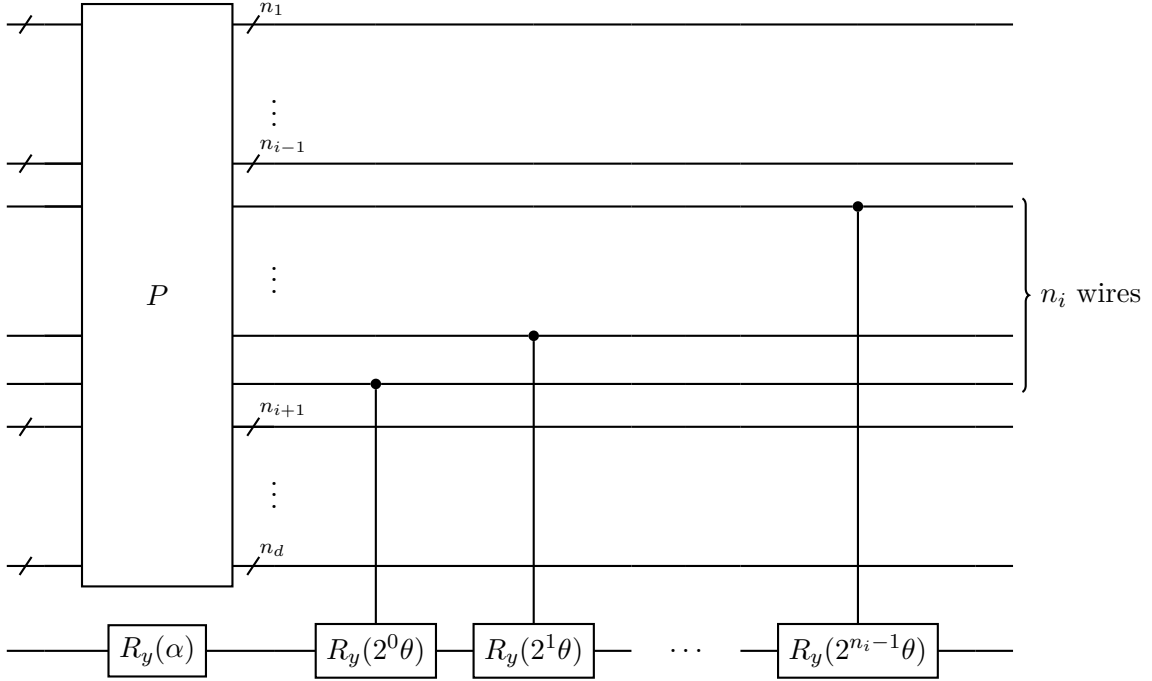
\begin{figure}[!t]
\centering
\begin{quantikz}
       & \qwbundle{} & \gate[wires=10, nwires={2,5,9}][2cm]{P} & \qwbundle{n_1} & \qw & \qw & \qw & \qw & \qw & \qw & \qw & \qw \\
       & & & \vdots & & & &  & & & &\\
      & \qwbundle{} & \qw & \qwbundle{n_{i-1}} &  \qw & \qw & \qw & \qw & \qw & \qw & \qw & \qw \\
      & \qw & \qw & \qw &  \qw & \qw & \qw & \qw & \qw & \ctrl{7} & \qw & \qw \rstick[wires=4]{$n_i$ wires}\\
       & & & \vdots  &  &  &  &  & & & &   \\
      & \qw & \qw & \qw & \qw &  \ctrl{5} & \qw & \qw & \qw & \qw & \qw & \qw   \\
      & \qw & \qw & \qw & \ctrl{4} &  \qw & \qw & \qw & \qw & \qw & \qw & \qw \\
     & \qwbundle{} & & \qw \qwbundle{n_{i+1}} & \qw  & \qw &  \qw & \qw & \qw & \qw & \qw & \qw \\
       & & & \vdots & & & & & & & & \\
     & \qwbundle{} & \qw & \qwbundle{n_{d}} & \qw & \qw & \qw & \qw & \qw & \qw & \qw & \qw \\
     \qw & \qw & \gate{R_y(\alpha)} & \qw  & \gate{R_y(2^0\theta)} & \gate{R_y(2^1\theta)} & \qw & \dots & & \gate{R_y(2^{n_i-1}\theta)} & \qw & \qw \\
    \end{quantikz}
\caption{Circuit diagram of $A(P, i, \beta, m, \omega)$, where $\alpha = m \omega x_\ell - \beta$ and $\theta = m \omega \Delta $, in which $x_\ell$ and $\Delta$ are the first point of probability mass and spacing (as defined in Section~\ref{subsect:qstates-as-prob}) for the $i^{th}$ dimension of $P$.}
\label{fig:FourierP}
\end{figure}

Fourier QMCI uses parameterised circuits of the form, $A(P, i, \beta, m, \omega)$, where:
\begin{itemize}
    \item $P$ is a circuit that loads a multivariate probability distribution;
    \item $i$ is a positive integer specifying the index of the dimension for which the marginal expectation will be estimated;
    \item $\beta$ is either 0 or $\pi / 2$, and enables respectively either the cosine or sine of the value in the qubit register to be computed;
    \item $m$ is a positive integer that corresponds to the index of a Fourier series harmonic\footnote{In Ref.~\cite{herbert1} this was $n$, and has been changed as $n$ is used throughout this paper to denote the number of qubits.};
    \item $\omega$ is the angular frequency of a periodic piecewise function that applies a desired function over the support of the random variable.
\end{itemize}
 Figure~\ref{fig:FourierP} details how the circuit $A$ is built from these parameters as well as: the value of the first point of non-zero probability mass, $x_{\ell}$, and spacing between points of probability mass, $\Delta$, for the dimension of interest. From this, we get:
\begin{proposition}\cite[Proposition 2]{herbert1}
\label{lem1}
Let $\mathtt{QAE}(A, q)$ be a QAE algorithm in the sense of Definition~\ref{def1}, which has RMSE convergence parameterised by $\lambda$ as in Eq.~\eqref{qae-converge}.
\begin{enumerate}[label=\roman*., leftmargin=*]
\item $1 - 2\mathtt{QAE}(A(P, i, 0, m, \omega),q)$ is an estimate of:
\begin{equation}
\sum_{x^{(i)} \in \Omega} p(x^{(i)})  \cos (m \omega x^{(i)} )
\end{equation}
with RMSE $\leq 2 \cqae q^{-\lambda/2}$.\\
\item $1 - 2\mathtt{QAE}(A(P, i, \pi/2, m, \omega),q)$ is an estimate of:
\begin{equation}
\sum_{x^{(i)} \in \Omega} p(x^{(i)})  \sin (m \omega x^{(i)}) 
\end{equation}
with RMSE $\leq  2\cqae q^{-\lambda/2}$.
\end{enumerate}
where $\Omega$ is the set over which the variable $x$ has been discretised\footnote{Note that we take the discretised random variable as the defined input and so are not concerned here with the work needed to actually do the discretisation.}. 
\end{proposition}

This proposition then allows us to bypass the circuit $R$: first we build a periodic piecewise function $\mathrm{g}(x)$ such that $\mathrm{g}(x) = g(x)$ over the support of $p(x)$. For example, some $\mathrm{g}(x)$ of the following form, which repeats with period $x_{\tilde{u}} - x_\ell$, is suitable in general:
\begin{equation}
\mathrm{g}(x) = \begin{cases} g(x)  &\mbox{if } x_\ell \leq x < x_u  \\
\tilde{g}(x) & \mbox{if } x_u \leq x < x_{\tilde{u}}  
\end{cases}
\end{equation}
where $x_{\tilde{u}} \geq x_u$ and $\tilde{g}(x)$ is itself sufficiently smooth and chosen such that the pieces join sufficiently smoothly. As $\mathrm{g}(x)$ is periodic, it has a Fourier series:
\begin{equation}
\mathrm{g}(x) = a_0 + \sum_{m=1}^{\infty} a_m \cos (m \omega x) + b_m \sin (m \omega x)
\end{equation}
where $\omega = 2 \pi / T$ and $T$ is the period of the periodic piecewise function.\\

We define $\mu$ as the expectation value we wish to approximate, and as $\mathrm{g}(x) = g(x)$ whenever $p(x^{(i)}) \neq 0$, we can express:
\begin{align}
    \mu = & \sum_{x^{(i)} \in \Omega} p(x^{(i)}) g(x^{(i)}) \nonumber \\
      = & \sum_{x^{(i)} \in \Omega} p(x^{(i)}) \mathrm{g}(x^{(i)}) \nonumber \\
     = & \sum_{x^{(i)} \in \Omega} p(x^{(i)}) \Bigg(  \sum_{m=1}^\infty  \left( a_m \cos (m \omega x^{(i)}) + b_m \sin (m \omega x^{(i)}) \right) + a_0 \Bigg) \nonumber \\
     \label{eqeq170}
     = & a_0 + \sum_{m=1}^\infty a_m \, \left( \sum_{x^{(i)}\in \Omega}  p(x^{(i)})  \cos( m \omega x^{(i)}) \right)     + b_m  \, \left( \sum_{x^{(i)}\in \Omega}  p(x^{(i)})  \sin ( m \omega x^{(i)}) \right)
\end{align}
To estimate $\mu$, we can thus estimate each of the parenthesised sums in the final line of \eqref{eqeq170} individually using the result of Proposition~\ref{lem1}. Upon this, the central result is based:
\begin{theorem}\cite[Theorem~3]{herbert1}.
The quantity, $\mu = \mathbb{E}(g(X))$, where $X \sim p(x^{(i)})$ and $g$ is a function that is continuous in value and first derivative and whose second and third derivatives are piecewise-continuous and bounded, can be estimated with RMSE $\leq \tilde{c}_g q^{-\lambda/2}$, where $q$ is the number of uses of a circuit preparing the quantum state $\ket{p}$, $\tilde{c}_g$ is a constant that depends on $\mathrm{g}(\cdot)$ and $\cqae$, and $\lambda$ is the convergence rate of some QAE subroutine (i.e., as defined in Eq.~\eqref{qae-converge}) which operates on circuits of the form defined in Fig.~\ref{fig:FourierP}.
\end{theorem}

Here, the smoothness conditions suffice to guarantee that the quadratic quantum advantage is always retained. This Fourier series decomposition approach to QMCI is extremely flexible in terms of the functions that can be applied, and a key element of the design of the QMCI engine is the construction of suitable periodic piecewise functions for statistical quantities of interest.

\subsection{Noise-aware quantum amplitude estimation}
\label{subsec:NoiseQAE}

QAE circuits -- in particular circuits for QPE-free QAE -- all have the same regular structure. They all consist of repeated instances of the same sub-circuit, $Q$, followed by the measurement of a single qubit. In our previous work, we showed how these properties can be exploited to model the inclement noise as a Gaussian random variable, regardless of the actual underlying noise characteristics -- and furthermore, that such a random variable can be accounted for as if it were an additional component of variation in the parameter estimation \cite{herbrolng21}. This we named \textit{noise-aware QAE}, and we showed how the noise parameter could be co-learned alongside the parameter (amplitude of interest), such that the additional component of variation introduced by the hardware noise can be exactly compensated for with additional shots of the circuit.\\

For practical purposes, one way to think about noise-aware QAE is as a means of application-specific error-mitigation, which allows the maximum executable circuit depth to be extended by additional sampling. For instance, observe that a simple depolarising noise model for the machine IBMQ \textit{Rome} running circuit $A_1$ has $\tilde{p}_{coh} = 0.9276$ according to Ref.~\cite[Table~1]{herbrolng21}, and a simple calculation therefore shows that (according to definition of $\tilde{p}_{coh}$ in Ref.~\cite[Eq.~(6)]{herbrolng21}) after $1/(1-0.9276) \approx 14$ Grover iterates, it is expected that the state has decohered. However, in Ref.~\cite[Fig.~3 \textit{Rome}]{herbrolng21} it can be seen that there is still a significantly higher than 0.5 probability of measuring 1 (as would be the case for a fully decohered state) when there are many more Grover iterates than 14. Noise-aware QAE therefore effectively allows the ``signal-to-noise ratio'' to be amplified in such cases in a manner that naturally combines with the outputted amplitude estimate. As such, for experimental purposes we can leverage noise-aware QAE to run deeper QMCI circuits than one may expect from simple ``rules of thumb''.
\newpage
\part{Principal novel methods}
\label{part:key-new-techniques}

\section{Enhanced $P$-builder}
\label{sec:enhanced-data-loader}

In principle it is always possible to offload the entire work of building the circuit $P$ (as defined in Section~\ref{subsect:qstates-as-prob}) onto the user (that is, they should either construct and submit their own circuit, or call a circuit from the standard library of distributions, as detailed in Section~\ref{sec:data-loaders}). However, our central philosophy is that the QMCI engine should allow users to build complex financial instruments from simple building blocks, and to this end, the circuit $P$ can be \textit{enhanced} by the following:
\begin{enumerate}
\item Taking simple functions of the random variables sampled from one or two dimensions of a multivariate probability distribution to define new random variables (which can therefore be represented in a multivariate distribution with the number of dimensions increased accordingly). One such function is ``summing'', and as it is easy to keep track of the number of random variables being summed, this operation also encompasses averaging, which is needed, for example, in pricing Asian options. 
\item Comparing the value of some random variable (i.e., the value sampled from some dimension of the multivariate probability distribution) to (i) that from another dimension or; (ii) some pre-defined threshold. These comparisons are set up as Boolean statements, and \textit{indicator} qubits are then added to represent the truth of these statements. Thresholds are required in, for example, pricing barrier options.
\item Indicator qubits may be combined in further Boolean expressions to give additional indicator qubits. For financial instruments whose payoff relies in a non-trivial way on various thresholds / comparisons, this allows the payoff to be conditioned on any logical function of these.
\end{enumerate}

In Section~\ref{subsect:ex-look-back} we give an explicit example of how some of these functions are used to construct \textit{look-back options}; and in Table~\ref{tab:all-calcs1} we sketch how these operations can be used to compute the payoff for some of the most widely-used financial derivatives, as well as enabling common portfolio risk metrics to be calculated\footnote{The enhanced $P$-builder is the subject of patent applications GB2210997.9 and GB2301481.4.}.

\subsection{Enhanced $P$-builder functions}

From user-specified commands, the enhanced $P$-builder automatically constructs efficient reversible circuits that accomplish the required function. In particular, by making mild (and practically reasonable) restrictions about the way that the binary registers encode numerical values, it is possible to use (reversible forms of) binary operations directly, hence avoiding the costly full arithmetic circuitry. Table~\ref{tab:enhanced-operations-1} summarises the functions that generate new random variables from existing ones (that is, those functions pertaining to the first item in the preceding enumeration). In each case the function takes either one or two random variables as its input, and it is worth noting that there is no loss in generality in this, as all of the operations are associative, and so it is trivial to extend to higher numbers of random variables by repetition.\\

The second and third items in the preceding enumeration may be grouped as \textit{logical} rather than \textit{binary} operations, and prepare single qubits whose truth value is determined by (and therefore correlated with) the values in the registers representing the relevant dimensions of the multivariate probability distribution. These qubits therefore encode Bernoulli random variables when measured in the computational basis, whose probability is equal to the probability that the encoded logical statement is true. Table~\ref{tab:enhanced-operations-2} summarises the logical operations included in the enhanced $P$-builder. In the case where the user wishes to add a further indicator qubit as a function of the existing indicator qubits, then this logical expression must be explicitly given in the form of an exclusive sum of products (ESOP - see \cite{mishchenko2001fast} and references therein) of the input indicator qubits (and/or their negations). There is no loss of generality in so doing, as every logical expression is equivalent to some ESOP, and in specifying this format it is easy to construct the corresponding quantum circuit.\\

\begin{table}[t!]
    \centering
    \begin{tabular}{c c c c c c}
         \hline
        \hline\\[-0.8ex]
         \textbf{Function} & &  \textbf{Inputs} & &   \textbf{Output} \\[1.5ex]
        \hline
         \multirow{2}{*}{Sum} & & $x^{(i)},  x^{(j)}$  & & $x^{(d+1)} = x^{(i)} + x^{(j)}$   \\
         & & $x^{(i)}, \, c$ & & $x^{(d+1)} = x^{(i)} + c$ \\[2ex]
         
         \multirow{2}{*}{Product} & & $x^{(i)},  x^{(j)}$  & & $x^{(d+1)} = x^{(i)} \times x^{(j)}$   \\
         & & $x^{(i)}, c$ & & $x^{(d+1)} = x^{(i)} \times c$   \\[2ex]
         
         \multirow{2}{*}{Maximum} & & $x^{(i)},  x^{(j)}$  & & $x^{(d+1)} = \max( x^{(i)}, x^{(j)})$   \\
         & & $x^{(i)},  c$ & & $x^{(d+1)} = \max(x^{(i)}, c)$ &  \\[2ex]
         \multirow{2}{*}{Minimum} & & $x^{(i)},  x^{(j)}$  & & $x^{(d+1)} = \min( x^{(i)}, x^{(j)})$   \\
         & & $x^{(i)},  c$ & & $x^{(d+1)} = \min(x^{(i)}, c)$ &  
         \\
         \hline 
         \hline
    \end{tabular}
    \caption[Enhanced $P$-builder: arithmetic operations]{Operations combining random variables contained in the registers pertaining to the dimensions of a $d$-dimensional multivariate probability distribution (encoded in a quantum state). For all $i$ it is necessary that $\Delta^{(i)} = 2^\alpha$ for some (positive or negative) integer $\alpha$. Strictly speaking this requirement is stronger than necessary for certain operations, however it is sufficient for all, and moreover it provides closure -- in the sense that the new dimension will certainly have $\Delta$ which is a integer power of 2. In each case the operation may combine two random variables, or a random variable and a classical constant, $c$, which is taken as an input. Observe that each operation prepares a new $(d+1)$th dimension of the probability distribution, and the reversible nature means that the original $d$ dimensions are also returned unaltered.}
    \label{tab:enhanced-operations-1}
\end{table}

\begin{table}[!t]
    \centering
    \begin{tabular}{c c c c c c c}
    \hline
        \hline\\[-0.8ex]
         \textbf{Function} & &  \textbf{Inputs} & &  \textbf{Output} & & \textbf{Notes} \\[1.5ex]
        \hline
         Compare & & $x^{(i)}, \, x^{(j)}$  & & $\text{IF}(x^{(i)} \geq x^{(j)})$ & & Require: $\Delta^{(i)} =  2^\alpha \Delta^{(j)}$ \\[2ex]
         \multirow{2}{*}{Threshold} & & \multirow{2}{*}{$x^{(i)}, \, c$, TYPE} && $\text{IF}(x^{(i)} \geq c)$ && When TYPE = lower  \\
         && && $\text{IF}(x^{(i)} < c)$ && When TYPE = upper  \\[2ex]
         ESOP && $\{ B \}$ && ESOP$(\{ B \})$, ESOP && -- \\
             \hline 
         \hline
    \end{tabular}
    \caption[Enhanced $P$-builder: logical operations]{Operations preparing indicator qubits from the values in the registers corresponding to the dimensions of the multivariate probability distribution (first two operations); or existing indicator qubits (third operation). $\alpha$ is a positive or negative integer and $\{ B \}$ is a set of Boolean variables; and for the third operation the specific ESOP is explicitly given by the user.}
    \label{tab:enhanced-operations-2}
\end{table}

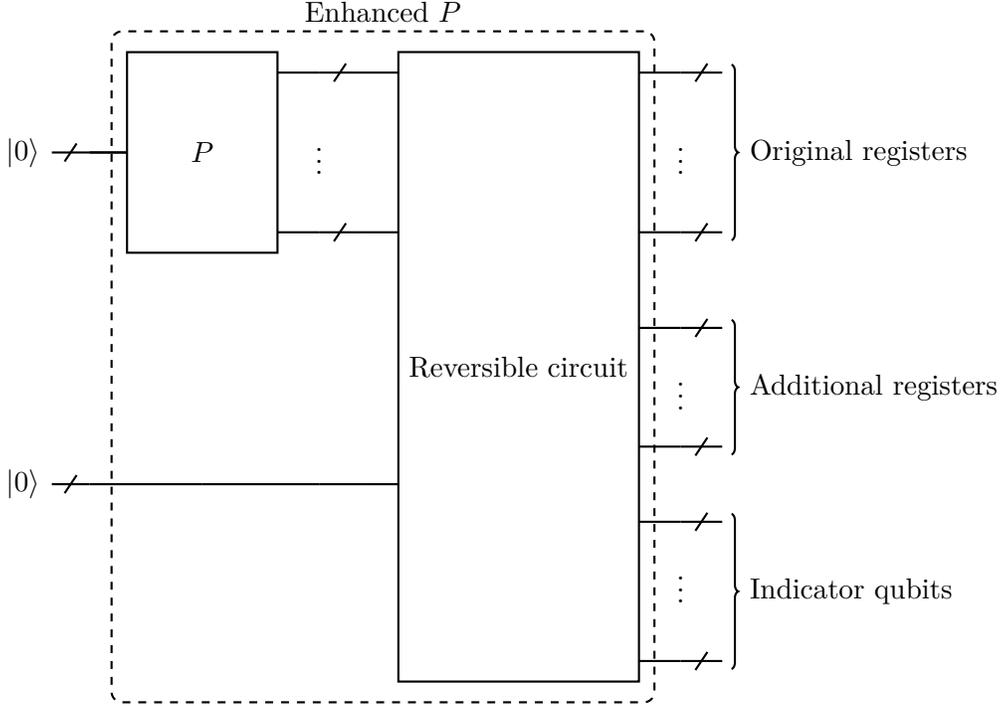
\begin{figure}[t!]
    \centering
\begin{quantikz}
     & &  \gate[wires=3, nwires={1,3}][2cm]{P}\gategroup[11,steps=4,style={dashed,
rounded corners, inner xsep=2pt},
background]{{Enhanced $P$}} & \qw & \qwbundle{} & \gate[wires=11, nwires={2,4,5,6,7,9,10,11}][2cm]{\text{Reversible circuit}} & \qw & \qwbundle{} \rstick[wires=3]{Original registers} \\
    \lstick{$\ket{0}$} & \qwbundle{} & \qw  & \vdots & & & \vdots  & \\
    & &  & \qw & \qwbundle{} & & \qw & \qwbundle{}\\
    & &  & & & & & \\
    & &  & & & & \qw & \qwbundle{} \rstick[wires=3]{Additional registers}\\
    & &  & & & & \vdots &  \\
    & &  & & & & \qw & \qwbundle{}\\
    \lstick{$\ket{0}$} & \qwbundle{} & \qw & \qw & \qw & & & \\
    & &  & & & & \qw & \qwbundle{} \rstick[wires=3]{Indicator qubits}\\
    & &  & & & & \vdots &  \\
    & &  & & & & \qw & \qwbundle{}
\end{quantikz}
    \caption{A general illustration of how the enhanced $P$-builder augments an input circuit $P$ with further multivariate probability distribution dimensions and indicator qubits.}
    \label{fig:enhanced-P}
\end{figure}

Figure~\ref{fig:enhanced-P} gives an illustration of how the enhanced circuit $P$ differs compared to that which was input. Notably, any of the output indicator qubits can be used to control the function applied as detailed in Section~\ref{sec:observables}, and in this way the enhanced $P$-builder provides a general framework for constructing quantities of interest in mathematical finance.

\subsection{Example: look-back options}
\label{subsect:ex-look-back}

\begin{figure}[t!]
\centering
    \begin{subfigure}[b]{0.8\textwidth}
    \centering
    \includegraphics[width=0.7\linewidth]{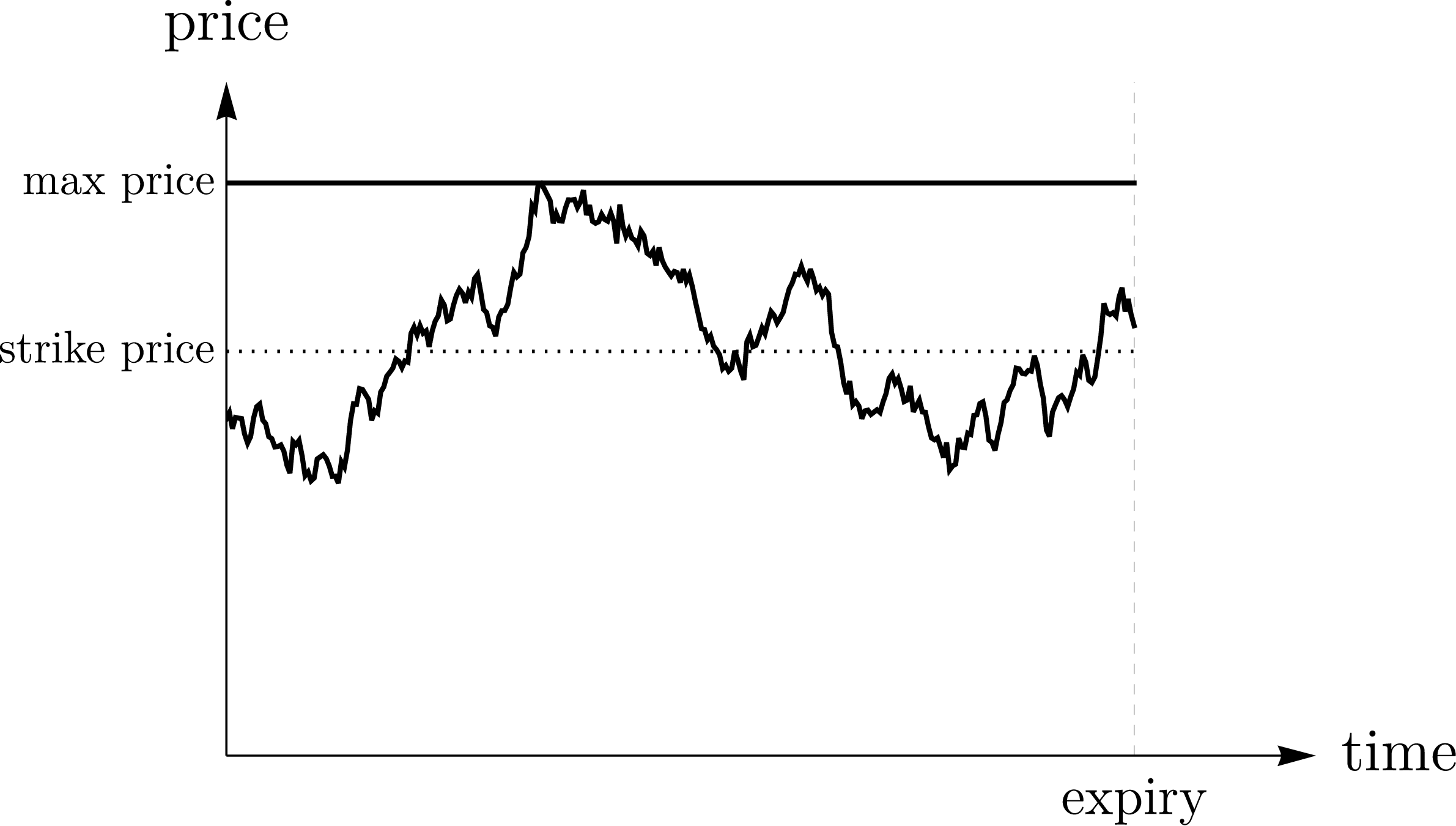}
    \caption{}
    \end{subfigure}\\
    \begin{subfigure}[b]{0.78\textwidth}
    \centering
    \begin{listing}[H]
\begin{minted}[fontsize=\footnotesize,
    linenos,
    style=manni,
    tabsize=4, 
    autogobble,
    numbersep=8pt]{python}
# The enhanced P-builder operates on a circuit encoding 4 time-slices stored as
# four dimensions of a  multivariate distribution

operations = [Max(1,2), Max(3,4), Max(5,6)]
# These max operations create new dimensions 5, 6 and 7 respectively

thresholds = [Threshold(dimension=7, value=1, type=BoundType.Lower),]
# The threshold is taken on the max of all 4 time-slices (encoded in dimension 7)
\end{minted}
\end{listing}
\caption{}
    \end{subfigure}
    \caption{ 
    (A) An illustrative sketch of a financial time-series, with strike price shown, and a look-back option payoff being (i) the maximum price reached in the lifetime of the option if this value is greater than the strike price; (ii) zero otherwise. (B) An example of how the enhanced $P$-builder can be used to construct a quantum circuit that encodes the payoff of a look-back option when the continuously varying time series is modelled as a discrete random process at four time slices. The threshold operation prepares an indicator qubit which can then be used to control the expectation, as detailed in Section~\ref{sec:observables}.}
    \label{fig:lookback}
\end{figure}

An example of the sort of calculation for which the enhanced $P$-builder can be used is look-back option payoffs. We can model a look-back call option as taking the maximum asset value over all of the time slices, and choosing to exercise the option if this maximum value exceeds the strike price. We can therefore use the max function of the enhanced $P$-builder, followed by a conditional expectation (see Section~\ref{sec:observables}), with the condition being that the maximum price exceeds the strike price (thus constituting a threshold operation). The instructions to programme the enhanced $P$-builder to build such a circuit are given in Fig.~\ref{fig:lookback}.

\subsection{Common financial calculations encompassed by the enhanced $P$-builder}

Let a suitable circuit, $P$, encode some financial time-series model for a number of (possibly correlated) assets that has been sliced appropriately such that a multivariate distribution is prepared with each dimension representing a unique asset / time pair. If such a $P$ is taken as an input, as is the case for example of look-back options given in Section~\ref{subsect:ex-look-back}, then the enhanced $P$-builder provides an extremely flexible framework to construct an appropriate circuit to compute the payoff for many of the financial instruments that are commonly traded. In Table~\ref{tab:all-calcs1} some of these are summarised, with high-level descriptions of how each instrument's payoff is programmed. Additionally, certain financial risk calculations can be programmed with the enhanced $P$-builder, in particular when the input circuit $P$ encodes a model for the prices of a large number of correlated assets composing a portfolio. For example, CVaR is essentially a European option payoff (on the total value of the portfolio) with the VaR value (once found) as strike price.  \\

\begin{table}[!t]
    \centering
    \begin{tabular}{p{2.5cm}  p{3cm} p{3.5cm} p{3.5cm}}
    \hline
        \hline\\[-0.8ex]
         \textbf{Option} & \textbf{Operations} & \textbf{Thresholds} & \textbf{Payoff conditions}   \\[1.5ex]
        \hline
        &&&\\[-0.8ex]
         European & - & Value = strike price; dimension = final time-slice & Threshold is met \\[2ex]
         &&&\\[-0.8ex]
         Asian & Sum time slices & Value = strike price $\times$ number of time-steps; dimension = sum  & Threshold is met \\
         &&&\\[-0.8ex]
         Look-back & Max time slices & Value = strike price; dimension = max  & Threshold is met \\
         &&&\\[-0.8ex]
         Barrier & - & (i) Value = knock in / out value; dimension = all time-slices; (ii) value = strike price; dimension = final time-slice &  
         All thresholds met\\[2ex]
         &&&\\[-0.8ex]
         Chooser & - & (i) Value = optimal decision threshold; dimension = decision time-slice; (ii) value = strike price; dimension = final time-slice & If call is chosen and strike price is exceeded; OR put is chosen and strike price is not exceeded.\\[2ex]
         &&&\\[-0.8ex]
         \hline 
         \hline
    \end{tabular}
    \caption[Common option payoff calculations using the enhanced $P$-builder]{Some of the common option payoff calculations that can be constructed with the enhanced $P$-builder. In the ``Thresholds'' column ``sum'' is the dimension containing the sum of the random variables for all the time-slices and ``max'' is the dimension containing the maximum of the random variables for all the time-slices. Note that the Asian option requires working in price space and ``$\times$ number of time-steps'' is required as the strike price applies to the average. We can easily elaborate to also encompass baskets of the included options, compound combinations of these options and options where the payoff is either binary or the actual final price.}
    \label{tab:all-calcs1}
\end{table}

\subsection{Using the enhanced $P$-builder to construct (geometric) Brownian motion}

The functions contained within the enhanced $P$-builder involve the addition of dimensions to multivariate probability distributions, and this feature can be put to additional use for constructing simple models of financial time-series.\\

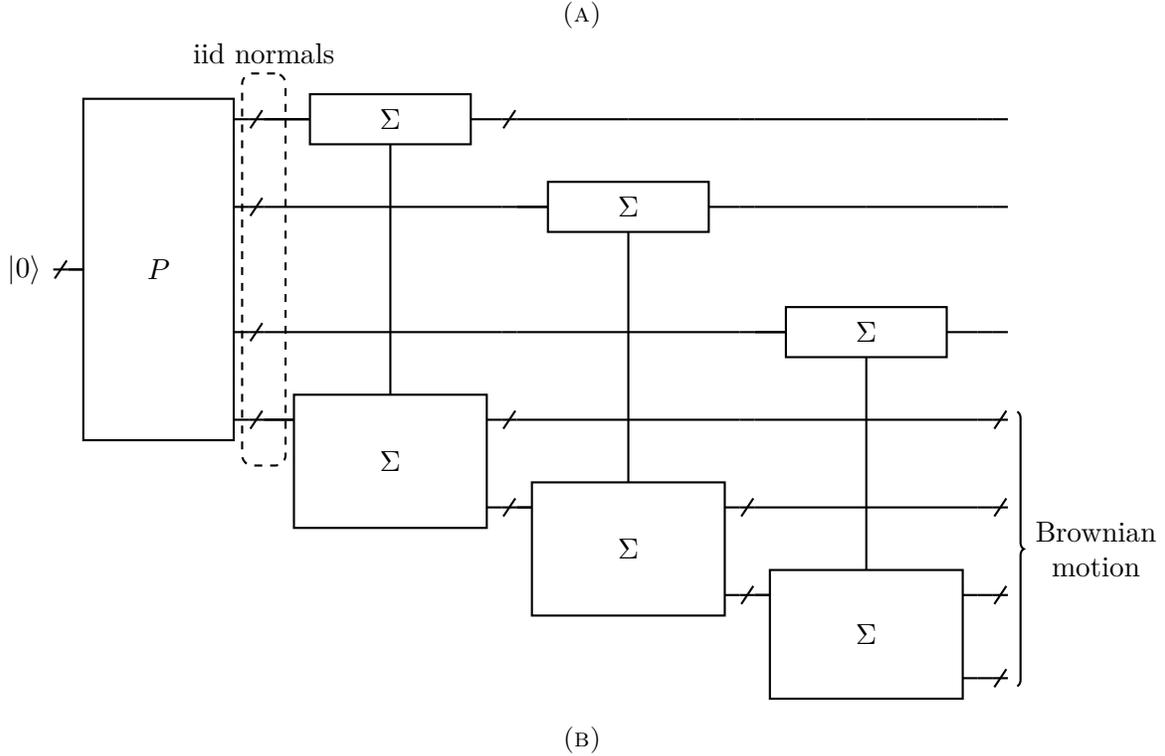
\begin{figure}[t!]
\centering
    \begin{subfigure}[b]{0.75\textwidth}
    \centering
    \begin{listing}[H]
\begin{minted}[fontsize=\footnotesize,
    linenos,
    style=manni,
    tabsize=4, 
    autogobble,
    numbersep=8pt]{python}
# The enhanced P-builder operates on a circuit encoding 4 iid. Normals

operations = [Sum(4, 1), Sum(5, 2), Sum(6, 3)]
# The Brownian motion is therefore in dimensions 4,5,6,7
\end{minted}
\end{listing}
    \caption{}
    \end{subfigure}\\
\begin{subfigure}[b]{\textwidth}
    \centering
\begin{quantikz}[column sep=0.2cm]
   & &  \gate[wires=5, nwires={1,2,4,5}][2cm]{P} & \qw &  \qwbundle{}\gategroup[5,steps=1,style={dashed,
rounded corners, inner xsep=2pt},
background]{{\iid~normals}} & [0.2cm] \gate{\Sigma}\ctrl{4} & \qw & \qwbundle{} & \qw & \qw  & \qw & \qw & \qw & \qw & \qw   \\
   & &  &  \qw & \qwbundle{} & \qw & \qw & \qw & \gate{\Sigma}\ctrl{4} & \qw & \qw & \qw & \qw & \qw & \qw \\
    \lstick{$\ket{0}$} & \qwbundle{} & \qw  & & & &   & & & \\
    & &  & \qw & \qwbundle{} & \qw & \qw & \qw & \qw & \qw & \qw & \gate{\Sigma}\ctrl{4} & \qw & \qw & \qw   \\
    & &  & \qw & \qwbundle{} & \gate[wires=2,nwires={2}]{\Sigma} \targ{} & \qw & \qwbundle{} & \qw & \qw & \qw & \qw & \qw & \qw & \qwbundle{} \rstick[wires=4]{Brownian \\ motion} \\
    & &  &  & & & \qw & \qwbundle{} & \gate[wires=2,nwires={2}]{\Sigma}\targ{} & \qw & \qwbundle{} & \qw & \qw & \qw & \qwbundle{} \\
    & &  &  & & & &  &  & \qw & \qwbundle{} & \gate[wires=2,nwires={2}]{\Sigma}\targ{} & \qw & \qw & \qwbundle{}    \\
    & &  &  & & & &  & & & & & \qw & \qw & \qwbundle{}
\end{quantikz}
\caption{}
    \end{subfigure}
    \caption{(A) Instructions to programme the enhanced $P$-builder to construct Brownian motion; (B) an illustration of the circuit generated. The $\Sigma$ boxes correspond to the summation of the input \iid~univariate Gaussian random variables, where the straight lines connecting boxes represent joint entangling operations.}
    \label{fig:GBM}
\end{figure}

If some time-series modelled by Brownian motion (as is a textbook model for financial time-series in return space) and is time-sliced at regular intervals, then the joint distribution between the random variables at the various time slices is a multivariate Gaussian distribution. However, the fact that the motion \textit{is} Brownian motion means that the correlations in the multivariate Gaussian are heavily structured, and hence a general purpose multivariate distribution loading circuit may incur unnecessary operations. Instead, we can see that using the enhanced $P$-builder, it is possible to load a circuit $P$ consisting of \iid~univariate Gaussians, and then to sum these random variables using the enhanced $P$-builder functionality, to construct additional dimensions corresponding to the Brownian motion. Similarly, if \iid~lognormal distributions are loaded then multiplication can be applied to yield \textit{geometric} Brownian motion (the corresponding textbook model when working in price space). Figure~\ref{fig:GBM} gives the instructions that programme the enhanced $P$-builder to construct Brownian motion, with geometric Brownian motion prepared analogously for an input of \iid~lognormals and \texttt{Sum} replaced by \texttt{Product}.

\subsection{Classes for common financial instruments}
\label{subsect:instrument-classes}

Although the main purpose of the enhanced $P$-builder is to provide users with a programming environment to construct arbitrary financial instruments, to aid usability we also provide classes covering simple financial instruments on single assets modelled by (geometric) Brownian motion. Noting that European and chooser options are easy to price when applied to single assets (as the former is path independent and the latter only depends on a single intermediate point in the path) we omit these, and so provide such classes for:
\begin{enumerate}
    \item barrier options;
    \item look-back options;
    \item autocallables (see Section~\ref{sec:benchmarks} for details).
\end{enumerate}

For each of these, the classes: (i) build the circuit $P$ by composing a suitable number of copies of a quantum circuit encoding an \iid~(log)normal distribution (which is taken as an input); (ii) use the enhanced $P$-builder to construct (geometric) Brownian motion; (iii) further use the enhanced $P$-builder to encode the expected payoff in the value of a register, controlled by a single indicator qubit (in the case of autocallables this is a little more complicated, as the whole of QMCI is run a number of times -- however we postpone discussion of this until Section~\ref{sec:benchmarks}). In particular, a further layer of abstraction is introduced, such that the expected payoff is computed according to user inputs\footnote{In time we will supplement these with other inputs, such the risk-free rate and the ability to discount according to risk.}:
\begin{itemize}
    \item whether to work in price or return space;
    \item the number of time-slices;
    \item the stock volatility;
    \item whether the option is a call or put, and the strike price;
    \item knock-in / knock-out barrier conditions (when applicable);
    \item autocallable conditions (when applicable);
    \item whether the payoff is binary or the asset's value;
    \item the desired RMSE / allowed number of uses of the input circuit $P$ (the quantum analogue of a limit on the number of random samples).\\
\end{itemize}

\section{Statistically robust quantum amplitude estimation}
\label{sec:robust-qae}

The original form of QAE \cite{brassard2000quantum} requires the quantum Fourier transform to perform QPE, and whilst this does not pose a problem in terms of asymptotic performance, it is likely to be prohibitively computationally costly in the near term; this has led to the development of alternative algorithms such as \textit{amplitude estimation without phase estimation} \cite{Suzuki_2020} (which we hereafter abbreviate as ``MLQAE'' -- owing to its alternative name ``maximum likelihood QAE'')  and other related proposals \cite{grinko2019iterative, Aaronson_2020}, all of which rely on classical post-processing to return the amplitude estimate. The headline quantum speed-up in QAE is in the RMSE convergence of the estimator as a function of the number of quantum queries / number of classical samples ($q$) -- for QAE the RMSE scales proportional to $q^{-1}$ compared to classical Monte Carlo methods, for which the RMSE scales proportional to $q^{-1/2}$. \\

Whilst RMSE captures the overall convergence, it does not fully describe the statistical properties and robustness of the estimator. In particular, the kurtosis -- a measure of the `tailedness' of the probability distribution for the estimator, and therefore for the propensity to produce outliers -- is an important property for potential applications of QAE that are sensitive to non-bulk performance of the estimator, such as in finance where rare events can carry great importance \cite{wachter20}. In practice the \textit{excess kurtosis} (relative kurtosis as compared to a unit Gaussian) is often studied, which is justified in this case because the classical MCI estimator has a Gaussian distribution. In addition, other related properties such as the skewness of the distribution for the estimator are important for gaining a general appreciation of the statistical character of the estimate returned by QAE.\\

We now define the quantities which together characterise the quality and robustness of any estimator. In the context of QAE, these should be of high quality for \textit{every} amplitude, and to be conservative it is necessary to consider the worst case across the full range of amplitudes. This is because, when QAE is used for MCI, the quantity of interest gets mapped to a certain (fixed) amplitude which is then estimated. So it follows, that if for some calculation this happens to be to an amplitude with relatively poor statistical properties, it is still necessary that the returned estimate adheres to the promised quality. This critical point has not always been given proper attention in the literature, and on occasion the RMSE convergence has either been aggregated across all amplitudes, or indeed is only given for some arbitrary specific amplitudes.

\subsection{Statistical robustness}
Let $\widehat{\theta}$ be the statistical estimator of a quantity $\theta$. Its central moments $\mu_n$ are defined as
\begin{equation}
\mu_n(\widehat{\theta})=\E\left( \left(\widehat{\theta}-\E(\widehat{\theta})\right)^n\right)
\end{equation}
To measure the quality of this estimator, we use four metrics.
\subsubsection{Bias}

\begin{figure}[t!]
    \centering
    \includegraphics[scale=0.75]{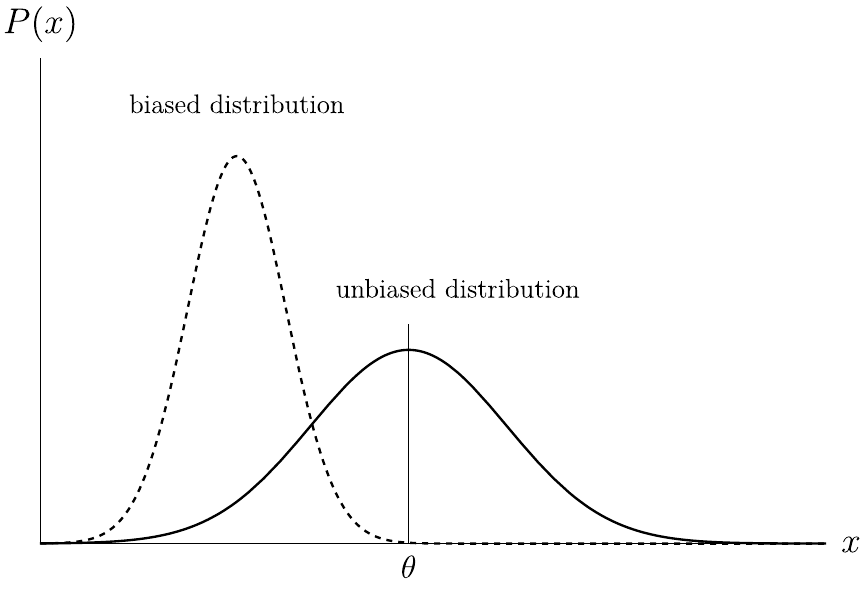}
    \caption{Example of biased and unbiased distributions.}
    \label{fig:bias-example}
\end{figure}

The bias of an estimator is given by
\begin{equation}
    \textrm{Bias}(\widehat{\theta})=\E ( \widehat{\theta}-\theta )
\end{equation}
and is therefore the amount that the mean of the estimator differs from the actual parameter. This is illustrated in Fig.~\ref{fig:bias-example}. Bias gives an indication of how much better some estimator could be -- in general, it should be possible to quantify and account for a bias, hence giving an unbiased estimator with better RMSE.

\subsubsection{(Root) mean-squared error}
As previously defined in Section~\ref{subsect:probability-prelims}, the mean-squared error of the estimator is given by
\begin{equation}
    \textrm{MSE}(\widehat{\theta})= \E ( (\widehat{\theta}-\theta)^2 )
\end{equation}
and the RMSE is the square-root of this value. The headline quantum advantage is given as an improved scaling of RMSE with the number of samples, and this is typically all that has been characterised in existing QAE algorithms.
\subsubsection{Skewness}
\begin{figure}[t!]
    \centering
    \includegraphics[scale=0.75]{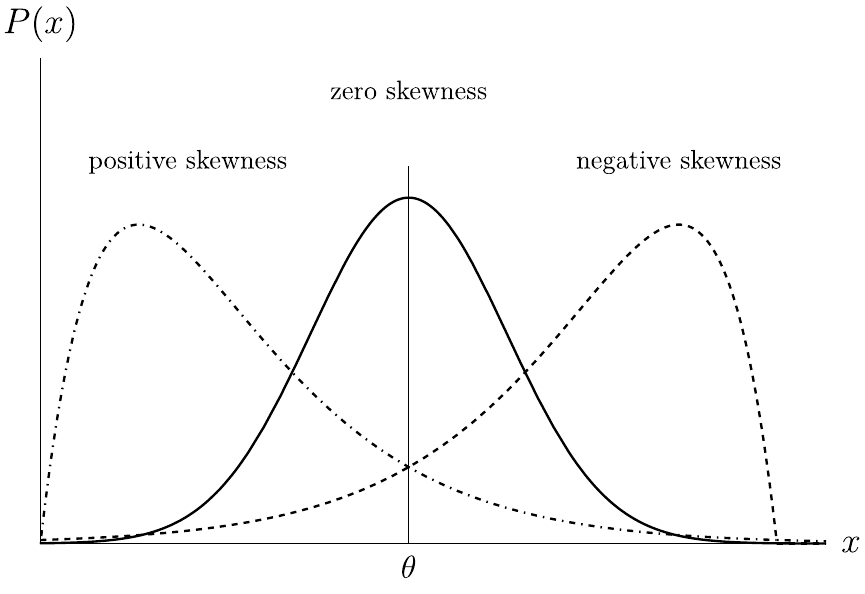}
    \caption{Examples of positive, negative and zero skewness.}
    \label{fig:skewness-example}
\end{figure}

The skewness of the estimator is 
\begin{equation}
    \textrm{Skew}(\widehat{\theta})=\frac{\mu_3(\widehat{\theta})}{\mu_2(\widehat{\theta})^{3/2}}
\end{equation}
See Fig.~\ref{fig:skewness-example} for an illustration of skewed distributions.

\subsubsection{Kurtosis and excess kurtosis} 
\begin{figure}[t!]
    \centering
    \includegraphics[scale=0.75]{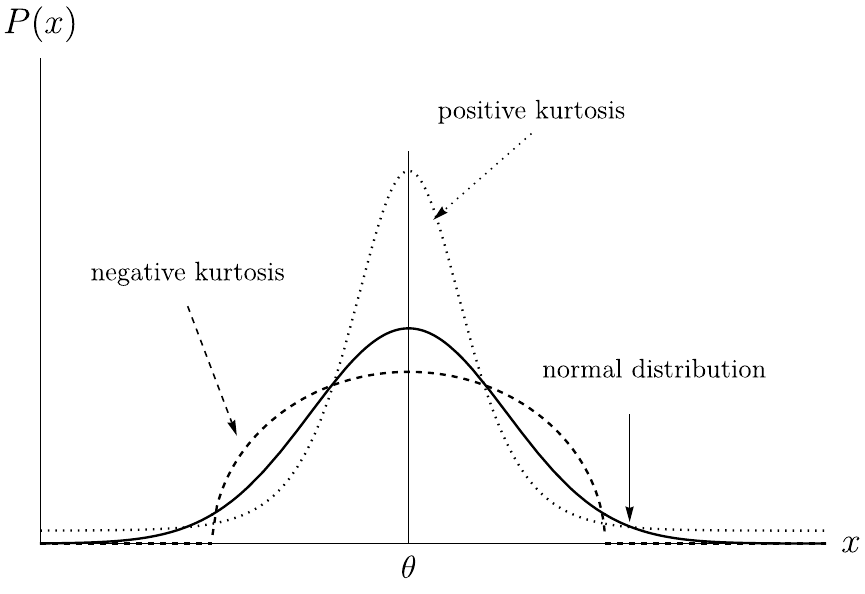}
    \caption{Examples of positive, negative and zero excess kurtosis.}
    \label{fig:kurtosis-example}
\end{figure}

The kurtosis of the estimator is 
\begin{equation}
    \textrm{Kurt}(\widehat{\theta})=\frac{\mu_4(\widehat{\theta})}{\mu_2(\widehat{\theta})^2}
\end{equation}
As the kurtosis of a Gaussian is equal to 3, the excess kurtosis is defined:
\begin{equation}
    \textrm{Ex-Kurt}(\widehat{\theta})=\frac{\mu_4(\widehat{\theta})}{\mu_2(\widehat{\theta})^2} - 3
\end{equation}
See Fig~\ref{fig:kurtosis-example} for an illustration of distributions with different kurtosis.

\subsection{Linear combination of unitaries QAE}

We conducted initial studies suggesting that existing forms of QPE-free QAE suffer from poor statistical robustness (a point substantiated in detail in our QAE statistical robustness benchmarks in Section~\ref{subsect:qae-robust-benchmarks}), which thus motivates us to propose a novel method for modifying \textit{any} QPE-free QAE algorithm to improve the robustness by making use of a linear combination of unitary (LCU) operations \cite{childswiebe12}. Whilst the method itself is generic, we give an explicit form of LCU QAE based on the exponentially increasing sequence (EIS) of MLQAE \cite{Suzuki_2020} that not only performs competitively when considering RMSE convergence, but also appears asymptotically unbiased and exhibits near-Gaussian kurtosis and skewness. This demonstrates the robustness of the general procedure as compared to standard QPE-free QAE implementations. As far as we are aware, this represents the first time that the statistical properties and robustness of a QAE algorithm have been analysed in this manner\footnote{This is the subject of patent applications GB2301482.2 and GB2211165.2.}.\\

To understand the operation of LCU QAE, and its improved statistical robustness, it is first necessary to introduce QAE more generally. QAE makes use of quantum amplitude amplification -- a generalisation of \textit{Grover's search algorithm} \cite{Groversearch} -- to provide quantum advantage in estimation convergence. Restating Def.~\ref{def1}, this is formulated by considering a unitary operator $A$ that prepares an initial $(n+1)$-qubit state 
\begin{equation}
\ket{\psi} \equiv A\ket{0^{n+1}} = \sqrt{1-a} \ket{\Phi_{0}}\ket{0} + \sqrt{a} \ket{\Phi_{1}}\ket{1}
\end{equation} where $a$ represents the amplitude of the state (the parameter to be estimated) and $\ket{\Phi_{0}}$ and $\ket{\Phi_{1}}$ are $n$-qubit states. Amplitude amplification then consists of amplifying the probability of measuring one on the final qubit by applying the following operator to $\ket{\psi}$
\begin{equation}\label{eq:grover}
Q \equiv -A S_{0} A^{\dagger} S_{\chi}
\end{equation}
where $S_{\chi}$ is a unitary operator which acts as 
\begin{align}\label{eq:schi}
&S_{\chi} \ket{\Phi_{1}}\ket{1} = - \ket{\Phi_{1}}\ket{1} \nonumber \\
&S_{\chi} \ket{\Phi_{0}}\ket{0} = \ket{\Phi_{0}}\ket{0}
\end{align}
and $S_{0}$ is a simple unitary operator that does not depend on $A$.\\ 

Defining the parameter $\theta \in [0, \pi/2]$ such that $\text{sin}^{2}\theta = a$, then
\begin{equation}
\label{eq:psi-def}
\ket{\psi} =  \cos \theta \ket{\Phi_{0}}\ket{0} + \sin \theta \ket{\Phi_{1}}\ket{1}
\end{equation}
so it follows that applying the operator $A$ can be considered as a rotation by an angle $\theta$ in the two-dimensional invariant subspace spanned by $\ket{\Phi_{1}}\ket{1}$ and $\ket{\Phi_{0}}\ket{0}$. Similarly, it can be shown that applying $Q^m$ (that is, applying $Q$ a total of $m$ successive times) to $\ket{\psi}$ yields
\begin{equation}
Q^{m} \ket{\psi} =  \cos((2m+1)\theta) \ket{\Phi_{0}}\ket{0} + \sin((2m+1)\theta) \ket{\Phi_{1}}\ket{1}
\end{equation}
such that each application of $Q$ further rotates the state by an additional $2\theta$. Thus the state can be rotated to any odd integer multiple of $\theta$ in this manner, providing the amplitude amplification. \\

However, the condition that the initial rotation angle is exactly $\theta$ limits the overall variation in the angles rotated to by the circuit $Q^m A$, and preliminary studies have shown that this leads to poor statistical robustness. In particular, for certain values of $\theta$ the rotation always results in a state that lies close to the principal axes of the invariant subspace, which leads to particularly poor performance. By contrast, further preliminary studies, in which we created the artificial situation where the initial starting angle could be set to a variety of values spread between 0 and $\pi / 2$, regardless of $\theta$, remedied these problems. Such an artificial situation cannot, of course, be exactly replicated in an actual quantum circuit -- however our proposed method for modifying any QPE-free QAE algorithm constructs the initial state using a linear combination of unitaries, and this does allow for some variation in the starting angle. Moreover, we show that statistical robustness can be obtained without drastically reducing the performance in terms of RMSE convergence. Specifically, we use LCU to initialise a new state $\ket{\tilde{\psi}}$ with a rotation to an angle $\alpha$ in the same subspace
\begin{equation}
\label{eq:alpha-angle-def}
    \ket{\tilde{\psi}} = \cos \alpha \ket{\Phi_{0}}\ket{0} + \sin \alpha \ket{\Phi_{1}}\ket{1}
\end{equation}
by exploiting the fact that using Eq. \eqref{eq:schi} it is easy to prepare
\begin{equation}
\label{eq:minus-theta-def}
S_{\chi}A\ket{0^{n+1}} =  \cos \theta \ket{\Phi_{0}}\ket{0} - \sin \theta \ket{\Phi_{1}}\ket{1}
\end{equation}
such that the state is instead rotated by an angle $-\theta$. That is, the states in Eq.~\eqref{eq:psi-def} and Eq.~\eqref{eq:minus-theta-def} lie in the same invariant subspace, and so a linear combination of the unitary operations that prepares these can be used to prepare a state of the form given in Eq.~\eqref{eq:alpha-angle-def}.\\

\begin{figure}[t!]
\begin{center}
\includegraphics[width=0.5\textwidth]{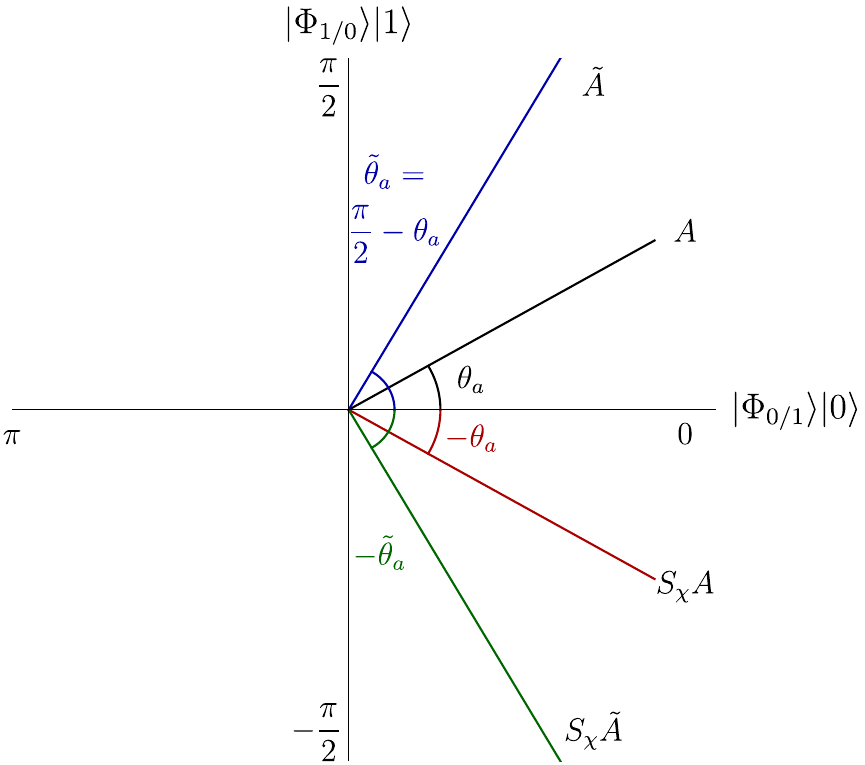}
\end{center}
\caption{Initial rotation angles in the two-dimensional invariant subspace spanned by either $\{ \ket{\Phi_{1}}\ket{1}, \; \ket{\Phi_{0}}
\ket{0} \}$ or $\{ \ket{\Phi_{0}}\ket{1} , \; \ket{\Phi_{1}}\ket{0} \}$, corresponding to applying the unitary operators $A$ and $S_{\chi}A$ or $\tilde{A}$ and $S_{\chi}\tilde{A}$, respectively.} 
\label{fig:lcu_angles}
\end{figure}

Additionally, to further enhance the variety of possible angles for the initially prepared state, we make use of the operator
\begin{equation}
\tilde{A} \equiv (I_{2^{n}} \otimes X) A
\end{equation}
which thus has the effect of applying a single Pauli-$X$ rotation to the $(n+1)^{\text{th}}$ qubit, such that state is rotated by the angle $\tilde{\theta} = \pi/2-\theta$ in a different invariant subspace spanned by $\ket{\Phi_{0}}\ket{1}$ and $\ket{\Phi_{1}}\ket{0}$. Following the above rationale, $S_{\chi}\tilde{A}$ thus rotates the state by an angle $-\tilde{\theta}$. Much of the following analysis applies to each case (rotated by $\theta$ or $\tilde{\theta}$) and so we introduce $\boldsymbol{\theta}$ to refer generally to either $\theta$ or $\tilde{\theta}$; and $\ket{\Phi_{0/1}}$ and $\ket{\Phi_{1/0}}$ refer to states that are $\ket{\Phi_{0}}$ or $\ket{\Phi_{1}}$ (when $\ket{\Phi_{0/1}}$ and $\ket{\Phi_{1/0}}$ are used as a pair, the implication is that one is $\ket{\Phi_{0}}$ and the other $\ket{\Phi_{1}}$). Figure~\ref{fig:lcu_angles} demonstrates the various initial angles of states that can be prepared using these simple operators.\\

A LCU -- corresponding to weighted linear sums of either $A$ and $S_{\chi}A$ or $\tilde{A}$ and $S_{\chi}\tilde{A}$ -- can thus be constructed, which when applied to the state $\ket{0^{n+1}}$ results in an initial state of the (un-normalised) form
\begin{equation}\label{eq:lcu_statevector}
\ket{\tilde{\psi}} =  \text{cos}(\boldsymbol{\theta}) \ket{\Phi_{0/1}}\ket{0} + F \text{sin}(\boldsymbol{\theta}) \ket{\Phi_{1/0}}\ket{1}
\end{equation}
where $-1 \leq F \leq 1$ and the definition of $\ket{\tilde{\psi}}$, from Eq.~\eqref{eq:alpha-angle-def}, has now been extended to correspond to states prepared in either the original and secondary subspaces. It thus follows that $\alpha$ in Eq.~\eqref{eq:alpha-angle-def} is given by:
\begin{equation}\label{eq:initial_angle} 
\alpha = \text{tan}^{-1}\left(F\text{tan}(\boldsymbol{\theta})\right)
\end{equation} 

\begin{figure}[t!]
\begin{center}
\begin{quantikz}[thin lines] 
            \lstick{$\ket{0}$}&  \gate{R_y(\beta)} & \octrl{1} & \qw & \ctrl{1} &  \gate{R^{\dagger}_y(\beta)} & \qw & \meter{} \\
            \lstick{$\ket{\omega}$} & \qw & \gate{U_{a}} & \qw & \gate{U_{b}} & \qw & \qw & \qw      
    \end{quantikz}
    \end{center}
\caption{Quantum circuit for (non-deterministically) preparing a LCU state. The desired state is prepared when the measurement
outcome of the ancilla qubit is post-selected to be zero (equivalently, in practice the procedure can just be repeated until successful).} 
\label{fig:lcu_circuit}
\end{figure}
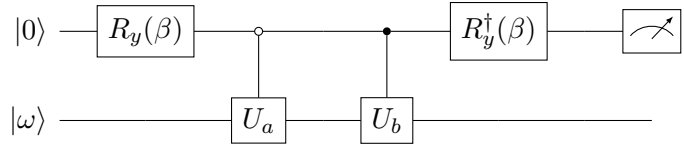

Figure~\ref{fig:lcu_circuit} demonstrates the form of the quantum circuit used to perform LCU state preparation\footnote{Note that the procedure requires an additional ancilla qubit, as is standard for LCU state preparation.},
 where $U_{a}$ and $U_{b}$ are either the pair $\{ A, S_{\chi}A \}$ or the pair $ \{\tilde{A},S_{\chi}\tilde{A} \}$. Such a circuit results in the state
\begin{equation}
\label{eq:lcu_state}
\ket{0} \ket{\omega} \to  \left[ \left(\text{cos}^{2}(\beta/2) U_{a} + \text{sin}^{2}(\beta/2) U_{b}\right) \right] \ket{0} \ket{\omega}+  \left[ \frac{1}{2} \text{sin}(\beta)(U_{b}-U_{a}) \right] \ket{1} \ket{\omega}
\end{equation}
which is thus such that, if the measurement outcome (for the first qubit) is zero, then the state collapses to $\left(\text{cos}^{2}(\beta/2) U_{a} + \text{sin}^{2}(\beta/2) U_{b}\right) \ket{\omega}$.\\

Our proposed method for modifying any QPE-free QAE algorithm thus utilises different LCU state-preparation circuits to prepare initial states with a variety of different starting angles. An important caveat of the LCU state preparation is that there is a probability that the LCU state preparation will fail:
\begin{equation}
p_{\text{fail}} = \frac{1}{4} \text{sin}^{2}(\beta) ||U_{b}-U_{a}||^{2} = \text{sin}^{2}(\beta)
\end{equation}
which thus tends to one as $\beta$ tends to $\pi/2$ -- i.e., when the LCU is designed to prepare an equal mixture of the two unitaries such that the prepared state is $\ket{\Phi_{0/1}}\ket{0}$. In order to avoid certain and highly probably failures, $p_{\text{fail}}$ can be upper-bounded by some $p_{\text{max-fail}}$, which is set by constraining the range of the rotation parameter $\beta$ as $0 \leq \beta \leq \text{sin}^{-1}(\sqrt{p_{\text{max-fail}}}) $. This corresponds to the weighting of the two terms $U_{a}$ and $U_{b}$ in Eq.~\eqref{eq:lcu_state}, defining a bound (either upper or lower) on the factor $F$ in Eq.~\eqref{eq:lcu_statevector}, $F_{\text{bound}}$, and thus indirectly the rotated angle of the initial state, which is then bounded (equivalently either upper or lower) by $\boldsymbol{\theta}_{\text{bound}} = \text{tan}^{-1}\left(\sqrt{1-p_{\text{max-fail}}}\text{tan}(\boldsymbol{\theta})\right)$.\\

To use the variety of starting angles that LCU provides to enhance the statistical robustness of QAE, we first note that there are four possible categories of LCU initial states that can be prepared (referred to as $\text{LCU}_{1-4}$), each corresponding to rotating the state to within a different angular range bounded by some $\boldsymbol{\theta}_{\text{bound}}$:  
\begin{align}\label{eq:lcutypes}
&\text{LCU}_{1}(\beta): \ \ \ \ \ \  \text{cos}^{2}(\beta/2) A + \text{sin}^{2}(\beta/2) S_{\chi}A, \ \ \ \ \ \ \theta_{\text{bound}} \leq \alpha \leq \theta  \nonumber \\
&\text{LCU}_{2}(\beta):  \ \ \ \ \ \  \text{cos}^{2}(\beta/2) S_{\chi}A + \text{sin}^{2}(\beta/2) A, \ \ \ \ \ -\theta \leq \alpha \leq -\theta_{\text{bound}} \nonumber \\
&\text{LCU}_{3}(\beta):  \ \ \ \ \ \  \text{cos}^{2}(\beta/2) \tilde{A} + \text{sin}^{2}(\beta/2) S_{\chi}\tilde{A}, \ \ \ \ \ \ \tilde{\theta}_{\text{bound}} \leq \alpha \leq \tilde{\theta} \nonumber \\
&\text{LCU}_{4}(\beta):  \ \ \ \ \ \  \text{cos}^{2}(\beta/2) S_{\chi}\tilde{A} + \text{sin}^{2}(\beta/2) \tilde{A}, \ \ \ \ \  -\tilde{\theta} \leq \alpha \leq -\tilde{\theta}_{\text{bound}}
\end{align} 
where $\beta = \text{cos}^{-1}(F)$.\\

\begin{figure}
\begin{center}
\begin{quantikz}[thin lines] 
              & \gate[wires=2]{\text{LCU}} & \ocontrol & \cw  & \cw & \cw  \\
             & & \gate{Q^{m}} \vcw{-1} & \qw & \meter{}
    \end{quantikz}
    \end{center}
    \caption{Quantum circuit for performing QAE using LCU state preparation. The LCU block stands for the circuit given in Figure~\ref{fig:lcu_circuit}, and the top wire is the post-measurement ancilla bit of this circuit.}
\label{fig:qae_lcu_circuit} 
    \end{figure}
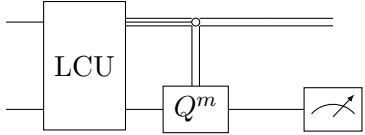

The circuits preparing these possible starting angles are then used in place of $A\ket{0^{n+1}}$ in QAE circuits, which thereafter proceed as usual. Figure~\ref{fig:qae_lcu_circuit} demonstrates the combined quantum circuit -- consisting of LCU state-preparation circuit and the amplitude amplification circuit, $Q^m$, -- required to realise this procedure. As the operations $Q^m$ only occur when the LCU post-selection is successful, LCU QAE can be seen to ``fail fast'' -- in the sense that the majority of the circuit need only be executed when the initial state preparation has been successful. This is the intended implication of the classical control on $Q^m$, although the principle of deferred measurement can be invoked to give an equivalent circuit where these measurements are performed in the final layer, if desired.\\

Our proposed method is a general method for modifying any QPE-free QAE algorithm by introducing LCU state preparation to improve the statistical properties and robustness of the estimator for the amplitude (or equivalently for $\theta$). Any QPE-free algorithm can be formulated as an instance of the generic framework given in Algorithm~\ref{alg:genlcuqae}. (Note that Algorithm~\ref{alg:genlcuqae} requires a classical probability distribution, $p(\theta)$ to be updated: we found that a sufficiently dense grid of point masses sufficed for this purpose, but more advanced techniques such as particle filtering \cite{ParticleFilter} may alternatively be used.) Going into further detail, \texttt{Stopping Criterion} could be, for example, the desired accuracy of the estimate, total wall-clock time, total number of uses of $A$ etc; the estimator $\hat{a}$ could be any estimator, for example, ML, MMSE etc (see Section~\ref{subsect:probability-prelims} for definitions of these). Optionally, following each shot $p(\theta)$ could be updated and the \texttt{Stopping Criterion} could also be assessed.\\ 

\begin{algorithm}[!t]
\caption{Generic algorithm for any QPE-free QAE.}\label{alg:genlcuqae}
\begin{algorithmic}[1]
\Require Quantum circuit $A$; probability distribution $p(\theta)$ initialised as the prior; \texttt{Stopping Criterion}
\State Set $m$, $n_{\text{shots}}$, (optional) additional parameters 
\State Prepare and measure $n_{\text{shots}}$ of $Q^{m}A\ket{0}$
\State Update $p(\theta)$ using measurement outcomes based on standard probabilistic procedures
\If{\texttt{Stopping Criterion}}
        	\State Return $\hat{a}$ (or $\hat{\theta}$)
	\Else{} Update $m$, $n_{\text{shots}}$; Goto Line 2
	\EndIf

\end{algorithmic}
\end{algorithm}

\begin{figure}[ht!]
    \centering
\includegraphics[scale=0.55]{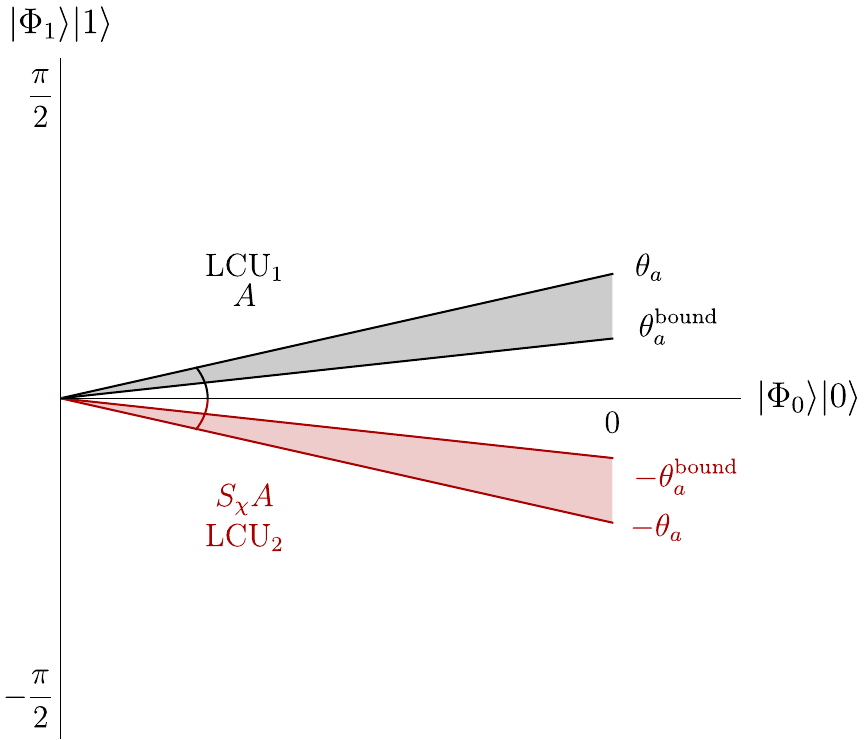}
    \includegraphics[scale=0.55]{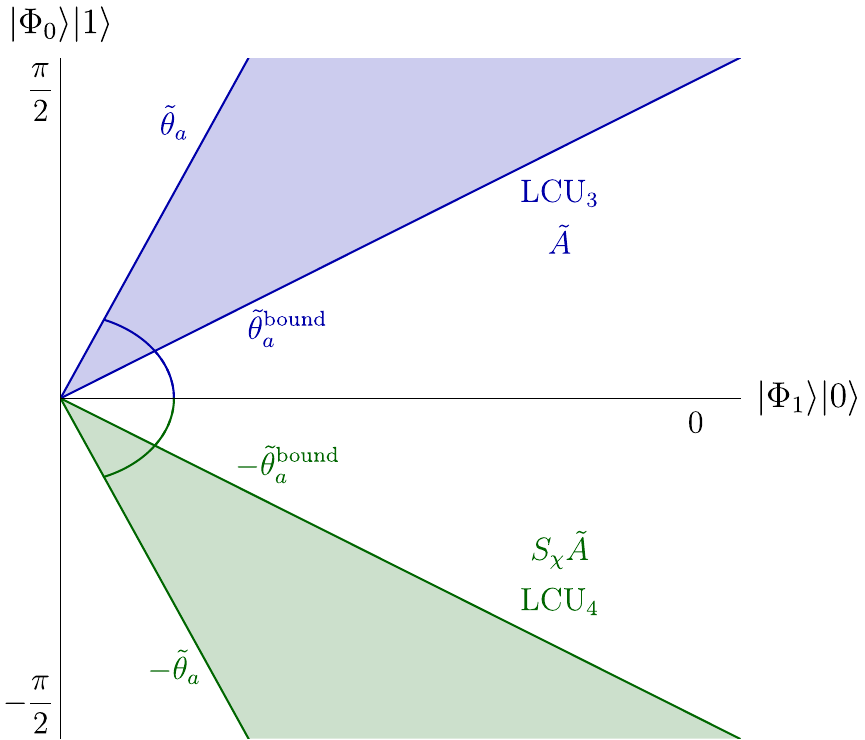}
\caption{Range of rotation angles in (left) the two-dimensional invariant subspace spanned by $\{ \ket{{\Phi_{1}}}\ket{1}, \ket{{\Phi_{0}}}\ket{0} \}$ for LCUs of the operators $A$ and $S_{\chi}A$, corresponding to LCU categories (black) $\text{LCU}_{1}$ and (red) $\text{LCU}_{2}$, and (right) the two-dimensional invariant subspace spanned by $\{ \ket{{\Phi_{0}}}\ket{1}, \ket{{\Phi_{1}}}\ket{0} \}$ for LCUs of the operators $\tilde{A}$ and $S_{\chi}\tilde{A}$, corresponding to LCU categories (blue) $\text{LCU}_{3}$ and (green) $\text{LCU}_{4}$, prepared given the specific algorithm described in this section. For illustrative purposes it is assumed that $F^{\text{bound}}$ and the true value of $\theta_{a}$ are such that $\theta^{\text{bound}}_{a} = 0.5 \theta_{a}$. The bold lines correspond to the boundaries of the sectors of the corresponding LCU categories. Note that if one subspace has a tight range of angles, the other will have a large range.} 
\label{fig:explicit_lcu_angles}
\end{figure}

This general framework can be enhanced to include LCU state preparation, as shown in Algorithm~\ref{alg:genlcuqae2}. In the general description, no preferred designation into the four regions (corresponding to $\text{LCU}_{1 - 4}$) is given, however there is good reason to divide these equally (or approximately equally if $n_{\text{shots}}$ is not a multiple of 4). This is because the way that the restriction on starting angle dictated by the choice of $p_{\text{max-fail}}$ works is such that, if $\text{LCU}_{1}$ and $\text{LCU}_{2}$ only provide a small range of different angles, then $\text{LCU}_{3}$ and $\text{LCU}_{4}$ will provide a large range. This is sketched in Fig.~\ref{fig:explicit_lcu_angles} and in this way, even without knowing $\theta$ (that is, after all, what we are trying to discover) a large variety of initial angles is guaranteed -- and this is the key ingredient for good statistical robustness. It is worth noting that, in Algorithm~\ref{alg:genlcuqae2}, the $n_{\text{shots}}$ where $m=0$ remain unchanged, as LCU state preparation is most useful when the state is subsequently rotated by amplitude amplification steps (i.e., $Q^m$ for $m>0$).

\begin{algorithm}[!t]
\caption{Generic algorithm for LCU QPE-free QAE.}\label{alg:genlcuqae2}
\begin{algorithmic}[1]
\Require Quantum circuit $A$; probability distribution $p(\theta)$ initialised as the prior; \texttt{Stopping Criterion}; $p_{\text{max-fail}}$
\State Initialise $m=0$, $n_{\text{shots}}$
\State Prepare and measure $n_{\text{shots}}$ of $Q^{m}A\ket{0}$
\State Set $m$, $n_{\text{shots}}$, (optional) additional parameters 
\For {it=1:$n_{\text{shots}}$}
\State Select $y \in \{1,2,3,4\}$
\State Select $\beta$ (choice restricted by $p_{\text{max-fail}}$)
\State Prepare and measure $Q^{m}[\text{LCU}_y(\beta)]\ket{0}$ (waiting for a successful LCU preparation)
\State Update $p(\theta)$ using measurement outcomes based on standard probabilistic procedures
\EndFor
\If{\texttt{Stopping Criterion}}
        	\State Return $\hat{a}$ (or $\hat{\theta}$)
	\Else{} Goto Line 3 
	\EndIf
\end{algorithmic}
\end{algorithm}

\subsection{QAE robustness benchmarks}
\label{subsect:qae-robust-benchmarks}

In order to benchmark the performance of LCU QAE, we define an explicit algorithm, based on the EIS ($m \in \{0,1,2,4,8,16,... \}$) of Ref.~\cite{Suzuki_2020}, with the following settings:
\begin{itemize}
    \item  For $m=0$, $n_{\text{shots}} = 66$; for all other $m$, $n_{\text{shots}} = 44$. The 44 shots are divided equally between the four LCU segments (11 into each), and a range of initial starting angles are chosen within each LCU segment.
    \item $q$, as selected by the user, is taken as the number of uses in circuits with \textit{successful} LCU post-selections. The overhead of failed LCU post-selections is then accounted for when reporting the overall converge as a function of total uses (failed and successful).
    \item The returned estimate is a MMSE estimate.\\
\end{itemize}

\begin{figure}[t!]
\begin{center}
\begin{quantikz}
  \qw & \gate{R_y(2\theta)} & \ctrl{1}& \qw \\
  \qw & \qw  & \targ{}& \qw\\
\end{quantikz}
    \end{center}
    \caption{The circuit used to prepare various amplitudes used to benchmark QAE. In particular, the parameterised angle is used to prepare the various amplitudes.}
\label{fig:qae-benchmark-circuit} 
    \end{figure}
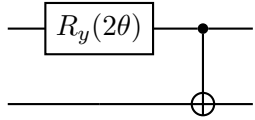

For each given value of $q$, a circuit $P$ whose form is given in Fig.~\ref{fig:qae-benchmark-circuit} is used (by varying $\theta$) to prepare states with 49 equally-spaced amplitude values in the range $[0.02,0.98]$. QAE is then run using a state-vector simulator, and the corresponding amplitude estimate stored. This process is repeated $10000$ times for each configuration (a configuration being a particular $q$ and amplitude value). The (absolute) bias, MSE/RMSE, (absolute) skewness, and excess kurtosis are then calculated for each configuration. Bootstrapping, implemented via the bias-corrected and accelerated bootstrap interval \cite{bootstrap1}, is used to determine a $68\%$ confidence interval on each metric, from which (asymmetric) uncertainties on the data points are determined. These values are then also averaged across all amplitude values, where the uncertainties are propagated to the average using the methods discussed in Ref.~\cite{barlow2003asymmetric}. For comparison, we also performed this process for two other prominent forms of QPE-free QAE, namely Iterative QAE (IQAE) \cite{grinko2019iterative} and MLQAE itself.\\

\begin{figure}[ht!]
    \centering
        \begin{tabular}{c c}
            \includegraphics[width=0.48\textwidth]{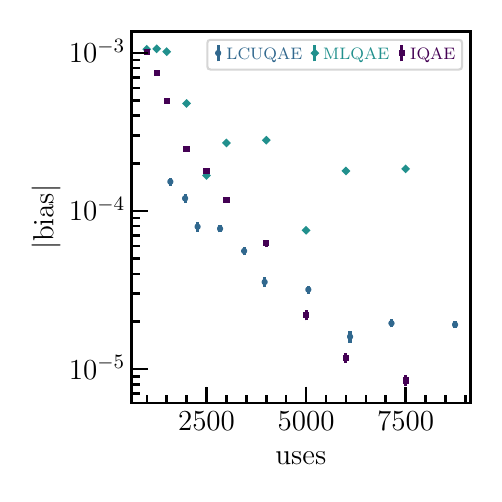} & \includegraphics[width=0.48\textwidth]{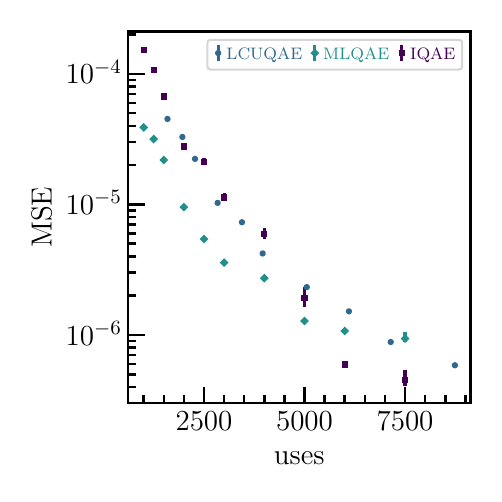} \\
            (a) & (b) \\
             \includegraphics[width=0.48\textwidth]{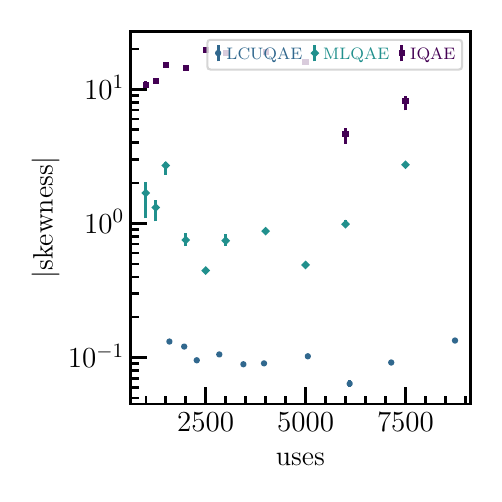} & \includegraphics[width=0.48\textwidth]{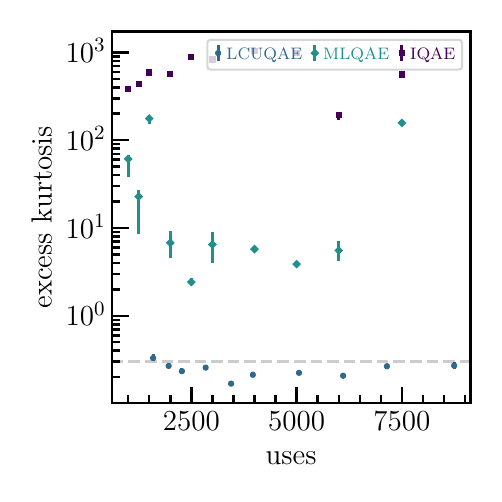} \\
            (c) & (d) \\
        \end{tabular}
    \caption{The statistical performance of the various forms of QAE averaged over the 49 amplitudes.}
    \label{fig:robustness-benchmarks}
\end{figure}

Figure~\ref{fig:robustness-benchmarks} shows the statistical benchmarks quantities for the various versions of QAE averaged over the 49 amplitudes. Even from this relatively coarse display of the results, it can be seen that LCU QAE has significantly better statistical robustness than the other two methods (which, in fairness, were not designed with this in mind). Crucially, this advantage is maintained when the results are viewed for each amplitude, as shown in Figs.~\ref{fig:bias-by-amp}--\ref{fig:kurt-by-amp} -- notably the statistical robustness is still good even in the worst case (that we must conservatively assume) confirming the superiority of LCU QAE in this regard. To give a little more detail, excess kurtosis is the single most important figure of merit when assessing whether extreme events occur with significantly greater (or smaller) probability than for a Gaussian random variable with the same mean and variance, and although there is no consensus about exactly what value constitutes sufficient approximate Gaussianity -- although excess kurtosis between $-2$ and $2$ is somewhat commonly taken as indicative of Gaussianity (see e.g., Ref.~\cite{Gauss-Kurtosis}). In our results we take the relatively aggressive stance of treating 0.3 as an acceptable value of excess kurtosis (i.e., $10\%$ of the kurtosis of a Gaussian) -- and we can see from Fig.~\ref{fig:kurt-by-amp} where the value of 0.3 is illustrated by the horizontal dashed line that, apart from the odd outlier, this is met by virtually all of the LCU QAE estimators, but never the IQAE and MLQAE estimators, across the various amplitudes and numbers of uses analysed. Indeed, IQAE and ML QAE would often not even meet the relaxed value of 2 as acceptable excess kurtosis. (Note we plot the excess kurtosis on a logarithmic scale -- which is made possible by the fact that the empirically-found values were always positive.)

\newpage
~
\newpage

\newpage
\part{Full technical specification}
\label{part:full-technical-specification}

\section{Overview of the QMCI engine architecture}
\label{sec:overview-engine}

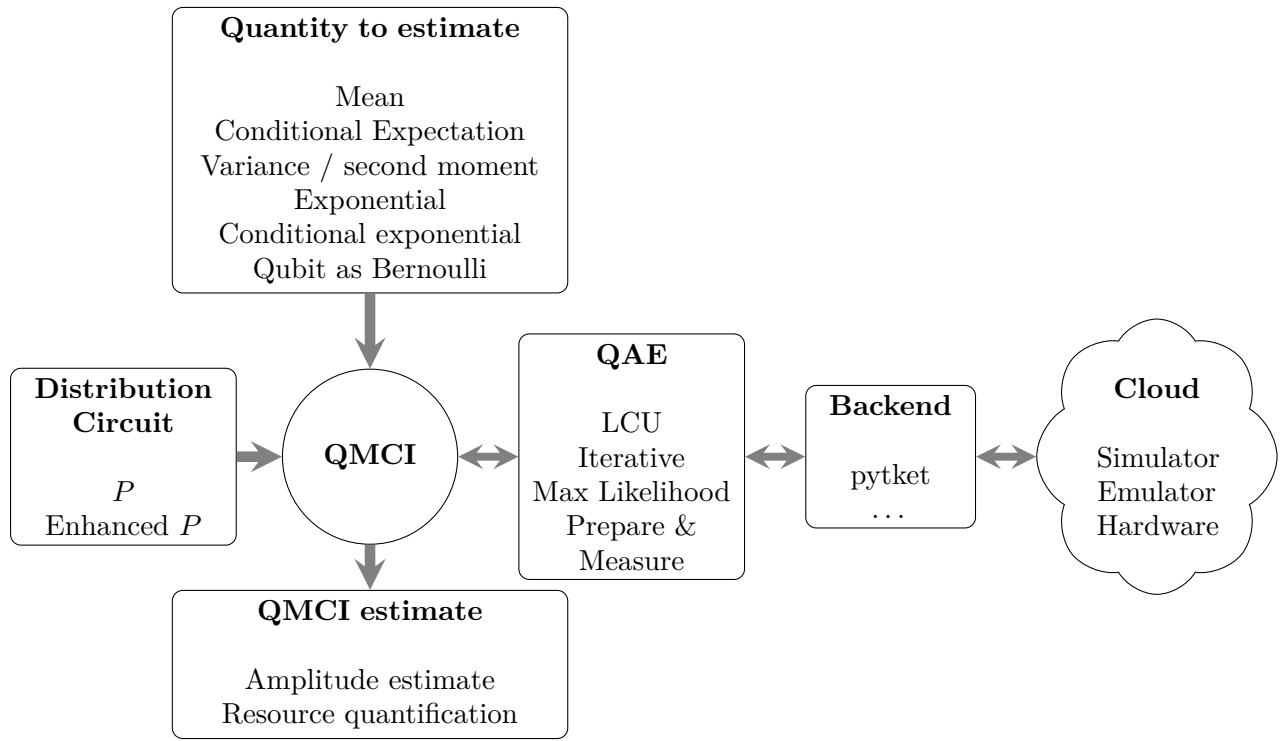
\begin{figure}[t!]
    \centering
    \begin{tikzpicture}[  module/.style={draw, rectangle, minimum width=3cm, minimum height=1.5cm, align=center, text width=2.7cm, rounded corners},  circlemodule/.style={draw, circle, minimum size=2cm, align=center, text width=2cm},  arrow/.style={->, >=stealth, line width=4pt, color=black!50},  doublearrow/.style={<->, >=stealth, line width=3pt, color=black!50},  ]

  \node[circlemodule] (qmci) {\textbf{QMCI}};

  \node[module, text width=5cm, above=1cm of qmci] (function) {\textbf{Quantity to estimate}\\~\\Mean\\Conditional Expectation\\Variance / second moment\\Exponential\\Conditional exponential\\Qubit as Bernoulli};

  \node[module, left=0.6cm of qmci] (distcircuit) {\textbf{Distribution Circuit}\\~\\ $P$\\ Enhanced $P$};

  \node[module, text width=5cm, minimum height=1cm, below=0.6cm of qmci] (amplitude) {\textbf{QMCI estimate}\\~\\Amplitude estimate\\ Resource quantification};

  \node[module, right=0.8cm of qmci] (qae) {\textbf{QAE}\\~\\LCU\\Iterative\\Max Likelihood\\Prepare \& Measure};

  \node[module, minimum width=2cm, text width=2cm, right=0.8cm of qae] (backend) {\textbf{Backend}\\~\\pytket\\ \dots};

  \node[cloud, draw, cloud puffs=8, cloud puff arc=120, aspect=0.8, align=center, text width=1.6cm, minimum width=2.4cm, minimum height=1.5cm, right=0.8cm of backend] (cloud) {\textbf{Cloud}\\~\\Simulator\\Emulator\\Hardware};

  \draw[arrow] (function) -- (qmci);
  \draw[arrow] (distcircuit) -- (qmci);
  \draw[arrow] (qmci) -- (amplitude);
  \draw[doublearrow] (qmci) -- (qae);
  \draw[doublearrow] (qae) -- (backend);
  \draw[doublearrow] (backend) -- (cloud);

\end{tikzpicture}
    \caption{Schematic showing the architecture of Quantinuum’s Quantum Monte Carlo Integration Engine. The Distribution Circuit module is described explicitly in Fig.~\ref{fig:flow-distrib-circuit-module}.}
    \label{fig:flow-qmci-engine}
\end{figure}

\begin{figure}[t!]
    \centering
\begin{tikzpicture}[  module/.style={draw, rectangle, minimum width=3cm, minimum height=3.5cm,  align=center, text width=4cm,rounded corners}, circlemodule/.style={draw, circle, minimum size=1.5cm, align=center, text width=1.5cm},  arrow/.style={->, >=stealth,line width=5pt, color=black!50},  ]

  \node[module, minimum width=2.3cm, text width=2.3cm] (distribution) {\textbf{Distribution}\\~\\Binomial\\Gaussian\\Lognormal\\Unstructured};

  \node[module, right=0.8cm of distribution, text width=5.4cm] (stateprep) {\textbf{State Preparation}\\~\\Quantum Walk\\PDEs\\PQCs};

  \node[module, right=0.8cm of stateprep] (distcircuit) {\textbf{Distribution Circuit}\\~\\Circuit\\Wires\\$\{x^{(i)}_\ell\}, \{ \Delta^{(i)} \}$\\Indicator qubits};

  \node[circlemodule, right=0.8cm of distcircuit] (qmci) {\textbf{QMCI}};

  \node[module, below=1.5cm of distcircuit] (enhancedP) {\textbf{Enhanced $P$-Builder}\\~\\Sum\\Product\\Min / Max\\Threshold \\ ESOP};

  \draw[arrow] (distribution.east) -- (stateprep.west);
  \draw[arrow] (stateprep.east) -- (distcircuit.west);
  \draw[arrow] (distcircuit.east) -- (qmci.west) ;
  \draw[arrow] (distcircuit) to[out=-45, in=45] (enhancedP);
  \draw[arrow] (enhancedP) to[out=135, in=-135] (distcircuit);

\end{tikzpicture}
\caption{Flow of the Distribution Circuit module.}
    \label{fig:flow-distrib-circuit-module}
\end{figure}
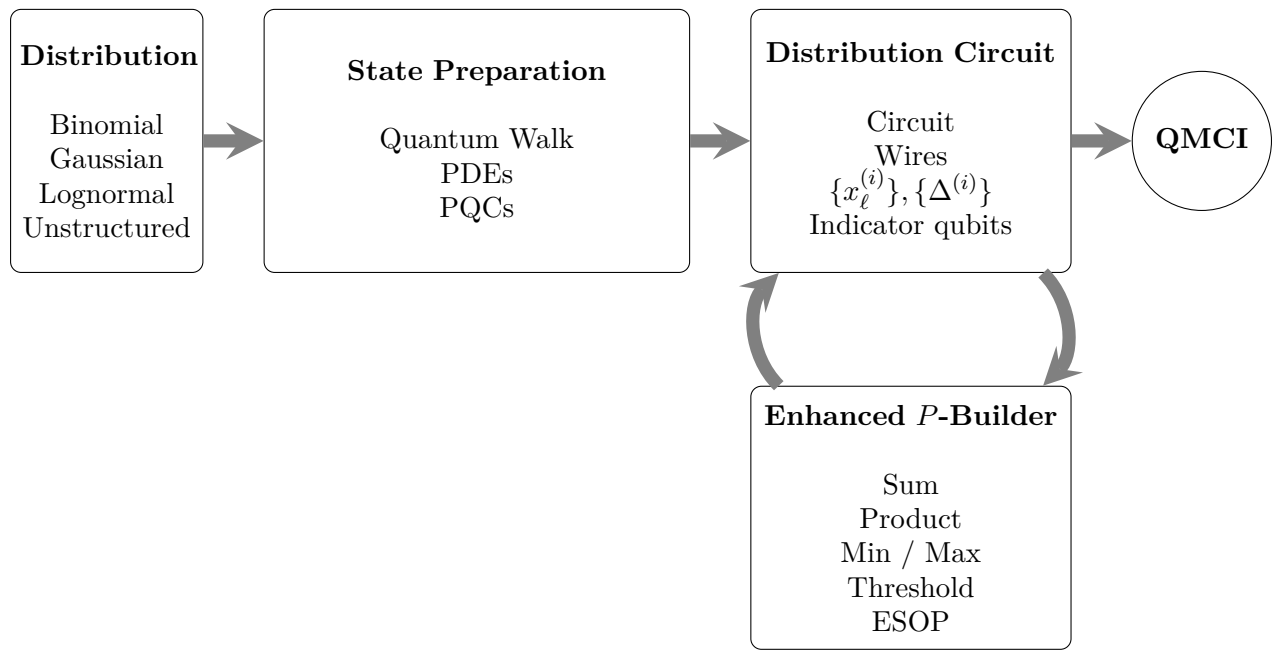

In this Section we present an overview of the modular architecture of Quantinuum's Quantum Monte Carlo Integration Engine. Fig.~\ref{fig:flow-qmci-engine} gives a high-level illustration, where the following six modules are included:
\begin{enumerate}
    \item Distribution circuit: the QMCI engine takes as an input the quantum circuit encoding the probability distribution of interest for the MCI. The precise architecture of this module is further broken down in Fig.~\ref{fig:flow-distrib-circuit-module}.
    \item Quantity to estimate: the QMCI engine takes as an input the function to apply to the random variables sampled in the MCI, i.e., the statistical quantity to estimate.
    \item Quantum amplitude estimation: the user can select which type of QAE algorithm is used. 
    \item Backend call: this compiles the required quantum circuits for the chosen backend before executing it. 
    \item Cloud call: this executes the quantum circuits required by QMCI on a simulator, emulator or quantum hardware.
    \item QMCI estimate: this returns the desired numerical MCI quantity and / or the resources needed to run the estimate. 
\end{enumerate}

Throughout the engine, careful attention is taken to ensure that the computational terms are adequately described. This is particularly well exemplified in the case of the additional data attached to the quantum circuit objects in order for them to be interpreted as preparing probability distributions (detailed in Section~\ref{subsect:qstates-as-prob}). Fig.~\ref{fig:flow-distrib-circuit-module} gives full details of how such distribution circuits are built:

\begin{enumerate}
    \item Distribution: this is a library of probability distributions of interest for the simulation coming e.g. from parametrised distributions (such as a Gaussian distribution) or stochastic models. This will eventually also contain multivariate distributions.
    \item State preparation: this is a library of quantum algorithms which can prepare the distribution chosen in \texttt{Distribution}. This library includes methods based on quantum walks and solving partial differential equations (PDEs) as well as pre-trained parameterised quantum circuits (PQCs).
    \item Enhanced $P$-builder: as detailed in Section~\ref{sec:enhanced-data-loader}, this module interacts with the distribution circuit to enhance it with circuitry related to basic operations on distributions such as sums, products, maxima / minimima and thresholds.
    \item Distribution circuit: this serves as an input of the QMCI engine and includes the quantum circuit to generate the distribution, the number of wires per dimension of the distribution and also the support and spacing of each dimension; as well as information about any indicator qubits introduced by the enhanced $P$-builder.
\end{enumerate}

The modular architecture has been designed to be extensible; for instance it is easy to add new forms of QAE without disturbing the essential operation of the QMCI engine. Fig.~\ref{fig:barrier-code} gives pseudocode for how the user can programme all of the options associated with these modules.

 \pagebreak

\section{Distribution loader}
\label{sec:data-loaders}

This Section describes the distribution loader\footnote{For our purposes, ``data loading'' and ``state preparation'' may be treated as synonyms of ``distribution loading''.} module of the Quantum Monte Carlo Integration engine. The design has versatility and extensibility at its heart: it encodes both univariate and multivariate probability distributions, and it is such that it will be easy to integrate future distribution loading methods. In general we may be interested in problems that are continuous or discrete in nature, and in the former case we require a suitable discretisation (as one does for classical computation). Indeed for a random continuous process described by a random variable $X$ with a PDF $f_X(x)$, to encode the probability distribution on a register, we first introduce its discretisation over a uniform grid of $2^n$ points $\{ x_1,\dots, x_{2^n} \}$ where points are at a distance $\Delta$ from each other, labelled by the binary encoding
\begin{equation}
\{     x_1 \equiv \underbrace{00\dots 00}_{n} , \; x_2 \equiv 00\dots 01, \; x_{2^n} \equiv 11 \dots 11 \} \
\end{equation}
The probability distribution is then evaluated on a discrete set $\Omega$ as in Proposition~\ref{lem1}, where the points are at distance $\Delta$ such that 
\begin{equation}
    p_i \approx  f_X(x_i)\Delta, \quad \forall i=1,\dots,2^n
\end{equation}
and following Eq.~\eqref{eq:def-prob-dist-state}, we prepare the following quantum state
\begin{equation}
\label{eq:data-loader-state}
    \ket{p} = P\ket{0^n} = \sum_{ x_i \in \{0,1\}^n} e^{2 \pi \I \varphi_i} \sqrt{p_{i}} \ket{x_i} \
\end{equation}
Notably the relative phases, $\{\varphi_i\}$, are arbitrary as we are only concerned with probabilities of computational basis state measurement outcomes. We envisage that the additional degrees of freedom that this property offers may help us to efficiently prepare a state encoding the required distribution in some cases.\\

To recap Section~\ref{sec:overview-engine}, the following information must be provided by the distribution loader in order for the downstream modules of the QMCI engine to correctly interpret the prepared state as an encoding of a probability distribution:
\begin{enumerate}
    \item the circuit $P$ preparing the quantum state;
    \item the list of the qubits encoding particular dimensions of the multivariate probability distribution (treating univariate distributions as a special case);
    \item the left endpoint of each dimension of the support of the distribution, denoted $x^{(i)}_\ell$ for the $i^{th}$ dimensions;
    \item the spacing between the points, denoted $\Delta^{(i)}$ for the $i^{th}$ dimension;
    \item the presence or one or more indicator qubits (as optionally included by using the enhanced $P$-builder).
\end{enumerate}
This information forms the attributes of the module \texttt{Distribution Circuit}, as presented in Fig.~\ref{fig:flow-distrib-circuit-module}.\\

In general there are three types of distribution loader:
\begin{enumerate}[label={(\roman*)}]
    \item classical methods to prepare some specific distribution;
    \item quantum approaches that aim to use actual quantum-mechanical phenomena to prepare some specific distribution;
    \item unstructured methods which consider $2^n$ positive real numbers that sum to one and prepare the corresponding PMF.
\end{enumerate}

Regarding ``classical methods'', Ref.~\cite{herbert2} shows that every classical sampling algorithm can be compiled as a Toffoli circuit applied to an initial uniformly random bitstring, and so every classical sampling algorithm can be turned into an appropriate circuit $P$ with about the same number of gates. This result was proposed principally to demonstrate that the quadratic reduction in sample complexity (of QMCI over classical MCI) can always be made to manifest as at least a quadratic reduction in computational complexity. However, using the \textit{q-marginal} construction of Ref.~\cite{herbert2} may also prove to be practically beneficial in certain circumstances. In particular, we envisage this approach potentially bearing fruit when preparing complicated random processes, that do not readily map into quantum circuits in any other manner. In the long-run we will supplement the enhanced $P$-builder to include all of the usual arithmetic operations needed to encode classical sampling algorithms into quantum circuits. For now, though, this is low priority and we have instead initially focused on included low-depth distribution-loading methods and the enhanced $P$-builder just has the functionality we need to construct simple (but still interesting) random processes (i.e., for Brownian and geometric Brownian motion) -- as well as the functionality to construct financial instruments.\\

``Quantum approaches'' is perhaps the most exciting of these, and certainly it is the one which has the most potential to complement the quantum advantage in QAE with a further advantage in the distribution loading. On this, we propose two approaches: firstly, using quantum walk to prepare a binomial distribution, detailed in Section~\ref{subsec:quantum-walk};  and secondly PDE-based methods to prepare Gaussian distributions, detailed in Section~\ref{subsec:pde}.\\

Whilst both of the above concern the situation where some analytically-defined probability distribution or random process is encoded in a quantum circuit, ``unstructured methods'' assumes no such structure (as its name implies). For this reason, it is the most general -- any analytically-defined distribution or process can be approximated as a suitable set of probability mass values -- however, it is also the least scalable: an $n$-qubit circuit requires $2^n$ classical values to be specified. For this reason, this method is only viable when relatively small resource states are required as inputs -- and are then turned into more complicated (multidimensional) random processes using the enhanced $P$-builder. This is exactly the setting that we are concerned with when constructing Brownian motion / geometric Brownian motion from \iid~Gaussian / lognormal probability distributions. For this reason, we have dedicated significant effort to unstructured data-loaders including implementing a simplification of the Grover-Rudolph method \cite{grover2002creating} and developing a method based on Ref.~\cite{Amy2020}. In each of these cases, we found that the circuits were far too deep to be viable for near- and medium-term use. We do, however, find that training PQCs as machine learning (generative) models to sample the desired probability distributions suffices as a means of unstructured data loading for the example benchmark problems we address in Section~\ref{sec:benchmarks}, and this is the focus of Section~\ref{subsec:param-quantum-circuit}.

\subsection{Quantum walk}
\label{subsec:quantum-walk}
The first method we present uses continuous quantum walks to prepare binomial distributions. The core of this method can be implemented using standard Hamiltonian simulation techniques \cite{qiang2016efficient, childs2003exponential, berry2009black}.\\

The binomial distribution is defined
\begin{equation}
\label{eq:binomial-distrib}
    p_i = \binom{n}{i}p^i (1-p)^{n-i}
\end{equation}

This method relies on the construction of a continuous quantum walk on a graph $\mathcal{G}$ whose transition matrix takes an initial state to a desired quantum state at a given time. In general if $\mathcal{G}$ is a graph with adjacency matrix $A$, then the continuous-time quantum walk on $\mathcal{G}$ is the quantum walk with transition matrix \cite{tutte2001graph, venegas2012quantum}
\begin{equation}
    U_{\mathcal{G}}(t) := e^{\I t A}
\end{equation}
To prepare the state of interest \eqref{eq:data-loader-state} we therefore require
\begin{equation}
    P \equiv U_{\mathcal{G}}(t)
\end{equation}

\begin{figure}[t!]
    \centering
\begin{tikzpicture}
  \fill (0,0) circle[radius=2pt] node[below=8pt] {0}; 
  \fill (1,0) circle[radius=2pt] node[below=8pt] {1}; 
  \draw (0,0) -- (1,0); 
\end{tikzpicture}
    \caption{Simplest complete graph with two vertices $K_2$}
    \label{fig:K2-graph}
\end{figure}
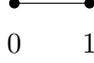

We begin by preparing the simplest probability distribution from a quantum walk
on a complete graph with two vertices, which we name $K_2$ and represent in Fig.~\ref{fig:K2-graph}. The adjacency matrix for this graph is the Pauli $X$ matrix \eqref{eq:K2-graph-matrices} and the transition matrix for the walk at time $t$, the matrix $U_{K_2}(t)$, can be written as

\begin{equation}
\label{eq:K2-graph-matrices}
  A=X=
  \begin{pmatrix}
0 & 1 \\
1  & 0
\end{pmatrix} \qquad 
  U_{K_2}(t) = \begin{pmatrix}
\cos(t) & \I \sin(t) \\
\I \sin(t) & \cos(t)
\end{pmatrix}
\end{equation}
Running the quantum walk until a time $\tau = \arccos{\sqrt{p}}$ induces the following transition matrix
\begin{equation}
    U_{K_2}(\tau) = \begin{pmatrix}
\sqrt{p} & \I \sqrt{1-p} \\
\I \sqrt{1-p} & \sqrt{p}
\end{pmatrix}
\end{equation}
This matrix acts on a 1-qubit state as a simple $R_x(.)$ rotation, i.e.
\begin{equation}
\label{eq:quantum-walk-k2}
    U_{K_2}(\tau)\ket{0} = \sqrt{p}\ket{0} + \I\sqrt{1-p}\ket{1}
\end{equation}
producing the Bernoulli distribution, which is the $n=1$ case of the Binomial distribution displayed in Eq.~\eqref{eq:binomial-distrib}.\footnote{The Bernoulli distribution can, of course, be prepared by a single $R_y$ gate, and so this is purely a motivational result for what will follow.} We now use this simple result to build a graph which prepares the Binomial distribution. 

\begin{definition}
If $\mathcal{G}_1$ and $\mathcal{G}_2$ are graphs with adjacency matrices $A(\mathcal{G}_1)$ and $A(\mathcal{G}_2)$, respectively, then
the Cartesian product of $\mathcal{G}_1$ and $\mathcal{G}_2$ is the graph $\mathcal{G}_1\boxtimes \mathcal{G}_2$ defined by the adjacency matrix
$A(\mathcal{G}_1\boxtimes \mathcal{G}_2) := A(\mathcal{G}_1) \otimes I + I \otimes A(\mathcal{G}_2)$. The following relationship between the transition matrices of $\mathcal{G}_1$, $\mathcal{G}_2$ and $\mathcal{G}_1\boxtimes \mathcal{G}_2$ is given by (see \cite[Eqs.~(34-35)]{christandl2005perfect} or \cite[Lemma 4.2]{godsil2012state})
\begin{equation}
\label{eq:product-morphism}
    U_{\mathcal{G}_1\boxtimes\mathcal{G}_2}(t) = U_{\mathcal{G}_1}(t) \otimes U_{\mathcal{G}_2}(t)
\end{equation}
\end{definition}

\begin{figure}[t!]
    \centering
   \begin{tikzpicture}[scale=2]
    \def\cubeSize{1}
    \foreach \x in {0,1}
        \foreach \y in {0,1}
            \foreach \z in {0,1}
                \node [circle,fill,inner sep=1.5pt,label={[xshift=0.4cm]below:$\x\y\z$}] at (\x*\cubeSize,\y*\cubeSize,\z*\cubeSize) {};
    
    \foreach \x in {0,1}
        \foreach \y in {0,1}
            \draw (\x*\cubeSize,\y*\cubeSize,0) -- (\x*\cubeSize,\y*\cubeSize,\cubeSize);
    
    \foreach \x in {0,1}
        \foreach \z in {0,1}
            \draw (\x*\cubeSize,0,\z*\cubeSize) -- (\x*\cubeSize,\cubeSize,\z*\cubeSize);
    
    \foreach \y in {0,1}
        \foreach \z in {0,1}
            \draw (0,\y*\cubeSize,\z*\cubeSize) -- (\cubeSize,\y*\cubeSize,\z*\cubeSize);
\end{tikzpicture}
    \caption{Representation of the hypercube $Q_3$ which is the 3-fold Cartesian product of $K_2$.}
    \label{fig:hypercube-q3}
\end{figure}
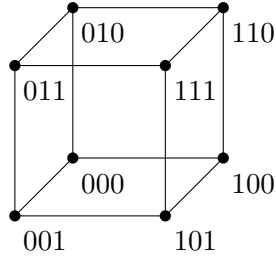

The $n$-fold Cartesian product $K_2\boxtimes K_2\boxtimes \cdots \boxtimes K_2:=K_2^{\boxtimes n}$ is equal to the $n$-dimensional hypercube $Q_n$ (see Fig.~\ref{fig:hypercube-q3} for an example with $n=3$), which has vertex set $\{0, 1\}^n$, and two vertices are adjacent if and only if their Hamming distance is $1$. (The Hamming distance between two bit strings of equal length is the number of different elements at the same position.) From Eqs.~\eqref{eq:product-morphism} and \eqref{eq:quantum-walk-k2}, we deduce the following result, which is a direct consequence of the tensor product structure. Let $u$ be a vertex in $Q_n$ of weight $n-k$, then we have the inner product
\begin{equation}
\label{eq:generation-bernoulli-distr}
    \abs{\bra{u}U_{Q_n}(\tau)\ket{0^n}}^2 =p^k (1-p)^{n-k}
\end{equation}
This construction is a product of Bernoulli states but not a Binomial state at this stage -- further work is needed to obtain the combinatorial factor $\binom{n}{k}$. 

\begin{definition}
For a given graph $\mathcal{G}$, a partition of $\mathcal{G}$ is a partition of its vertex set. We say a
partition $\pi = \{C_0, \ldots, C_n \}$ of $\mathcal{G}$ is equitable if for any two classes $C_i$, $C_j \in \pi$, the number
of neighbours of $v \in C_i$ in $C_j$ is a constant $b_{ij}$ which depends only on $i$ and $j$.
\end{definition}

For the $n$-dimensional hypercube $Q_n$, the partition $\pi = \{C_0, \ldots, C_n\}$ where $C_k$ is the
set of vectors in $\{0, 1\}^n$ of weight $k$, is equitable \cite{ge2011perfect}. The integers $b_{ij}$ are computed as
\begin{equation}
b_{ij} =
\begin{cases}
n-i & j=i+1 \\
i   & j=i-1 \\
0 & \text{otherwise}
\end{cases}
\end{equation}

\begin{definition}
Let $\hat{S}$ be
the matrix with
\begin{itemize}
    \item rows labelled by the vertices of $\mathcal{G}$,
    \item columns labelled by the classes of $\pi$,
    \item and whose $(v, C_j)$ entry is $1/\sqrt{|C_j|}$ if $v \in C_j$ and $0$ otherwise.
\end{itemize}
 
 We call the graph with adjacency matrix $\hat{A}:=\hat{S}^\intercal A \hat{S}$ the symmetrized quotient of $\mathcal{G}$. It has rows and columns
labelled by the classes of $\pi$. The $(C_i, C_j)$-entry of $\hat{S}^\intercal A \hat{S}$ is
$\sqrt{b_{ij} b_{ji}}$.
\end{definition}
Let $\hat{U}(t):= \exp(\I t\hat{A})$ be the transition matrix of the symmetric quotient of $\mathcal{G}$ for some
equitable partition $\pi$. Following Ref.~\cite[Theorem 1]{bachman2011perfect}, if $\pi$ contains a class $C_0$ with only one vertex, say $C_0 = \{a\}$, then
for any class $C_i \in \pi$ and any vertex $b \in C_i$, we have 
\begin{equation}
\label{eq:inner-product-symmetric-quotient}
 \bra{C_i}\hat{U}(t)\ket{C_0} = \sqrt{|C_i|}\bra{b}U(t)\ket{a}   
\end{equation}
The overall construction can be seen as a generalisation of Ref.~\cite{christandl2005perfect}. We denote the symmetrised quotient
of $Q_n$ by this partition as $\hat{Q}_n$, whose adjacency matrix of $\hat{A}$ has entries
\begin{equation}
\hat{A}_{i,j} =
\begin{cases}
\sqrt{(n-i)(i+1)} & j=i+1 \\
\sqrt{(i-1)(n-i+1)} & j=i-1 \\
0 & \text{otherwise}
\end{cases}
\label{eq:adjacency_matrix_entries}
\end{equation}
for $0\leq i,j\leq n$. Note that $\hat{A}$ is the adjacency matrix of a weighted path. Since the class $C_k$ of $\pi$ has size $|C_k|=\binom{n}{k}$ and considering $C_0=\{\ket{0^n} \}$, we have our final result presented in Theorem~\ref{thm:binomial-state}. 

\begin{theorem}
\label{thm:binomial-state}
Given $\hat{U}(t)$, the transition matrix of $\hat{Q}_n$, we have at time $\tau = \arccos \sqrt{p}$
\begin{equation}
\begin{split}
\abs{\bra{ C_k}\hat{U}(\tau)\ket{C_0} }^2 &=\abs{\bra{ C_k}\hat{U}(\tau)\ket{0}}^2\\
&=\binom{n}{k}p^{k}(1-p)^{n-k}
\end{split}
\label{eq:binomial_distribution}
\end{equation}
for any $0\leq k\leq n$. That is, the graph $\hat{Q}_n$ prepares the Binomial distribution $\mathcal{B}(n,p)$.
\end{theorem}

The implementation of such procedure is then done by encoding a Hamiltonian representing the adjacency matrix
\begin{equation}
    \mathcal{H}_{\mathcal{B}}:=\hat{A}
\end{equation}
and running Hamiltonian evolution. Motivated by the general result of Theorem~\ref{thm:binomial-state}, we derive explicitly in Example~\ref{example:quantum-walk-3} the construction for $n=3$: this is special case of the above theorem that is small enough to follow explicitly, but large enough to be non-trivial.

\begin{example}[Example with $Q_3$, $n=3$]
\label{example:quantum-walk-3}
    Let us study in details the case of the hypercube $Q_3$ depicted in Fig.~\ref{fig:hypercube-q3}. The vertex set reads 
    \begin{equation}
      \{ (000), (001), (010),(100), (110), (101), (011), (111)\}  
    \end{equation}
We denote this vertex set in parenthesis because these are intermediate objects not representing our final qubits. We then partition the vertex set by Hamming weight in four classes as
 \begin{equation}
    C_0 = \{ (000) \} , \; C_1 = \{ (001),  (010), (100) \},    C_2 = \{ (110), (101), (011)\}, \;C_3 = \{ (111) \} 
 \end{equation}
	Each class of this partition will thereafter represent a different quantum state.\\

	Note that the size of each subset is given by $|C_i|={3 \choose i}$. To confirm that this gives an equitable partition, we check that the number of neighbours of a vertex $v \in C_i$ in the class $C_j$ is a constant $b_{ij}$ which only depends on $\{i,j \}$, and not on the vertex we choose from $C_i$.\\
 
 For example, let's look at $C_1,C_2$: for any vertex $v$ we choose from $C_1$, it will have $2$ neighbours in $C_2$. For example, $(001)$ has neighbours $(101)$ and $(011)$ in $C_2$; $(100)$ has neighbours $(110)$ and $(101)$ in $C_2$. We can check this for any two classes $C_i$ and $C_j$ and see that the same property holds. From considering $C_1, C_2$, we can see that $b_{1,2} = 2$. In general, we get the constants:
 \begin{equation}
     b_{0,1} = 3, \; b_{1,2} = 2, \; b_{2,3} = 1,  \; b_{3,2} = 3, \; b_{2,1}= 2, \; b_{1,0}=1
 \end{equation}
	and $0$ otherwise. The matrix $\hat{A}$ whose entries are computed by $\sqrt{b_{ij}b_{ji}}$ therefore reads
\begin{equation}
\begin{split}
    	 \hat{A} 
 &=  \begin{pmatrix}
	0 & \sqrt{3} & 0 & 0 \\
	\sqrt{3} & 0 & 2 & 0 \\
	0 & 2 & 0 & \sqrt{3} \\
	0 & 0 & \sqrt{3}  & 0 \\
	\end{pmatrix} 
 \end{split}
\end{equation}

	We choose to label the rows and columns of $\hat{A}$  by new basis states $\ket{00}, \ket{01}, \ket{10}, \ket{11}$, which corresponds to the classes $C_0,C_1,C_2,C_3$, respectively. From Eq.~\eqref{eq:inner-product-symmetric-quotient}, the coefficients of the transition matrix $\hat{U}(t):= \exp(\I t\hat{A})$ are given by 
 \begin{equation}
     \langle i | \hat{U}(t)|00\rangle = \sqrt{|C_i|}\langle{b_i}|U(t)|(000)\rangle 
 \end{equation}
 for any element $b_i$ in the class $C_i$, and where we used the mapping
\begin{equation}
    \{ C_0,C_1,C_2,C_3\} \mapsto \{ \ket{00}, \ket{01}, \ket{10}, \ket{11} \}
\end{equation}
 
	We can write the basis states on the left side in binary representation to make it clearer how to interpret this on the quantum computer: 
 \begin{equation}
 \begin{split}
     & \bra{ 00 } \hat{U}(t)\ket{00} = \sqrt{{3 \choose 0}}\langle{(000)}|U(t)|(000)\rangle , \quad \langle 01 | \hat{U}(t)|00\rangle = \sqrt{{3 \choose 1}}\langle{(001)}|U(t)|(000)\rangle , \\
     &\langle 10 | \hat{U}(t)|00\rangle = \sqrt{{3 \choose 2}}\langle{(011)}|U(t)|(000)\rangle  , \qquad \langle 11 | \hat{U}(t)|00\rangle = \sqrt{{3 \choose 3}}\langle{(111)}|U(t)|(000)\rangle 
\end{split}
 \end{equation}

	We already showed in Eq.~\eqref{eq:generation-bernoulli-distr} that $|\langle{j}|U(t)|(000)\rangle|^2 = p^k(1-p)^{n-k}$ when $j$ is a vertex that has weight $k$, allowing us to obtain the  the binomial distribution.
\end{example}

We have studied here a particular graph -- the Cartesian product of $K_2$ -- to generate the Binomial distribution. More generally, we expect continuous-time quantum walk to be an important way of preparing quantum states. In particular, the fact that this method uses the intrinsically quantum-mechanical nature of quantum walk to prepare a classically-defined probability distribution suggests that it may be an effective means of preparing large quantum states, where classical (Toffoli circuit) approaches prove to be expensive. In the immediate future, however, the reliance on Hamiltonian simulation puts it out of reach for all but the most trivial probability distributions.\\

\subsection{Partial differential equations}
\label{subsec:pde}

The second method we present is based on the solution of PDEs. Indeed, one way to generate a probability distribution is to solve a PDE called the Fokker-Planck equation for stochastic systems \cite{risken1996fokker}.
As a simple example let us consider the following heat equation in $1+1$ dimensions for a field $f(x, t)$ depending on one spatial coordinate $x \in \R$ and one time coordinate $t > 0$:
\begin{equation}
 \p_t f = \p_x^2 f .
\end{equation}
If one starts from the initial condition $f(x,t=0) = \delta(x)$ where $\delta$ is the Dirac delta function, then the standard normalised solution is given by the heat kernel
\begin{equation}
 f(x, t)=\frac{1}{\sqrt{4 \pi t}} e^{-\frac{x^2}{4t}}, \quad \int_\R  f(x,t) \, \rmd x = 1, \quad \forall t > 0
\end{equation}
It is then possible to obtain a Gaussian with an arbitrary variance depending on how long we let the heat equation evolve.
We can change the mean of the Gaussian by inducing a translation in the initial condition.\\

More generally, the Fokker-Planck equation can generate arbitrary probability distributions that arise from stochastic dynamics such as the following one-dimensional stochastic differential equation
\begin{equation}\label{eq:stochastic-dynamics}
 \rmd X_t = \mu(X_t, t) \rmd t + \sigma(X_t, t) \rmd W_t
\end{equation}
Here $\mu$ is the drift, $\sigma$ defines the so-called diffusion coefficient $D = \sigma^2 / 2$ and $W_t$ is a standard Wiener process \cite{ross1995stochastic}.
The PDF, $f_X(x, t)$, of observing the process $X_t$ at position $x$ and time $t$ then obeys the PDE
\begin{equation}
\label{eq:one-dim-fokkerplanck}
 \p_t f = -\p_x [\mu(x, t) f] + \p_x^2 [D(x,t) f].
\end{equation}
Note that equations~\eqref{eq:stochastic-dynamics} and~\eqref{eq:one-dim-fokkerplanck} can be extended to higher-order spatial dimensions allowing a potential encoding of multivariate distributions.\\

There exist quantum algorithms which can solve linear and nonlinear PDEs and encode their solution in the amplitudes of a wave function \cite{Wi96, Za98, LeOs08, CaEtAl13, Be14, MoPa16, BeEtAl17, CoJoOs19, lubasch2020variational, ChLi20, lloyd2020quantum, ChLiOs21, liu2021efficient, pool2022solving, JiLiYu22, JiLiYu22b}.
Common approaches to solving linear differential equations, such as the Fokker-Planck equation, are Hamiltonian simulation~\cite{Wi96, Ll96} and the quantum linear systems algorithm (widely known as ``HHL'' -- the initials of the authors)~\cite{HaHaLl09}.
In the context of Hamiltonian simulation it is convenient to rewrite the differential equation considered as a Schr\"{o}dinger equation and then use quantum algorithms for simulating that Schr\"{o}dinger equation.
Based on that idea a versatile framework, so-called \textit{Schr\"{o}dingerisation}, for differential equations was proposed recently~\cite{JiLiYu22, JiLiYu22b}.
HHL is a quantum algorithm for solving linear equations and can be applied to time-dependent differential equations by discretising space and time and expressing the entire evolution as a single system of linear equations~\cite{Be14}.
For the Fokker-Planck equation it was argued that HHL has advantages over Hamiltonian simulation via Schr\"{o}dingerisation~\cite{GnSuSa23}.
Both HHL and Hamiltonian simulation applied to the Fokker-Planck equation are, in principle, efficient in the number of variables and, therefore, should not suffer from the curse of dimensionality encountered by classical solvers.
However, these quantum algorithms lead to deep circuits in general and can require a large number of ancilla qubits.\\

One possibility to reduce the quantum hardware requirements of solving PDEs is to make use of variational quantum algorithms such as~\cite{lubasch2020variational, BeFiLu21} or to classically optimize quantum circuits as in~\cite{McLu23, KiEtAl23}.
Additional variational approaches to state preparation are presented in the following section.

\subsection{Parameterised quantum circuits}
\label{subsec:param-quantum-circuit}

The third state preparation method we present here uses PQCs~\cite{benedetti2019parameterized}.
 We explore this method to prepare Gaussian and lognormal distributions. In general PQCs are trained to minimise the $L_\alpha$ norm for some $\alpha$. Specifically, for a target state $\ket{\psi_{\text{target}}}$ and an approximate version thereof prepared by a trained PQC, $\ket{\psi_{\text{PQC}}}$, the $L_\alpha$ norm is defined:
 \begin{equation}
     ||\ket{\psi_{\text{target}}} - \ket{\psi_{\text{PQC}}}||_\alpha = \left( \sum_{i=1}^N  \left( \left[\ket{\psi_{\text{target}}} - \ket{\psi_{\text{PQC}}} \right]_i \right)^n \right)^{1/\alpha}
 \end{equation}
 where $\ket{\psi_{\text{target}}}$ and $\ket{\psi_{\text{PQC}}}$ are $N$-element vectors. $L_1$, $L_2$ and $L_\infty$ norms are most commonly used as the distances to be minimised in PQC training, and in the last case this amounts to the greatest absolute discrepancy (for any element) between the target and prepared states. We now consider two classes of PQC ansatz.

\subsubsection{Hardware-efficient ansatz} 

In the following we consider a PQC ansatz $U(\vec{\theta})$ composed of $L+1$ layers for a register of $n$ qubits which overall defines $n \times (L+1)$ variational parameters according to

\begin{equation}
U(\vec{\theta}) =  U_R(\vec{\theta}^{ L+1})~ \overbrace{
U_{\rm ENT} U_R(\vec{\theta}^{ L}) \ldots U_{\rm ENT} U_R(\vec{\theta}^{ 1})}^{L\rm{-times}}
\label{eq:trial_ansatz_u_theta}
\end{equation} where the free parameters are the angles of the $R_y(.)$ rotations, i.e.
\begin{equation}
U_R (\vec{\theta}^k) = \bigotimes_{i=0}^{n-1} R_y({\theta}_{i}^k)
\end{equation}
and the entanglement is applied by a (fixed) set of gates $U_{\rm ENT}$. For clarity, we summarize our different notations:
\begin{enumerate}
    \item $\vec{\theta}$ stands for the vectors of all parameters
    \item $\vec{\theta}^k$ stands for the vector of parameters in the $k$-th layer of the PQC
    \item $\theta^k_i$ is the scalar angle of rotation of the $i$-th qubit in the $k$-th layer of the PQC.
\end{enumerate}

Starting from the state $\ket{0^n}$, the state obtained using our ansatz is then
\begin{equation}
\ket{\psi_{\text{PQC}}} = \ket{\psi(\vec{\theta})} = U(\vec{\theta}) \ket{0^n}
\label{eq:trial_ansatz}
\end{equation}
\begin{figure}[t!]
\centering
    \begin{subfigure}[a]{\textwidth}
    \centering
    \includegraphics[width=94.848mm]{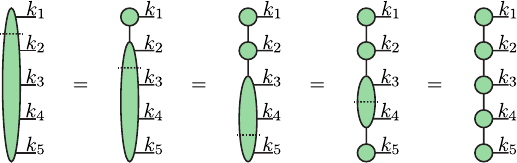}
    \caption{}
    \end{subfigure}\\
        \begin{subfigure}[b]{\textwidth}
    \centering
    \includegraphics[width=94.848mm]{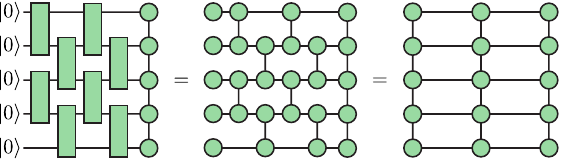}
    \caption{}
    \end{subfigure}\\
            \begin{subfigure}[c]{\textwidth}
    \centering
    \includegraphics[width=94.848mm]{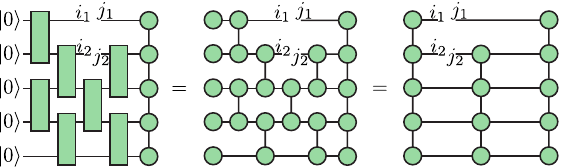}
    \caption{}
    \end{subfigure}\\
\caption{
Five-qubit quantum state preparation using the TN approach in Ref.\cite{lubasch2020variational}.
(A) The vector representing $\sqrt{p(x_{k})}$ for a probability distribution $p(x)$ evaluated on an equidistant grid of $2^{5}$ grid points, $x_{k}$, is written as a tensor with $5$ indices $(k_{1}, k_{2}, k_{3}, k_{4}, k_{5}) = \text{binary}(k)$.
An exact matrix product state (MPS)~\cite{VeMuCi08, Or14, Ba23} representation is obtained by performing several QR or polar decompositions at the dotted lines.
An approximate MPS representation can be obtained by using truncated SVD instead of the QR or polar decompositions.
(B) The quantity $\mathcal{F} = \langle \psi_{\text{MPS}} | \psi_{\text{PQC}} \rangle$ for a variational quantum brickwall circuit preparing the state denoted $| \psi_{\text{PQC}} \rangle$ of circuit depth $4$ is written as a TN.
To achieve this, the vector $|0\rangle$ is written as a tensor with one index and every two-qubit gate is QR- or polar-decomposed into two tensors with three indices each.
Finally we multiply adjacent tensors to turn $\mathcal{F}$ into a TN consisting of MPS and a matrix product operator (MPO)~\cite{VeGaCi04, ZwVi04}.
(C) The TN required for the update of a variational two-qubit gate is obtained by removing that gate from the $\mathcal{F}$ TN.
Then the same steps as in (B) are used to transform the resulting TN into one containing only MPS and MPO.
Contracting the resulting TN gives a matrix $A_{(i_{1}, i_{2}), (j_{1}, j_{2})}$ with row-multi-index $(i_{1}, i_{2})$ and column-multi-index $(j_{1}, j_{2})$.
The optimal unitary $\tilde{U}$ maximizing $\mathcal{F} = \text{tr}(A \tilde{U})$, where $\text{tr}(\cdot)$ denotes the trace over $(\cdot)$, follows from the SVD of $A = U S V$: $\tilde{U} = V^{\dag} U^{\dag}$~\cite{Ke75}.
We loop over all of the variational two-qubit gates and for each compute the optimal gate and insert it in the TN.
This procedure is iterated until the cost function (here the $L_2$ norm) converges or until a set maximum number of iterations is exhausted.
}
\label{fig:TN}
\end{figure}
This type of ansatz is part of the class of hardware-efficient (HWE) ansatze \cite{kandala2017hardware}, as the entangling gates $U_{\rm ENT}$ can be chosen differently for different physical devices. In order to train and optimise the rotation angles we minimise the $L_1$, $L_2$ and $L_\infty$ norms\footnote{The $L_\infty$ is also useful for bounding the discrepancy between continuous PDFs and the discretised version thereof. 
}. Other cost functions which are physics-inspired can also be introduced \cite{chakrabarti2020threshold}.\\

\subsubsection{Tensor network methods with brickwall ansatz}
\label{subsect:MPS}

Another approach is to use a brickwall ansatz circuit composed of generic two-qubit gates. This is then used for quantum state preparation based on tensor network (TN) methods~\cite{VeMuCi08, Or14, Ba23} as was used in Ref.~\cite{lubasch2020variational}. In this case, the $L_2$ norm is chosen to measure the distance between the target and trained state, and the training uses singular value decomposition (SVD) of the environment TN \cite{Ke75}, as detailed in Fig.~\ref{fig:TN}.

\subsubsection{Training a PQC to prepare a Gaussian distribution}

\begin{figure}[t!]
    \centering
    \includegraphics{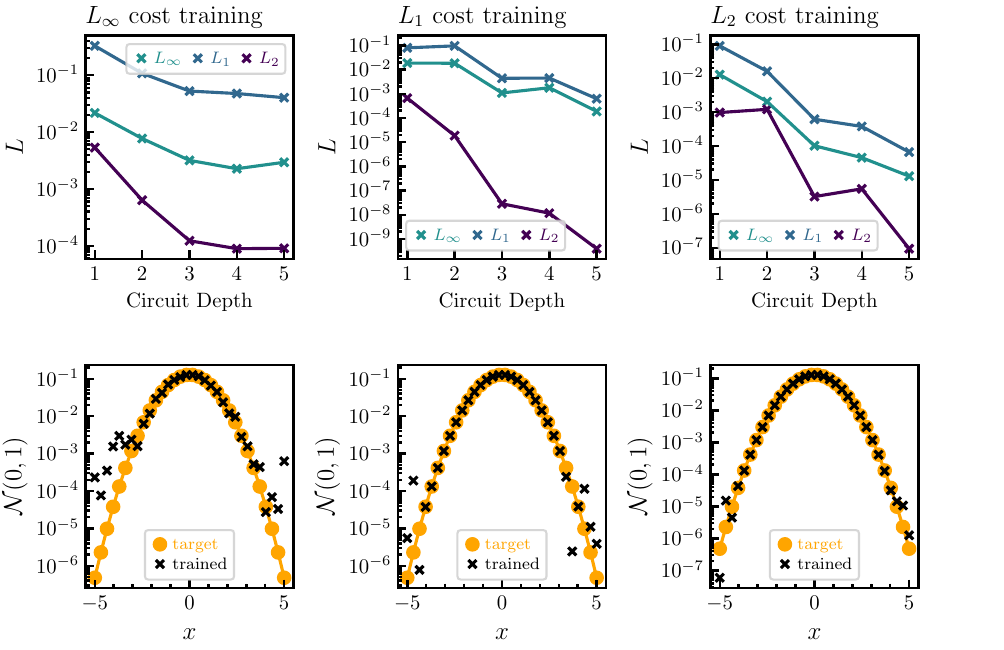}
    \caption{Training of a Gaussian random variable $\mathcal{N}(0,1)$ with respect to three cost functions ($L_\infty$, $L_1$, $L_2$) as a function of the circuit depth for a HWE ansatz. The second row shows the prepared 32-point probability mass functions versus the target unit Gaussian for the maximum circuit depth of five layers. The quality of the training is measured by the $L$ norm discrepancy of the prepared amplitudes with respect to the ground truth. We observe that the $L_2$ cost function provides the best training performance for the various metrics measured.}
    \label{fig:normal-training-cost-comparison}
\end{figure}

\begin{figure}[t!]
    \centering
    \includegraphics[scale=0.9]{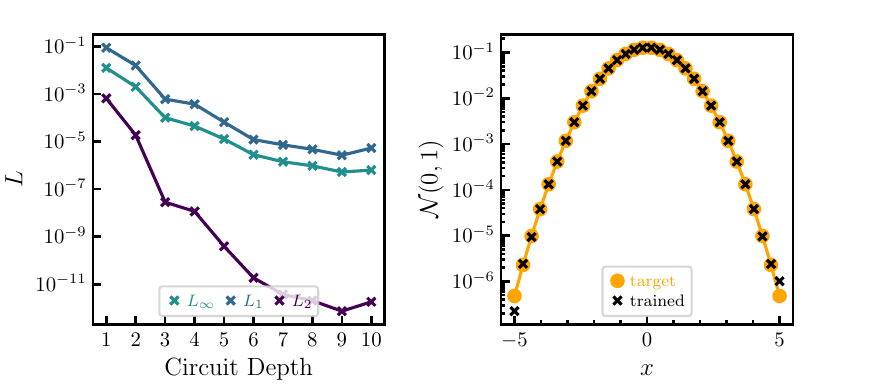}
    \caption{Training of a Gaussian random variable $\mathcal{N}(0,1)$ with respect to the $L_2$ norm cost function for a HWE ansatz. We can see in the left-hand subfigure the benefit of adding additional layers, and the right-hand subfigure shows the prepared 32-point probability mass functions versus the target unit Gaussian for the maximum circuit depth of nine layers (i.e., the number of layers with best performance).}
    \label{fig:2_normal_training_L_2_cost}
\end{figure}

\begin{figure}[t!]
    \centering
    \includegraphics[width=0.8\textwidth]{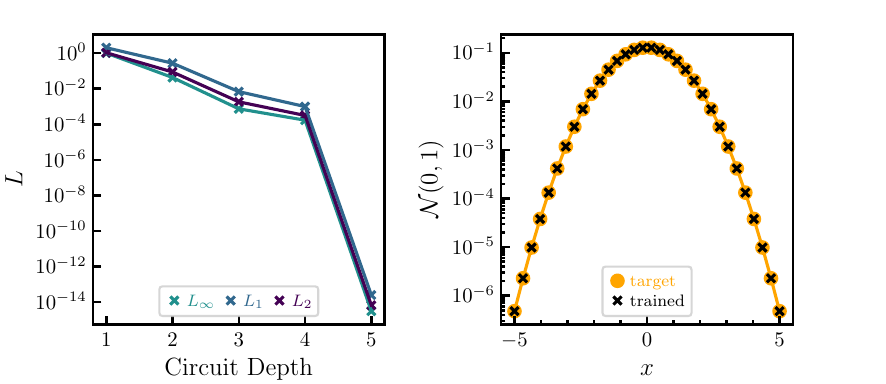}
    \caption{Training of a Gaussian random variable $\mathcal{N}(0,1)$ with respect to the $L_2$ norm cost function for TN training with a brickwall ansatz. We can see in the left-hand subfigure the benefit of adding additional layers, and the right-hand subfigure shows the prepared 32-point probability mass functions versus the target unit Gaussian for the maximum circuit depth of five layers (i.e., the number of layers with best performance).}
    \label{fig:2_normal-MPS}
\end{figure}

We first train a PQC to prepare a Gaussian random variable with 0 mean and variance 1, i.e.
\begin{equation}
    X \sim \mathcal{N}(0,1)
\end{equation}
which is known as a ``unit Gaussian''. By shifting and re-scaling this variable, we can then obtain samples from any $\mathcal{N}(\mu,\sigma^2)$. In terms of the distribution loader this means that the circuit is untouched, and the classical data describing how to interpret the quantum circuit as preparing an encoding of a probability distribution (namely $x_\ell$ and $\Delta$) is updated accordingly. Hence the training only has to be done for the unit Gaussian. Figure~\ref{fig:normal-training-cost-comparison} gives the results when a HWE-ansatz PQC with 5 qubits is used to prepare a $2^5 = 32$ point discrete approximation of the Gaussian distribution over the range $[-5,5]$. In particular, it shows results for the cases where each of the $L_\infty$, $L_1$ and $L_2$ norms are minimised, and we observe that overall $L_2$ norm provides the best performance. Henceforth we therefore only train to minimise the $L_2$ norm, and focus on deeper ans\"{a}tze to improve the overall performance. In particular, Fig.~\ref{fig:2_normal_training_L_2_cost} shows a 5-qubit approximation of the unit Gaussian with nine layers. We also obtain good results for preparing a 32-point discrete approximation of the unit Gaussian using the TN training with the brickwall ansatz, as shown in Fig.~\ref{fig:2_normal-MPS}.\\

\subsubsection{Training a PQC to prepare a lognormal distribution}

\begin{figure}[t!]
    \centering
    \includegraphics[scale=0.9]{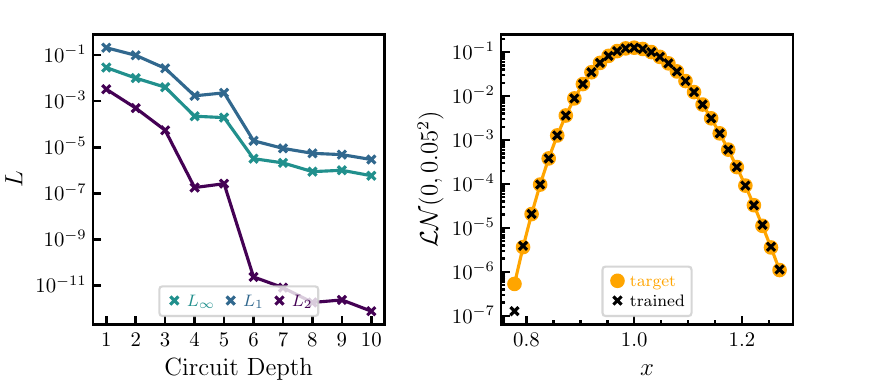}
    \caption{Training of a five-qubit approximation of the lognormal distribution with parameters $\sigma=0.05$ and $\mu=0$ with respect to $L_2$ cost function using a HWE ansatz.}
    \label{fig:3_lognormal_training_L_2_cost}
\end{figure}

\begin{figure}[t!]
    \centering
    \includegraphics[width=0.8\textwidth]{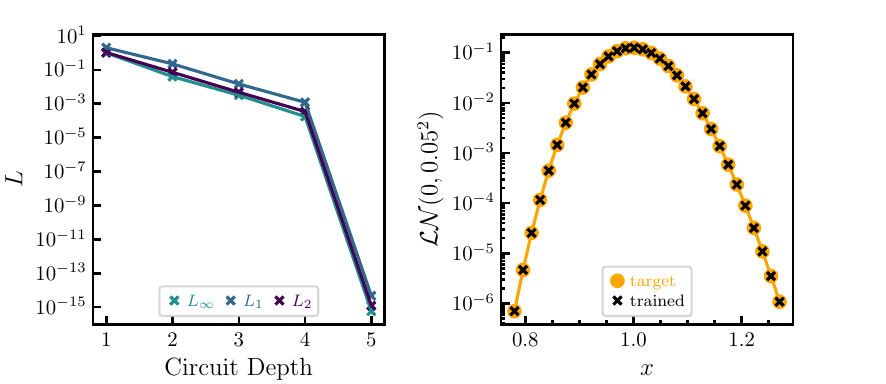}
    \caption{Training of a five-qubit approximation of the lognormal distribution with parameters $\sigma=0.05$ and $\mu=0$ with respect to $L_2$ cost function using a brickwall ansatz with TN training.}
    \label{fig:3_lognormal-MPS}
\end{figure}

Unlike for the Gaussian distribution, the lognormal distribution does not enjoy the location-scale property. This means that it is not possible to simply train a generic Lognormal, $\mathcal{LN}(\mu_0,\sigma^2_0)$  for a fixed set of parameters $\{\mu_0, \sigma^2_0 \}$ to obtain the samples for any Lognormal. More precisely, if $\sigma^2$ is fixed, then it is possible to obtain samples for any $\mu$ by re-scaling, and hence we focus on Lognormals of the form $\mathcal{LN}(0,\sigma^2)$. This still means that we must select $\sigma$, and here we set $\sigma=0.05$ -- which matches the low-volatility setting we address in the benchmarks in Section~\ref{sec:benchmarks}. For HWE and brickwall (with TN training) ansatz respectively, we show in Figs.~\ref{fig:3_lognormal_training_L_2_cost} and \ref{fig:3_lognormal-MPS} the training of the quantum circuit using 5 qubits with the $L_2$ cost function for the lognormal distribution with these parameters.

\subsection{Benchmarks}
\label{subsec:benchmark-data-loading}

For \textit{every} quantum circuit that encodes a probability distribution, we must ask two crucial questions: how well do samples from the quantum state prepared match those from the desired probability distribution, and what quantum resources are needed to achieve this. In general there will be a trade-off between these two desiderata -- higher quality approximations will require more quantum resources. In this section, we define a set of metrics to measure how well a certain distribution loading method performs, according to these two criteria. We then evaluate these for five- and six-qubit Gaussian and lognormal distributions prepared with the two PQC training methods described in Section~\ref{subsec:param-quantum-circuit}. Pseudocode giving an example to illustrate how these circuits can be loaded to prepare the desired probability distribution is given in Fig.~\ref{fig:distr-load-code}.

\subsubsection{Statistical figures of merit to assess the quality of a distribution loading routine}

We introduce common metrics used in statistics to compare two discrete probability distributions $\mathtt{P}$ (resp. $\mathtt{Q}$) with probability masses $\{p_i\}_{i \in \Omega}$ (resp. $\{q_i\}_{i \in \Omega}$). The first is the Kullback-Leibler (KL) divergence, or relative entropy\footnote{Here we use the natural logarithm in the definition -- even though $\log_2$ is more conventional in information theoretic circles.}
\begin{equation}
    \KL{\mathtt{P}}{\mathtt{Q}} = \sum_{i \in \Omega} p_i \log\left(\frac{p_i}{q_i}\right)
\end{equation}
The second is the Jensen-Shannon (JS) divergence, which is a symmetrised version of the KL divergence
\begin{equation}
    \JS{\mathtt{P}}{\mathtt{Q}} = \frac{1}{2}\left(\KL{\mathtt{P}}{\frac{\mathtt{P}}{2}+\frac{\mathtt{Q}}{2}}+\KL{\mathtt{Q}}{\frac{\mathtt{P}}{2}+\frac{\mathtt{Q}}{2}}\right)
\end{equation}
The third is the total-variation distance
\begin{equation}
    \delta(\mathtt{P},\mathtt{Q})=\frac{1}{2}\sum_{i \in \Omega} \, \abs{p_i-q_i}
\end{equation}
The fourth is the infidelity, which is one minus the fidelity, defined as
\begin{equation}
\label{eq:def-infidelity}
    F(\mathtt{P},\mathtt{Q}) =  \left(\sum_{i \in \Omega} \sqrt{p_i q_i}\right)^2
\end{equation}
The $L_1$, $L_2$ and $L_\infty$ norms also give measures on the closeness of fit, and in particular the $L_\infty$ norm is useful for upper-bounding how close the final Monte Carlo estimate can be to the true solution.

\begin{figure}
    \begin{subfigure}{0.8\textwidth}
    \centering
\begin{minted}[fontsize=\footnotesize,
    linenos,
    style=manni,
    tabsize=4, 
    autogobble,
    numbersep=8pt]{python}
state_preparation = StatePreparation()
state_preparation_parameters = StatePreparationParameters(
    distribution=Normal(mean=0, variance=1),  # Load the univariate unit Gaussian
    n_qubits=6,                               
    xl=-5,                                    
    delta=10/63,                              
    ansatz="HWE",                             # (or "TN") Specify the ansatz/loader
    n_layers=6,                               # Number of ansatz layers
    )                              
distribution_circuit = state_preparation.get_circuit(state_preparation_parameters)
\end{minted}
\end{subfigure}
\caption{Example pseudocode for distribution loading.}
\label{fig:distr-load-code}
\end{figure}

\subsubsection{Quantum resource figures of merit}

In order to assess the resources required for some distribution loading method, we first need to compile the circuit into a standard form. Compiling to the gateset $\{\text{TK}1, \text{CNOT} \}$, where 
\begin{equation}
\label{eq:def-tk1}
 \text{TK}1 = R_z(\alpha) R_x(\beta) R_z(\gamma)   \, .
\end{equation}
(for some $\alpha, \, \beta, \, \gamma$) captures all single-qubit unitaries is a suitably generic choice for this. We then count the following resources: the number qubits in the prepared state (``state qubits'') and ancilla qubits required; the circuit, CNOT and TK1 depths; as well as the number of total gate, CNOT and TK1 gate counts as obtained from the \texttt{tket} methods \texttt{n\_gates\_of\_type(OpType.CX)} and \texttt{depth\_by\_type(OpType.CX)} (and similarly by replacing CX by TK1).

\subsubsection{Numerical experiments} 

We now evaluate these figures of merit for trained PQCs encoding five- and six-qubit Gaussian and lognormal distributions (for both straightforward $L_2$-norm minimisation with a HWE ansatz and MPS training with a brickwall ansatz). (Note that each of these have been studied before in the literature, for the Gaussian distribution, see Refs.~\cite{chakrabarti2020threshold, mcardle2022quantum, rattew2021efficient, herman2022survey, holmes2020efficient, gundlapalli2022deterministic, zoufal2019quantum}, and for the lognormal distribution, see Refs.~\cite{holmes2020efficient, gundlapalli2022deterministic, zoufal2019quantum}.) The six-qubit Gaussian and lognormal distributions encoded in HWE are subsequently used in the benchmarks detailed in Section~\ref{sec:benchmarks}, and the full circuits are included in Appendix~\ref{sec:benchmark-lognormals}. The results are displayed in Tables~\ref{tab:gaussian-stats} and \ref{tab:lognormal-stats}~\footnote{All circuits were rebased using \texttt{RebaseTket} and were then optimised using the \texttt{FullPeepholeOptimise} of pytket, the documentation of pytket can be found at \url{https://cqcl.github.io/tket/pytket/api/}.}.\\

\begin{table}[t!]
\centering
\begin{tabular}{l c c c c }
\hline
\hline\\[-0.8ex]
 & HWE & TN & HWE & TN  \\[1ex]
 \hline
Number of state qubits  &   5 & 5 & 6 & 6\\
Number of ancilla qubits  & 0 & 0 & 0 & 0 \\
Circuit depth             & 18 & 21 & 22 & 37 \\
CNOT depth                  &12 & 10 & 15 & 18 \\
TK1 depth                 &6 & 11 & 7 & 19 \\
Number of gates           &50 & 65 & 72 & 142 \\
Number of CNOT gates        &20 & 20 & 30 & 46 \\
Number of TK1 gates       &30 & 45 & 42 & 96 \\
KL divergence(target, trained) & $1.85 \times 10^{-5}$ & $1.25 \times 10^{-14}$ & $1.14 \times 10^{-4}$ & $7.65 \times 10^{-15}$ \\
KL divergence(trained, target)  & $2.84 \times 10^{-5}$ & $1.25 \times 10^{-14}$ & $2.14 \times 10^{-4}$& $6.86 \times 10^{-15}$ \\
JS divergence                   & $5.30 \times 10^{-6}$ & $1.38 \times 10^{-17}$ & $2.96 \times 10^{-5}$& $1.19 \times 10^{-16}$ \\
Total-variation distance        & $3.31 \times 10^{-5}$ & $1.26 \times 10^{-14}$ & $2.17 \times 10^{-4}$ & $3.10 \times 10^{-9}$ \\
$L_\infty$                      & $1.29 \times 10^{-5}$ & $3.14 \times 10^{-15}$ & $2.58 \times 10^{-5}$ & $7.37 \times 10^{-10}$ \\
\hline
\hline
\end{tabular}
\caption[Performance of Gaussian distribution loaders]{Numerical data for the circuits preparing a Gaussian random variable $\mathcal{N}(0,1)$ on the interval $[-5,5]$. }
\label{tab:gaussian-stats}
\end{table}

\begin{table}[t!]
\centering
\begin{tabular}{l c c c c  }
\hline
\hline\\[-0.8ex]
 & HWE & TN & HWE & TN  \\[1ex]
 \hline
Number of state qubits            & 5 & 5 & 6 & 6 \\
Number of ancilla qubits         & 0 & 0 & 0 & 0 \\
Circuit depth                   & 21 & 21 & 34 & 37 \\
CNOT depth                        & 14 & 10 & 23 & 18 \\
TK1 depth                        & 7 & 11 & 11 & 19 \\
Number of gates                  & 59 & 64 & 116 & 142 \\
Number of CNOT gates               & 24 & 20 & 50 & 46 \\
Number of TK1 gates              & 35 & 44 & 66 & 96 \\
KL divergence(target, trained)  & $1.08\times 10^{-5}$ & $4.12 \times 10^{-15}$ & $9.39\times 10^{-6}$  & $9.04 \times 10^{-11}$ \\
KL divergence(trained, target)  & $2.12\times 10^{-6}$ & $4.10 \times 10^{-15}$ & $1.54\times 10^{-5}$ & $9.05 \times 10^{-11}$ \\
JS divergence                   & $6.85\times 10^{-7}$& $3.34 \times 10^{-17}$ & $1.67\times 10^{-3}$ & $2.26 \times 10^{-11}$ \\
Total-variation distance        & $9.57 \times 10^{-6}$& $3.97 \times 10^{-15}$ & $2.78\times 10^{-6}$ & $4.74 \times 10^{-7}$ \\
$L_\infty$                       & $3.20 \times 10^{-6}$ & $1.86 \times 10^{-15}$ & $3.96\times 10^{-6}$ & $6.68 \times 10^{-8}$\\
\hline
\hline
\end{tabular}
\caption[Performance of lognormal distribution loaders]{Numerical data for the circuits creating a lognormal random variable with mean $0.0$ and variance $0.05^{2}$  on the interval $[49/63, 81/63]$.}
\label{tab:lognormal-stats}
\end{table}

For the purposes of our benchmarks (see Section~\ref{sec:benchmarks}), achieving $L_\infty$ of around $10^{-5}$ -- $10^{-6}$ is sufficient (as in this case the error incurred by the training is significantly less than that of the QMCI itself) and we can see that the HWE PQC method achieves this with relatively modest quantum resources (note that the most important resource quantity is the CNOT gate count). For this reason, we find that PQC methods are suitable, at least for simple financial calculations and this corroborates the findings of Ref.~\cite{chakrabarti2020threshold}. We also note that in the TN PQC benchmarks, where we allowed more quantum resources to be used, extremely close fit was achieved -- further validating the PQC methods in general. It is, however, worth observing that PQC methods are likely to suffer during fault-tolerant operation, where continuously-parameterised rotation gates are typically approximately synthesised by gates from a finite set. Furthermore, it is only by training generic ``resource'' states, that can be re-used, that we can reasonably omit mention of the training cost itself. Should it be the case that a bespoke trained PQC were needed for a single problem, then this would likely negate the quantum advantage once the overall cost (including training) is accounted. \\

\section{Statistical quantities estimated by the QMCI engine}
\label{sec:observables}

One of the strengths of the Fourier series decomposition of the Monte Carlo integral, as in Ref.~\cite{herbert1}, is that it is extremely flexible in terms of allowing different functions to be applied to the random variables prior to taking the expectation. For the purposes of the QMCI engine, we need only consider a relatively modest collection of statistical quantities to be estimated (i.e., functions to be applied), although it is easy to bring further functions online as needed.\\

To apply a function, $g$, to a random variable $X$, it is necessary to construct a periodic (in general piecewise) function such that the piece covering the support of the random variable corresponds to the desired function, and the overall function is sufficiently smooth to achieve the required convergence condition. A further important requirement, at least from an operational perspective, is to have a clear characterisation of the overall convergence of the returned estimate. This can be achieved by noting that the output is a straightforward affine combination of estimated amplitudes, where each amplitude has RMSE proportional to $\cqae/q$ (here $q$ is the number of uses assigned to that harmonic). In particular, a general form of the Fourier series approximation of the Monte Carlo integral value is (from Eq.~\ref{eqeq170})):
\begin{equation}
    \mu = a_0 + \sum_{m=1}^\infty a_m \, \left( \sum_{x^{(i)} \in \Omega}  p(x^{(i)})  \cos( m \omega x^{(i)}) \right) + b_m  \, \left( \sum_{x^{(i)} \in \Omega}  p(x^{(i)})  \sin ( m \omega x^{(i)}) \right)
\end{equation}
so it follows that, according to Proposition~\ref{lem1}, each of the parenthesised terms can be estimated using QAE operating on circuits of the form shown in Fig.~\ref{fig:FourierP} Furthermore, letting $q_m^{(a)}$ and $q_m^{(b)}$ be respectively the uses assigned to estimate the cosine and sine terms of the $m^{th}$ harmonic, then the overall estimate has RMSE given by:
\begin{equation}
    \text{RMSE}(\hat{\mu}) = 2 \cqae \sqrt{ \sum_{m=1}^\infty a_m^2 \left(q_m^{(a)}\right)^{-2} + b_m^2 \left(q_m^{(b)}\right)^{-2}}
\end{equation}
which follows directly from Proposition~\ref{lem1}.\\

With knowledge of the specific Fourier series, as well as the way that a given number of uses is spread amongst the various harmonics, the RMSE of the estimate can always be expressed in the following form:
\begin{equation}
\label{eq910}
    \sqrt{\mathbb{E}( (\widehat{\mathbb{E}_{g(x)}(X)} -\mathbb{E}_{g(x)}(X))^2)}  \leq c_g \cqae \frac{\max_{\substack{x \in [x_\ell,x_u]}}(g(x)) - \min_{\substack{x \in [x_\ell,x_u]}}(g(x))}{q} 
\end{equation}
where $c_g$ is a constant that depends only on the function being applied.\\

With these requirements in mind, we include in the QMCI engine Fourier series decompositions that allow the following six quantities to be calculated:
\begin{enumerate}
    \item the mean -- i.e., $g(x) = x$;
    \item the conditional expectation (defined in Eq.~\eqref{eq:def-condi-expec});
    \item the variance -- which uses the mean, as above, and the second moment, i.e., $g(x) = x^2$;
    \item expectation of the random variable exponentiated -- i.e., $g(x) = \exp(x)$;
    \item the conditional expectation of the random variable exponentiated (defined in Section~\ref{subsect:exp-exponentiated});
    \item the expectation of a single qubit treated such that (computational basis) measurements thereof are samples of a Bernoulli random variable.
\end{enumerate}

\subsection{Estimating the mean}
\begin{figure}[t!]
    \centering
    \includegraphics[scale=0.75]{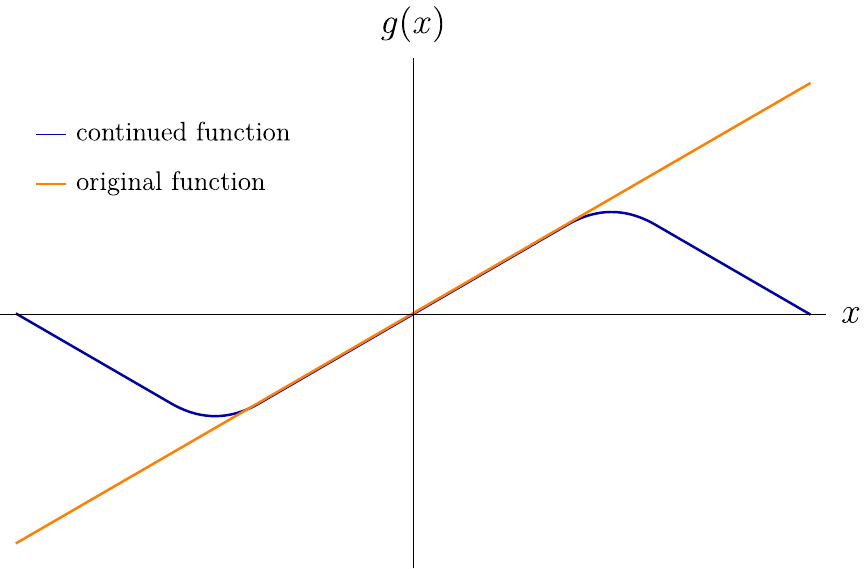}
    \caption{A sketch of the periodic piecewise function for estimating the mean. The location-scale property of the expectation is used to transform the support to the piece $x \in [-1,1]$, as shown here, and the remainder of the function is constructed such that it is sufficiently smooth and has even symmetry.}
    \label{fig:piecewise-x}
\end{figure}

To estimate the mean, we use the location-scale property of the mean to always assume that the random variable is supported over $x \in [-1,1]$ within the QMCI engine, then shifting and re-scaling the returned estimate accordingly. This means that we always use exactly the same periodic piecewise function to construct the Fourier series, as sketched in Fig.~\ref{fig:piecewise-x}, and in this way we obtain overall convergence:
\begin{equation}
\label{eq8p20}
    \sqrt{\mathbb{E}( (\widehat{\mathbb{E}_x(X)} -\mathbb{E}_x(X))^2)}  = 1.68 \cqae \frac{x_u - x_\ell}{q} 
\end{equation}
i.e., $c_{x} = 1.68$ as in this case $\max_{\substack{x \in [x_\ell,x_u]}}(f(x)) - \min_{\substack{x \in [x_\ell,x_u]}}(f(x)) = x_u - x_\ell$.\\

\subsection{Estimating the conditional expectation}

The enhanced $P$-builder allows \textit{indicator} qubits to be included, which encode a superposition of Boolean values, whose truth varies with each point supporting the multivariate probability mass function. The indicator qubit is therefore (as the name implies) an indicator function of the random variable, $\mathbb{1}(X)$, from which we obtain:
\begin{equation}
    \mathbb{E}_{x|\mathbb{1}(x)}(X) = \sum_X X \mathbb{1}(X) p(X)  + x^* \sum_X (1-\mathbb{1}(X)) p(X)
\end{equation}
where $x^*$ is a user-defined value such that $x_\ell \leq x^* \leq x_u $. By default, if no value is given by the user:
\begin{equation}
    x^* = 
\begin{cases}
0 & \, \text{if } x_\ell \leq 0 \leq x_u \\ 
x_\ell  & \, \text{otherwise} 
\end{cases}
\end{equation}
In all of these cases, the convergence for the conditional expectation is equal to that given in Eq.~\eqref{eq8p20}.\\

Calculating the payoff of a European call option gives a (very) simple example of how we may use indicator qubits in this way. In this case, the threshold conditioning in the enhanced $P$-builder is set-up such that $\mathbb{1}(X)$ returns the value 1 for the subset of prices above the strike price. More generally (for path-dependent derivatives), the indicator qubit may ``gather up'' conditioning on various logical statements throughout the history of the derivative instrument.\\

\subsection{Estimating the variance \& second moment}
\begin{figure}[t!]
    \centering
    \includegraphics[scale=0.75]{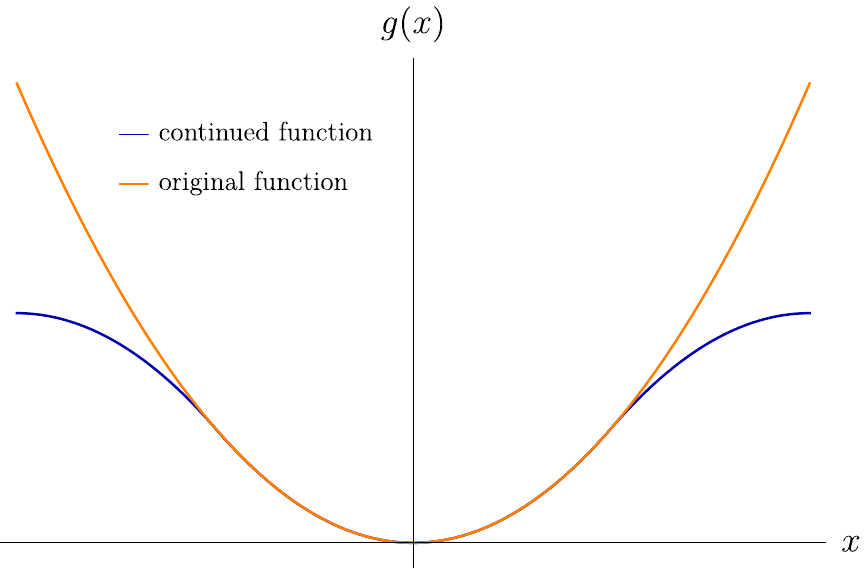}
    \caption{A sketch of the periodic piecewise function for estimating the second moment. The scale property of the expectation is used to transform the support to the piece $x \in [-1,1]$, as shown here, and the remainder of the function is constructed such that it is sufficiently smooth and has even symmetry.}
    \label{fig:piecewise-x2}
\end{figure}

The variance of a random variable is defined as $\E[(X - \E[X])^2]$, which is estimated by first estimating the mean, and (classically) subtracting this from the support of $X$ to ``centralise''. The second moment of this centralised version of $X$ is then estimated. For this, we require $g(X) = X^2$, and use the fact that zero will now certainly be contained within the support to treat the support as $-x_b \leq x \leq x_b$, where $x_b = \max(x_u, -x_\ell)$. To construct the corresponding piecewise periodic function, the scale property of the second moment is used such that the support of the random variable is treated as $[-1, 1]$, with $x_b$ then used to re-scale the returned estimate. Once again, in this way we can use the same periodic piecewise function every time the second moment is to be estimated, regardless of the support of the random variable. In particular, we design a periodic piecewise function which is equal to $x^2$ in the range $[-1,1]$, as sketched in Fig.~\ref{fig:piecewise-x2}, and in this way we obtain overall convergence:
\begin{equation}
\label{eq:variance-estimate}
    \sqrt{\mathbb{E}( (\widehat{\mathbb{E}_{x^2}(X)} -\mathbb{E}_{x^2}(X))^2)}  = 2.82 \cqae \frac{(\max(x_u,-x_\ell))^2}{q} 
\end{equation}
i.e., $c_{x^2} = 2.82$ as in this case $\max_{\substack{x \in [x_\ell,x_u]}}(g(x)) - \min_{\substack{x \in [x_\ell,x_u]}}(g(x)) = (\max(x_u,-x_\ell))^2$.\\

The QMCI engine allows to user to directly request the variance, in which case the mean and second moment are each estimated -- in such a way that the overall desired accuracy is obtained with minimal total computational load.

\subsection{Estimating the expectation of the random variable exponentiated}

\begin{figure}[t!]
    \centering
    \includegraphics[scale=0.75]{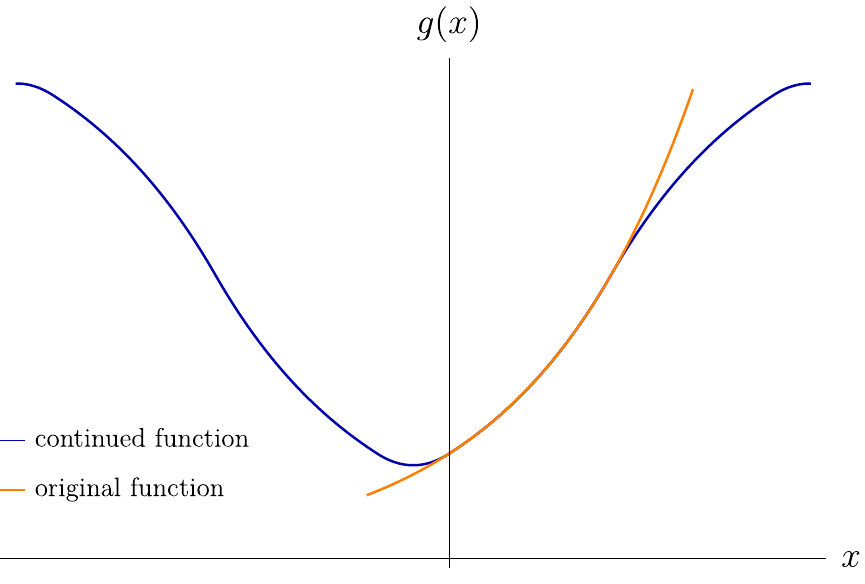}
    \caption{A sketch of the periodic piecewise function for estimating the expectation of a random variable exponentiated. Here, the highlighted piece is $\exp(x)$ with the support shifted accordingly, and the remainder of the function is constructed such that it is sufficiently smooth and has even symmetry.}
    \label{fig:piecewise-exp}
\end{figure}

\begin{figure}[t!]
    \centering
    \includegraphics[scale=0.7]{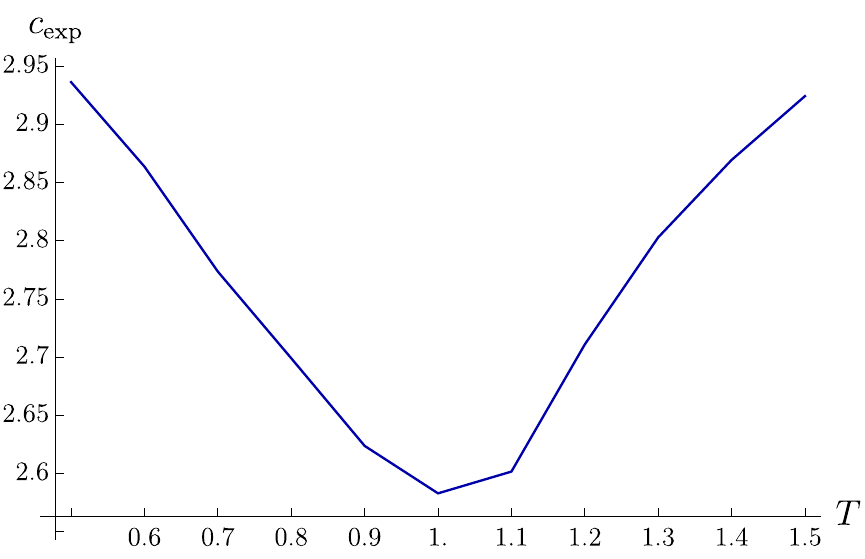}
    \caption{An illustration of $c_{\exp}$ as a function of $T\equiv\exp(x_u)-\exp(x_\ell)$, for the periodic piecewise function that has a minimum around $\exp(x_u)-\exp(x_\ell) = 1$.}
    \label{fig:cexp}
\end{figure}

Estimating $\E(\exp{X})$ is especially relevant when working in return (log-price) space in financial applications. In the case of the exponential, we can shift the random variable in a manner that can be accounted for according to:
\begin{equation}
    \mathbb{E}_{\exp(x)}(X) = \exp(a)\mathbb{E}_{\exp(x)}(X-a)
\end{equation}
however we cannot re-scale the random variable. This means that we cannot use exactly the same periodic piecewise function, regardless of the user-defined support, in the way we can for the mean and second moment. Our approach is to optimise the design of the periodic piecewise function by shifting the random variable, as shown in Fig.~\ref{fig:piecewise-exp}, however it is unavoidably the case that a different Fourier series approximates the estimated quantity as the range of the support varies. This property, it turns out, means that when the RMSE of the estimator is divided by the number of uses, $q$, and the range of the support $\max_{\substack{x \in [x_\ell,x_u]}}(g(x)) - \min_{\substack{x \in [x_\ell,x_u]}}(g(x))= \exp(x_u) - \exp(x_{\ell})$, the result is not constant (although it does not depend on $\exp(x_u)$ and  $\exp(x_{\ell})$ individually, just $\exp(x_u) - \exp(x_{\ell})$). In order to adhere to the canonical form given in Eqn.~(\ref{eq910}), it is therefore necessary to specify the support of $x$. For this, we choose to focus on the low-volatility setting\footnote{It is possible to design an alternative periodic piecewise extension of $\exp(x)$ optimised for greater ranges of support, should high-volatility applications be needed, however this is not our focus for the benchmarks we give in this paper.}, specifically, the standard deviation, $\sigma =0.1$, which means that for five standard deviations\footnote{Stretching out to five standard deviations is probably unnecessarily conservative in most applications, however we follow the precedent set in Ref.~\cite{chakrabarti2020threshold}.} we have to set $x_{\ell} = -0.5$, $x_u=0.5$, thus giving a total range of 1. With this in mind, we tailor the periodic piecewise function such that the RMSE of the estimator is divided by the number of uses, $q$, and the range of the support $ \exp(x_u) - \exp(x_{\ell})$ is as shown in Fig.~\ref{fig:cexp}, and from this we get that $c_{\exp} \leq 2.59$, i.e.,
\begin{equation}
\label{eq:exp-estimate}
    \sqrt{\mathbb{E}( (\widehat{\mathbb{E}_{\exp(x)}(x)} -\mathbb{E}_{\exp(x)}(x))^2)}  \leq 2.59 \cqae \frac{\exp(x_u)-\exp(x_\ell)}{q} 
\end{equation}

\subsection{Estimating the conditional expectation of the random variable exponentiated}
\label{subsect:exp-exponentiated}

As was the case for the mean, it is also important to be able to exponentiate a random variable subject to some condition given by an indicator function. In this case, we again allow the user to input some value $\exp(x_\ell) \leq \exp(x^*) \leq \exp(x_u)$ (set to $\exp(x_\ell)$ by default), and then estimate
\begin{equation}
\label{eq:def-condi-expec}
    \mathbb{E}_{\exp(x)|\mathbb{1}(x)}(X) = \sum_X \exp(X) \mathbb{1}(X) p(X) + \exp(x^*) \sum_X (1-\mathbb{1}(X)) p(X)
\end{equation}
which has the same convergence properties as the straightforward exponential, given in Eqn.~(\ref{eq:exp-estimate}).

\subsection{Estimating a single qubit as sampling a Bernoulli random variable}
By definition, this is just a restatement of QAE itself, and so we get convergence:
\begin{equation}
\label{eq:bernoulli-estimate}
    \sqrt{\mathbb{E}( (\widehat{\mathbb{E}_{\text{Bernoulli}}(X)} -\mathbb{E}_{\text{Bernoulli}}(X))^2)}  =  \cqae \frac{1}{q} 
\end{equation}
i.e., $c_{\text{Bernoulli}}=1$ and in this case $\max_{\substack{x \in [x_\ell,x_u]}}(g(x)) - \min_{\substack{x \in [x_\ell,x_u]}}(g(x)) = 1$ ($x_\ell = 0$; $\Delta=1$). 

\subsection{Applying the function to part of the support}
\label{subsect:func-part-support}

As each dimension of the multivariate probability distribution has explicitly defined support, by default the periodic piecewise function is such that the desired function is applied to the entire support (i.e., $x_\ell \dots x_u$), however the user can optionally choose for this section to cover only part of the support. This is useful if, for example, it is known that the entire probability mass is concentrated in this region. In this case, the convergence would be correspondingly improved, as the numerator in Eq.~(\ref{eq910}) would now only take the maximum and minimum function evaluation over the restricted range.

\subsection{Summary of QMCI convergences}

Values of $\cqae$ for the three forms of QAE included in the QMCI engine are given in Table~\ref{tab:qae-convergence} which can thus be used to perform a direct comparison between QMCI and classical MCI in terms of convergence rates, using also the bounds on the convergence of classical MCI estimators, in Section~\ref{subsect:probability-prelims} :
\begin{equation}
    \sqrt{\mathbb{E}( (\widehat{\mathbb{E}_f(x)} -\mathbb{E}_f(x))^2)}  \leq \begin{cases}
        \frac{(\max(x_u,-x_\ell))^2}{2\sqrt{q}} & \text{ if } f=x^2 \\
        \frac{f(x_u)-f(x_\ell)}{2\sqrt{q}} & \text{ otherwise }
    \end{cases}
\end{equation}
Here the condition that $x_\ell \leq 0 \leq x_u$ for the second moment estimation has again been used, as well as the property that all of the other functions are monotonically increasing. Table~\ref{tab:QMCI-convergence} compares the classical and quantum MCI convergences in terms of the number of samples, however this is a fair (or even conservative) way to do so, given that a classical sample is at least as hard to prepare as a quantum sample \cite{herbert2}, and with Fourier QMCI we have little circuitry other than that encoding the sampling circuit \cite{herbert1}. (Note that hardness of preparing a sample is equivalent to sampling speed.) \\

\begin{table}[!t]
    \centering
    \begin{tabular}{l c c c c c c c c}
        \hline
\hline\\[-0.8ex]
         \multirow{2}{*}{\textbf{Quantity}}  & & \multirow{2}{*}{$c_f$}  & & \textbf{Crossover}  & & \textbf{Crossover} & & \textbf{Crossover}  \\
         && && \textbf{(LCU QAE)} && \textbf{(ML QAE)} && \textbf{(IQAE)} \\[0.5ex]
         \hline\\[-0.8ex]
         Mean \& & &   \multirow{2}{*}{1.68}  & &  \multirow{2}{*}{1.12\%} & &  \multirow{2}{*}{1.86\%} & &  \multirow{2}{*}{1.03\%}\\
         Conditional expectation & & & & & & \\[1.5ex]
         
         Second moment  & &  2.82 &&  0.67\% && 1.11\% && 0.61\% \\[1.5ex]
         Exponential expectation \&  & &  \multirow{2}{*}{2.59}  & &  \multirow{2}{*}{0.72\%} & &  \multirow{2}{*}{1.20\%} & &  \multirow{2}{*}{0.67\%} \\
         Conditional exponential expectation & & & & & &  \\[1.5ex]
         Single qubit (Bernoulli)  && 1 & &  1.88\% && 3.12\% && 1.73\%\\
         \hline 
         \hline
    \end{tabular}
   \caption[Convergence of QMCI calculations]{Comparison between (upper-bounds on) classical and quantum convergences for various statistical quantity estimates in terms of the number of samples, $q$. ``Crossover'' is the error (RMSE), given as a percentage of the range of the support, at which it becomes more efficient to use the quantum algorithm. }
    \label{tab:QMCI-convergence}
\end{table}

\begin{figure}[t!]
    \centering
        \begin{tabular}{c c c}
            \includegraphics[width=0.3\textwidth]{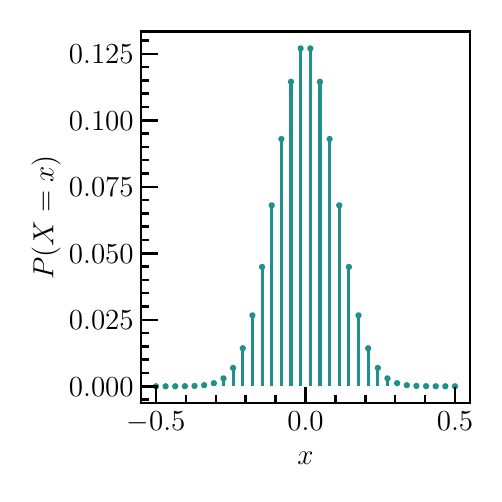} & \includegraphics[width=0.3\textwidth]{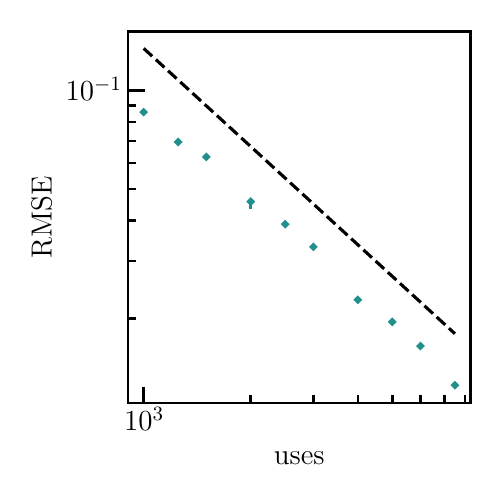} & \includegraphics[width=0.3\textwidth]{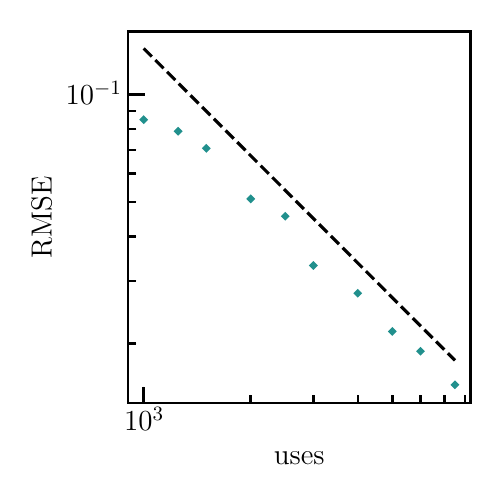} \\
            {~~~~~~}Distribution & {~~~~~~~~}Mean & {~~~~~~}Conditional expectation\\
              \includegraphics[width=0.3\textwidth]{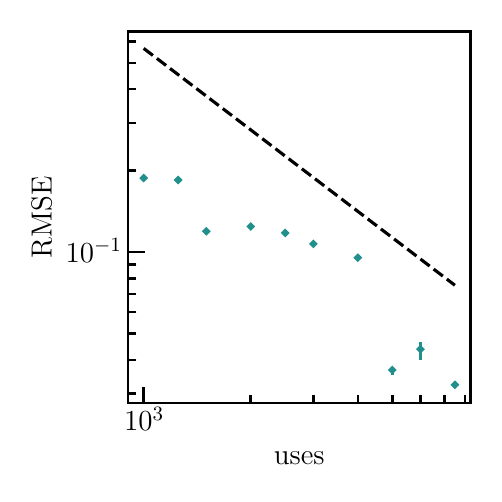}  & \includegraphics[width=0.3\textwidth]{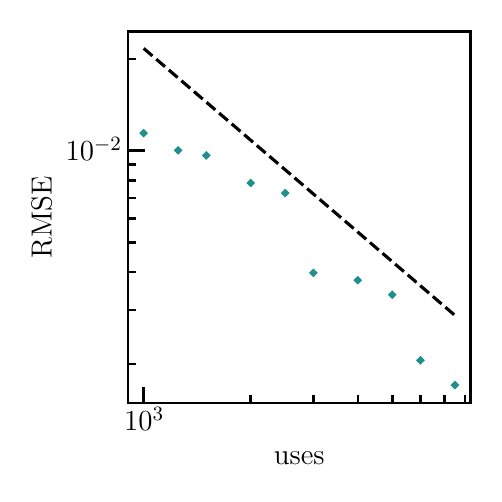} & \includegraphics[width=0.3\textwidth]{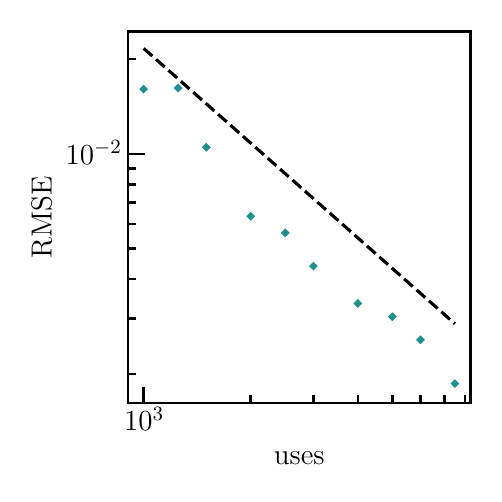} \\
            {~~~~~~}Second moment &
            {~~~~~~~}Exponential & {~~~~~~}Conditional exponential \\
        \end{tabular}
    \caption{Theoretical upper-bounds (dashed lines) on QMCI convergence for the five quantities of interest, against numerically simulated data (single points). Thus we can see that the theoretical upper-bounds indeed hold in all cases.}
    \label{fig:QMCI-benchmarks}
\end{figure}

In order to illustrate the validity of these bounds in practice, we give explicit examples for the case where the quantity is being calculated for a 32 point (5 qubit) discrete approximation of a Gaussian random variable distributed as $\mathcal{N}(0,0.1^2)$. For the conditional expectation / conditional exponentiated expectation we set $x^* = 0$, and in all cases MLQAE was used. Figure~\ref{fig:QMCI-benchmarks} shows that the upper-bounds hold for all five quantities when compared to numerical data obtained by running the calculations 10000 times on a state-vector simulator (notice that the upper-bound on \textit{qubit as Bernoulli} follows directly from the extensive analysis on QAE in Sections~\ref{sec:robust-qae} and \ref{sec:qae-description}, and hence we omit repeating this here). 

\section{Quantum amplitude estimation}
\label{sec:qae-description}

The QMCI engine includes three QAE subroutines, our LCU QAE, MLQAE and IQAE; additionally we include \textit{prepare and measure} (PAM), in which the state $A\ket{0}$ is simply prepared and measured the specified number of times. Definition~\ref{def1} dictates the functional form of a QAE subroutine, and crucially the relationship $\text{RMSE}(\hat{a}) \leq \cqae q^{-\lambda/2}$ needs to hold for \textit{any} positive integer $q$. This is not necessarily the case for ``off-the-shelf'' implementations of QAE, in which $q$ is effectively only permitted to be a value that aligns with the needs of the QAE algorithm in question. Moreover, in the case of the quadratic advantage-achieving exponentially increasing sequence (EIS) of MLQAE, these numbers of allowed uses spread out exponentially. It is therefore necessary to enhance these algorithms to accept any number of uses -- and to use those uses productively.\\

The constant $\cqae$ is defined to upper-bound the RMSE convergence, and hence it is necessary to consider the worst case for all amplitudes. In the case of  PAM this is easy, as we automatically have $\text{RMSE}(\hat{a}) \leq 0.5 q^{-1/2}$ for any positive integer $q$ -- this is just classical MCI, and the bound is saturated when $a = 0.5$. However, here $\lambda = 1$ and for the forms of QAE that actually use the quantum mechanical effects to achieve $\lambda=2$ (the full quadratic advantage), some work is needed to adapt the algorithms to work for any number of uses before thereafter evaluating $\cqae$ numerically.

\subsection{LCU QAE}

Our LCU QAE algorithm, as described in Section~\ref{sec:robust-qae}, is based on the EIS of Ref.~\cite{Suzuki_2020}. Even when the input number of uses is interpreted as the successful uses, we are still faced with the fact that only certain values of $q$ exactly align with the number needed to complete the EIS for some value of $m$. As this will not, in general, be the case, it is necessary to curtail the EIS at the final value of $m$ for which the necessary circuit executions can be completed (i.e., as bounded by $q$), and find some suitable way of deploying the remaining uses. \\

Letting $m^*$ be the maximum value in the EIS for which it is possible to run all $n_{\text{shots}}$, then the idea is to find greatest $m' > m^*$ such that it is possible to run $n_{\text{shots}}$ with $m'$ applications of $Q$. Once the shots for $m'$ have been accounted for, any further remaining uses are allotted to  the existing values of $m$ (including $m=m'$) such that the higher values of $m$ are prioritised, but ultimately all uses are taken up (this can always be guaranteed, as adding a single shot to $m=0$ consumes just a single use). This approach is motivated by two principles:
\begin{enumerate}
    \item As the quadratic advantage is obtained by rotating the state to ever higher values of $\theta$, we aim to use up the remaining uses by running the full round of $n_{\text{shots}}$ for the highest possible value of $m$ -- even if this is not the next value of $m$ in the EIS.
    \item We may as well always exhaust the assigned uses, and following principle 1, we do this in such a way that the uses are prioritised for the largest values of $m$.
\end{enumerate}
In LCU QAE, we have two notions of the number of uses of the circuit $P$, namely \textit{successful uses} (assuming post-selection on the LCU condition) and expected total uses (including LCU post-selection failures). LCU QAE has been designed to fail fast, and so a failed LCU state preparation costs the same regardless of the circuit depth. This means that, as the circuit depths (i.e., the desired accuracy) increases, the contribution of the failed uses to the total uses is ever-diminishing. For this reason, counting only successful uses indeed gives us the asymptotic rate of convergence. However, for any real (finite) problem, we should count the failed uses, and to do so in a manner that is consistent with the overall scaling of RMSE with uses, we choose the minimum upper-bounding line for all of the empirical data.

\subsection{MLQAE}

To enable MLQAE to accept any number of uses, we can use the same strategy as for LCU QAE. 

\subsection{IQAE}

Our implementation of the IQAE algorithm follows the original presentation given by Grinko \textit{et al}~\cite{grinko2019iterative}, with some modifications to allow us to directly compare this method with the other forms of QAE we consider. \\

The original algorithm takes as input a circuit $A$, a \textit{desired error} $\epsilon$ and a \textit{confidence value} $\alpha$ and proceeds as follows. Start with the interval $[{\theta}^0_\ell, {\theta}^0_u]\vcentcolon= [0,\pi/2]$, which trivially contains $\theta$. The \textit{iterations} of the algorithm then produce increasingly smaller $\alpha$-confidence intervals for $\theta$ $$[{\theta}^0_\ell, {\theta}^0_u] \supset [{\theta}^1_\ell, {\theta}^1_u] \supset [{\theta}^2_\ell, {\theta}^2_u] \supset \ldots $$ The algorithm terminates after $T$ iterations, where $T$ is the smallest value such that ${\theta}^T_u - {\theta}^T_\ell \leq 2\epsilon$. The algorithms then returns as output $[a_\ell, a_u] = [\sin^2({\theta}^T_\ell), \sin^2({\theta}^T_u)]$, an $\alpha$-confidence interval for the amplitude $a$ of $A$. \\

The functional form of this algorithm is different from that presented in Definition \ref{def1}. Indeed, instead of taking $A$ and a number of uses $q$ as input it takes $A$ and the pair $(\alpha, \epsilon)$ and instead of returning a point estimate it returns an interval, giving it the following form:
$$ [a_\ell, a_u] = \texttt{IQAE}(A, (\epsilon, \alpha))$$
Furthermore, for a given $A, \alpha$ and $\epsilon$, it is not defined from the outset how many uses of $A$ will be required to run the IQAE algorithm. So, we define a function $\texttt{optAE}(q)$ which chooses a suitably optimal pair $(\alpha, \epsilon)$ such that  $\texttt{IQAE}(A, (\epsilon, \alpha))$ uses at most $q$ calls to $A$. This can be done by using the following upper-bound which can be derived from the original IQAE paper~\cite{grinko2019iterative}:
$$ q_A \leq q(\alpha, \epsilon) \vcentcolon= \left( \frac{100}{\epsilon} + \frac{32}{(1-2\sin(\pi/14))^{2}}\right)\log\left(\frac{2}{\alpha}\log_2\left(\frac{\pi}{4\epsilon}\right)\right)  $$
Noting also that the RMSE of the estimate $\hat{\theta} = (\theta^T_\ell + \theta^T_u)/2$ is upper-bounded as 
$$\text{RMSE}(\hat{\theta}) \leq R(\alpha, \epsilon) \vcentcolon= (1-\alpha) \epsilon^{2} + \alpha \left(\frac{\pi}{2}\right)^{2}$$
 we implement $\texttt{optAE}$ as the function which takes input $q$ and returns $\alpha$ and $\epsilon$ which minimise $R(\alpha, \epsilon)$ subject to $q(\alpha, \epsilon) = q$. Thus, we can define the following version of $\texttt{IQAE}$ which is of the desired form:
$$\texttt{IQAE}^{\ast}(A, q) \vcentcolon= \texttt{midpoint}(\texttt{IQAE}(A, \texttt{optAE}(q)))$$
where $\texttt{midpoint}$ is a simple function that takes the mean of the two values returned by $\texttt{IQAE}$ (i.e., interpreted as their ``midpoint'' -- as the two values are real numbers).\\

Empirically, we found that this implementation rarely uses the total number of uses $q$, indicating that the bound $q(\alpha, \epsilon)$ is quite loose. In order to provide a fair comparison of this algorithm with other forms of QAE we modify $\texttt{IQAE}^{\ast}$ by first running $\texttt{IQAE}(A, \texttt{optAE}(N))$ and, then, given the final confidence interval $[{\theta}^T_\ell, {\theta}^T_u]$ we continue to perform the iterations of the IQAE algorithm until almost all $q$ uses are exhausted. This implementation is the one used to generate the data for example in Table \ref{tab:qae-convergence}.

\subsection{Prepare and measure}

As already noted, for PAM we have the analytical result $\cqae = 0.5$ which automatically holds for any number of uses, but now this is a constant of proportionality for RMSE convergence proportional only to $q^{-1/2}$.

\subsection{Summary of QAE convergences}

\begin{table}[t!]
\centering
\begin{tabular}{l ccc }
\hline
\hline\\[-0.8ex]
  & \textbf{Maximum} & \textbf{Median} &\textbf{Minimum}  \\[1ex]
 \hline
LCU QAE (upper-bound) & 13.3 &11.4 &3.74  \\
LCU QAE (asymptotic) & 7.82 & 6.80 & 2.15  \\
MLQAE & 8.02 & 6.18 & 2.13 \\
IQAE & 14.4 & 7.48 & 1.49  \\
PAM & 0.5 & 0.44 & 0.14  \\
\hline
\hline
\end{tabular}
\caption[Convergence of QAE methods]{$\cqae$ values for the different forms of QAE. By definition, $\cqae$ is taken as the value shown in the column ``maximum''. 
}
\label{tab:qae-convergence}
\end{table}

We used the numerical data of the simulations described in Section~\ref{sec:robust-qae} to find $\cqae$ for the various QAE sub-routines included in the QMCI engine. As discussed in the opening of this section, it is necessary to conservatively select the worst case (largest) value of $\cqae$ in each case, and Figs.~\ref{fig:lcu-successful-rmse-fit} -- \ref{fig:iqae-rmse-fit} show these for each of the 49 amplitudes for which we simulated QAE. Table~\ref{tab:qae-convergence} summarises the convergences for the three featured forms of QAE, and PAM. By definition, $\cqae$ is taken as the value shown in the column ``maximum'', however also included are ``median'' -- which gives an indication of the typical convergence one would expect for an arbitrarily chosen angle; and ``minimum'' -- which shows the best case convergence.\\ 

It is also worth briefly comparing to ``canonical QAE'' -- using the quantum Fourier transform \cite{brassard2000quantum}. According to Ref.~\cite[Thm 12]{brassard2000quantum} the absolute error can be upper-bounded by a term that is at least $2\pi \frac{\sqrt{a(1-a)}}{q/2}$ -- but that this is only achieved with probability $8/\pi^2$. Neglecting the failure probability and assuming that the estimate is uniformly distributed within the margin bounded by the maximum absolute error, we then get $\cqae \approx \frac{2}{\sqrt{3}}\pi = 5.44$. This figure is comparable to those in Table~\ref{tab:qae-convergence}, however the deepest circuits in canonical QAE are about 60 times as deep as those of MLQAE and LCU QAE (for the same value of $q$). Furthermore, the fact that canonical QAE only outputs the estimate within the confidence bounds with a certain (fixed) probability means some repeats are necessarily going to be required to boost this probability (i.e., to ensure that the overall RMSE -- when the non-zero failure probability is also included -- is at the  required level). Grinko~\textit{et al} find that when canonical QAE is run with a suitable number of repeats, it actually performs worse (in terms of RMSE convergence) than QPE-free forms of QAE \cite[Fig. 3]{grinko2019iterative}. These facts justify our focus on QPE-free forms of QAE, and the need to find a statistically robust form thereof.\\

\section{Resource mode}
\label{sec:resource-mode}

There is a significant discrepancy between the sizes of circuits which can be executed on present-day hardware (at the time of writing), and those which will be needed to realise a quantum advantage. It is, however, the case that the techniques we describe in the preceding sections do allow the explicit construction of those circuits which will eventually realise a useful quantum advantage -- and whilst there is no current prospect of executing or simulating these, there is great value in knowing some broad properties of these circuits. This is because knowing the \textit{resources} that these circuits require will enable more precise predictions about when certain calculations will see the impact of quantum computing and hence, in turn, will enable forward planning for deployment and timelines for necessary infrastructural developments \textit{etc}. To facilitate this we have enabled the QMCI engine to run in \textit{resource mode}, wherein rather than executing the circuits, instead the resource quantification is returned to the user. Resource mode enables the QMCI engine to be used as a prototyping and experimentation tool, wherein the impact on resource requirements of different designer-choices about how to construct various calculations is quantified precisely. For instance, there is often a range of appropriate models for financial time-series, each with myriad possible ways of being encoded in a quantum circuit; additionally, the user has a certain degree of freedom as to how to slice up the time-series and model the derivative instrument payoff. These design choices may well lead to vastly differing resource requirements, and so it is only through research and development using resource mode that the most direct pathway to quantum advantage will be revealed.\\

Resource mode comes with two sub-modes, namely \textit{NISQ resource mode} and \textit{fault-tolerant resource mode}. In each case, resources are counted for \textit{all} of the component QMCI circuits -- although this constitutes something of an over-count, as in practice certain parts of the QMCI could be run classically (indeed this is a further strength of the Fourier series decomposition: the quantum computer need only be used for those parts of the Monte Carlo integral that are truly hard classically):
\begin{itemize}
    \item circuits corresponding to $m=0$ \textit{are} nothing more than classical Monte Carlo samples with sine / cosine applied;
    \item indeed, even for low values of $m \neq 0$ it is likely that shallow-circuit (tensor-contraction based) classical simulators could readily be deployed to further reduce the load on the quantum computer;
    \item harmonics that are awarded only a small number of uses by Fourier QMCI could be wholly estimated using classical MCI instead.
\end{itemize}

\subsection{NISQ resource mode}

To count gates in resource mode, it is necessary to fix a gateset to which the circuit is rebased\footnote{In the future, we may extend this to allow user-defined gatesets, however for now -- as our focus is on a generic quantification of resources, primarily aimed at problem sized that cannot yet be executed -- having just a single standard gateset meets the requirements.}. Even though the native gateset varies from machine to machine, an appropriate choice is the universal gateset $\{ \text{TK}1, \text{ CNOT} \}$, where the parameterised TK1 gate introduced in Eq.~\eqref{eq:def-tk1} is universal for single-qubit operations (up to a global phase). After the circuit has been thusly rebased, NISQ resource mode proceeds to construct all of the circuits for the QMCI, but rather than running, instead reports, for each circuit:
\begin{enumerate}
    \item the number of qubits;
    \item the total gate count and depth;
    \item the CNOT count and depth,
\end{enumerate}
i.e., the total dimensions of the circuit; and those of the usual bottleneck (the CNOT gate).

\subsection{Fault-tolerant resource mode}

As many of the eventual applications of QMCI will require error-corrected quantum computers, it is useful to provide an alternative set of resource quantities based on compilation to some fault-tolerant gateset. The standard choice is to compile to the gateset $\{\text{CNOT}, H, S, T \}$ -- known as ``Clifford+$T$'' and denoted $\{ \mathrm{Clifford }, T\}$ -- however, this is a discrete gateset, and so it is also necessary to determine the required closeness of approximation. To do this, we first note that in general, the circuits built by the QMCI engine use gates from the set:

\begin{equation}
    \mathsf{G}_{\text{full}} =\{\mathrm{Clifford}, T \} \cup \{ \text{Toffoli},  R_x(\cdot), R_y(\cdot), R_z(\cdot), CR_x(\cdot), CR_y(\cdot), CR_z(\cdot) \}  
\end{equation}
However, for fault-tolerant (error-corrected) operation we need to (approximately) compile instead to a gate-set:

\begin{equation}
    \mathsf{G}_\text{FT} =\{\mathrm{Clifford}, T \}
\end{equation}
This, in turn, means realising the following decomposition to an appropriate approximation:
\begin{equation}
    \{\text{Toffoli},  R_x(\cdot), R_y(\cdot), R_z(\cdot), CR_x(\cdot), CR_y(\cdot), CR_z(\cdot) \} \longrightarrow \{\mathrm{Clifford}, T \}
\end{equation}
In the case of Toffoli, this is straightforward as an exact decomposition exists; and we also note that controlled rotations can exactly decomposed into a circuit with Clifford+$T$ gates and precisely six (uncontrolled) rotation gates \cite[Corollary 4.2 \& Fig. 4.6]{nielsenchuang2010}. It is therefore possible to exactly rebase the original circuit to a circuit containing just gates from the set:

\begin{equation}
\mathsf{G}_{\text{reduced}}= \{\mathrm{Clifford }, T \} \cup \{  R_x(\cdot), R_y(\cdot), R_z(\cdot) \}   
\end{equation}
However, Clifford+$T$ compilation of the (uncontrolled) rotation gates can only be achieved up to a specified approximation. This approximation can be optimised -- by which we mean the total $T$-gate count can be minimised, as this is typically the bottleneck in fault-tolerant circuit execution. This optimisation uses the fact that each $R_{x}(.)$, $R_y(.)$ and $R_z(.)$ gate can be approximated with a circuit containing $3 \log_2 (1/ \epsilon)$ $T$ gates, where $\epsilon$ is an upper-bound on the operator norm distance between the exact and approximate unitary \cite{ross-selinger}.\\

If a positive operator-valued measurement (POVM) is made for a circuit containing a total of $n_R$ gates from the set $\{  R_x(\cdot), R_y(\cdot), R_z(\cdot) \}$, each of which is approximated to at most $\epsilon$ operator norm distance, then the maximum absolute discrepancy in probability of measuring any given POVM element is upper-bounded by \cite[Eq. (4.62)]{nielsenchuang2010}
\begin{equation}
    \epsilon_{\text{tot}} \leq 2 n_R \epsilon
\end{equation}

To see how we can use this to optimise the choice of $\epsilon$ for a certain instance of QMCI, first, we imagine that all of the circuits to be executed are arranged parallel, such that the whole algorithm is a tensor product of the component circuits. The output is then a partial computational basis measurement, to give an output bitstring $x$, which will then be mapped to $\hat{\kappa}$ -- the estimate of the quantity to be estimated, $\kappa$. This mapping from $x$ to $\hat{\kappa}$ need not be 1-to-1 (that is, different bitstrings may map to the same $\hat{\kappa}$), however, using the fact that the upper-bound on the discrepancy in measurement outcome probability holds for \textit{any} POVM, we can choose a POVM such that bitstrings which map to the same value of $\hat{\kappa}$ are grouped in the same POVM element. In this way we label the possible measurement outcomes by $\hat{\kappa}$ without repetition, and hence we can consider the measurement to be directly supported over some discrete values of $\hat{\kappa}$. Thus we have that the final measurement is such that each possible outcome is observed with probability whose magnitude differs at most by $\epsilon_{tot}$, and we can use this to bound the probability of obtaining some $\hat{\kappa}$ under approximate gate synthesis:

\begin{equation}
\label{eqn140}
    p(\hat{\kappa}) \leq p_{\text{exact}}(\hat{\kappa}) + \epsilon_{tot} 
\end{equation}
where $p_{\text{exact}}(\hat{\kappa})$ is the probability had all of the QMCI circuits been executed exactly rather than approximately. Therefore, using simple statistics:
\begin{align}
    \text{MSE}(\hat{\kappa}) & = \int (\hat{\kappa} - \kappa)^2 p(\hat{\kappa}) \mathrm{d} \hat{\kappa} \nonumber \\
    & \leq \int (\hat{\kappa} - \kappa)^2 p_{\text{exact}}(\hat{\kappa}) \mathrm{d} \hat{\kappa} + \epsilon_{tot} \int (\hat{\kappa} - \kappa)^2  \mathrm{d} \hat{\kappa} \nonumber \\ 
    & \leq \text{MSE}_{\text{exact}}(\hat{\kappa}) + \epsilon_{tot} \left[ \frac{(\hat{\kappa}-\kappa)^3}{3}  \right] \nonumber \\
    \label{eq130}
    & \leq \text{MSE}_{\text{exact}}(\hat{\kappa}) + \frac{\epsilon_{tot}}{3} \left(\max_{\substack{x \in [x_\ell,x_u]}}(g(x)) - \min_{\substack{x \in [x_\ell,x_u]}}(g(x))\right)^3 
\end{align}
where $\text{MSE}_{\text{exact}}(\hat{\kappa})$ is the MSE had all of the QMCI circuits been executed exactly rather than approximately.\\ 

From Eq.~\eqref{eq8p20} we also have that the convergence using QMCI is
\begin{equation}
    \text{MSE}_{\text{exact}}(\hat{\kappa}) \leq c^2_g \cqae^2 \left( \max_{\substack{x \in [x_\ell,x_u]}}(g(x)) - \min_{\substack{x \in [x_\ell,x_u]}}(g(x)) \right)^2 q^{-2}
\end{equation}
and thus
\begin{equation}
   \text{MSE}(\hat{\kappa}) \leq c^2_g \cqae^2 \left( \max_{\substack{x \in [x_\ell,x_u]}}(g(x)) - \min_{\substack{x \in [x_\ell,x_u]}}(g(x)) \right)^2 q^{-2} + \frac{\epsilon_{tot}}{3} \left( \max_{\substack{x \in [x_\ell,x_u]}}(g(x)) - \min_{\substack{x \in [x_\ell,x_u]}}(g(x))\right)^3
\end{equation}
If we make the slightly conservative simplifying assumption that every time the circuit $P$ is used it is done so to construct the circuit $Q$, then we get an upper-bound on the number of $T$ gates as a function of the number of uses of $P$. Specifically, we let $n^{(Q)}_{R}$ and $n^{(Q)}_{T}$ be respectively the number of rotation and $T$ gates in the circuit $Q$ (prior to the approximate $T$ synthesis for the rotation gates, but after every other gate has been rebased to Clifford+T) and thus the total amount of $T$ gates is $\frac{q}{2} \left( 3 n^{(Q)}_R \log_2 (1/\epsilon) + n^{(Q)}_T  \right)$ where $q$ is the number of uses, and hence $q/2$ upper-bounds the number of instances of $Q$. Putting this together, we can obtain an optimisation problem, where the aim is to satisfy the user specified MSE ($\text{MSE}_{\text{user}}$), whilst minimising the $T$ gate count:
\begin{align*}
    \text{Minimise:  } &  \frac{q}{2} \left( 3 n^{(Q)}_R \log_2 (1/\epsilon) + n^{(Q)}_T \right) \\
    \text{such that:  } & \text{MSE}_{\text{user}} = c^2_g \cqae^2 \left(\max_{\substack{x \in [x_\ell,x_u]}}(f(x)) - \min_{\substack{x \in [x_\ell,x_u]}}(f(x))\right)^2 q^{-2} \nonumber \\
    & \,\,\,\,\,\,\,\,\,\,\,\,\,\,\,\,\,\,\,\,\,\,\,\,\,\,\,\,\,\,\,\,\,\,\,\,\,\,\,\, + \frac{q n^{(Q)}_R \epsilon}{3}  \left(\max_{\substack{x \in [x_\ell,x_u]}}(f(x)) - \min_{\substack{x \in [x_\ell,x_u]}}(f(x))\right)^3
\end{align*}
which uses $\epsilon_{\text{tot}} \leq 2 n_R \epsilon = 2 \frac{q}{2} n^{(Q)}_R \epsilon = q n^{(Q)}_R \epsilon$.\\

In the QMCI engine, this is solved to give the optimal $q$ and $\epsilon$ for the user-defined overall MSE. (If the user specifies the number of uses rather than (R)MSE, the relationship between number of uses and RMSE convergence, for the selected statistical quantity and choice of QAE, as detailed in Section~\ref{sec:observables}, is used to obtain the users implied desired (R)MSE.) Once this optimisation has been performed, resource mode proceeds to quantify the resources assuming the synthesis has been performed to the specified accuracy and reports for each circuit:
\begin{enumerate}
    \item the number of qubits;
    \item the $T$ count and depth,
\end{enumerate}
where $T$ is now the most relevant single ``bottle-neck'' gate. 

\subsection{Tightening the fault-tolerant resource estimate}
\label{subsec:tightening}

The bound $\epsilon_{tot} \leq 2n_R \epsilon$ is, in reality, extremely conservative -- in the sense that it presumes that all of the errors line up such that the discrepancy between the ideal state and that prepared by the imperfect synthesis is a coherent sum of the individual discrepancies. A more reasonable assumption would be that the discrepancies are uniform \iid, in which case the large dimension of the Hilbert space $2^n$ relative to the number of $T$ gates means that the errors will be quasi-orthogonal with very high-probability \cite{lecture-high-dimensional}
and hence a more useful approximation is given by:
\begin{equation}
    \epsilon_{tot} \approx 2\sqrt{n_R} \epsilon
\end{equation}
Using this would enable the Clifford+T synthesis to be optimised further, and there is good reason to believe that further improvements will be possible in time:
\begin{itemize}
    \item effective fault-tolerant circuit compilation is not yet a mature field, and it is reasonable to expect that in due course there will be techniques to greatly reduce rotation and T gate counts;
    \item Clifford+$T$ synthesis performed by simply replacing each rotation gate by a string of Hadamard and $T$ gates is sub-optimal -- a better approach is to re-write small sub-circuit blocks, and if ancillas are available the $T$ gate counts can be heavily reduced in some cases;
    \item the bank of rotation gates used in Fourier QMCI is highly structured -- each rotation is a power of 2 multiplied by the same constant -- and this structure could be exploited to synthesise the controlled rotations with many fewer $T$ gates;
    \item the permutational subcircuits that the enhanced $P$-builder adds have not yet been fully optimised for fault-tolerant execution.
\end{itemize}
As well as these general reasons to suppose that fault-tolerant resource requirements will be greatly reduced, for the specific benchmarks given in Section~\ref{sec:benchmarks}, it is worth observing that the distribution loading circuits are variationally trained and hence have a high proportion of rotation gates. When expressly designed for fault-tolerant execution (the benchmarks are for both NISQ and fault-tolerant execution), such circuits could be alternatively constructed to have more favourable resource requirements.

\subsection{Resource boxes}

In resource mode it is possible to include resource boxes which specify the number of qubits and number and types of gates rather than the explicit circuits. This is included to allow for resource quantification when prototyping new ideas; for instance, it may be that some application requires a certain arithmetic or logical operation that is not currently included in the enhanced $P$-builder. In this case, the user can approximate the resources needed for the operation, include these as a resource box, and thus estimate the total resources for the calculation of interest. This also further aids the development in terms of ascertaining which functions should be prioritised for future inclusion in the enhanced $P$-builder.

\newpage
\part{Benchmarks and Discussion}
\label{part:examples-results}

\section{Benchmarks }
\label{sec:benchmarks}

To demonstrate the capability of the quantum Monte Carlo integration engine, and in particular its ability to output resource estimates, we propose a set of benchmarks relevant to finance. Our motivation is to provide a series of calculations that are hard enough to be of interest -- they are all path-dependent options where MCI would be likely to be the most effective method classically -- but simple enough to be comprehensible and of relatively near-term interest. With this in mind, we give resource estimates for computing the expected payoff for:
\begin{enumerate}
    \item barrier options;
    \item look-back options;
    \item autocallables;
\end{enumerate}
where each of these derivative instrument depends on a single low-volatility underlying asset.

\subsection{Modelling the financial time-series}

In order to provide a simple and re-usable set of benchmarks, we propose a slightly simplified problem that is inspired by a financial time-series. In particular, any digital computer (classical or quantum) can only work with a discrete approximation of a continuously varying financial time-series, and for our purposes we take this discrete approximation as the ground truth. \\

We use the standard textbook model of geometric Brownian motion as the starting point for our discrete financial time-series, however, as we choose to work in return space, it is a discretisation of Brownian motion that is required. Specifically, at each time step an \iid~random variable that is a 64-point (i.e., six-qubit) discrete approximation of a Gaussian random variable is added to the logarithm of the price. Owing to the location-scale property of the Gaussian distribution it suffices to generate a single quantum circuit which encodes a Gaussian distribution, and to then interpret the support according to the needs of the problem. The specific circuit we use to do so is given in Fig.~\ref{fig:normal-discrete}; the resulting distribution approximates a unit Gaussian, with support extending out to $\pm 5 \sigma$. From this, we consider two discrete time-series, each with total volatility $10 \%$:
\begin{enumerate}
    \item a time-series with four time-slices -- thus each \iid~Gaussian is $\mathcal{N}(0,1/400)$ (i.e., the total volatility is thus $\sqrt{4 \times 1/400} = 10 \%$);
    \item a time-series with eight time-slices -- thus each \iid~Gaussian is $\mathcal{N}(0,1/800)$ (i.e., the total volatility is thus $\sqrt{8 \times 1/800} = 10 \%$);
\end{enumerate}
We choose to work in return space as, even though $c_{\text{exp}} > c_{x}$ (hence the rate of convergence is slower by a constant factor), constructing Brownian motion rather than geometric Brownian motion is especially advantageous in the low-volatility regime. In this setting, when working in price space the random variable takes a value close to one (when normalised by the asset's starting price) with high probability, and therefore the most efficient use of the qubit registers representing the random variable is a grid of values closely spread around one (so with $x_\ell$ slightly smaller than one, and small $\Delta$). However, multiplication of such lognormal random variables (as is required to construct geometric Brownian motion) is not straightforward, and the necessary register size effectively grows as if the random variables had instead been initialised with $x_{\ell} =0$, thus leading to vast numbers of qubits and gates. No such problem exists when working in return space, where the Brownian motion is constructed simply by summing zero-mean Gaussian random variables. For these reasons, even though the sample complexity is higher working in return space, we expect the overall gate complexity (and qubit number) to be significantly lower when working therein.\\

Even though our preference \textit{is} to work in return space, for completeness we also give circuits to sample the lognormal random variables, as would be required to work in price space, and all of the circuits are displayed in Appendix~\ref{sec:benchmark-lognormals}\footnote{Pytket code for the circuits can also be supplied upon request.}.

\subsection{Details of the selected instruments}

As we are interested in resource estimates, the exact settings are of little consequence as the resources do not vary significantly with these; however, for the sake of definiteness we use the following settings\footnote{Note that the barriers have all been set at approximately one standard deviation -- and this is a conservative choice, in terms of the resource counts, as the barriers need to be checked at every time slice in this case.}.
\begin{itemize}
    \item \textbf{The barrier option} is a knock-out call option, with (constant) barrier of 1.1 times the initial price; and strike price of 1.05 times the initial price.
    \item \textbf{The look-back option} is a call option with strike price of 1.05 times the original price.
    \item \textbf{The autocallable} is defined by two components. 
    \begin{enumerate}
        \item a series of binary call options $\{(K_i, t_i , b_i)\}_{i= 1, \ldots m}$, which pay out $b_i$ times the initial price if the price at $t_i$ exceeds $K_i$ times the initial price. In our examples we include calls on every odd numbered timestep with linearly increasing values for $K_i$ and $b_i$.
        \item  A knock-out put option with barrier of 0.9 times the initial price and strike price 0.95 times the initial price.
    \end{enumerate}
    If any of the call options are satisfied the payoff of the autocallable is the payoff of the earliest option (smallest $t_i$) that is satisfied. If none of the call options are fulfilled, the payoff is determined by the knock-out put and will be negative in this case. This is done in practice to reduce the overall price of an autocallable option, see for example Ref.~\cite{chakrabarti2020threshold} from which we have adapted this example.
\end{itemize}

\begin{figure}
\begin{center}
\begin{minted}[fontsize=\footnotesize,
    linenos,
    style=manni,
    tabsize=4, 
    autogobble,
    numbersep=8pt]{python}
# USING STATE PREPARATION MODULE TO LOAD THE DISTRIBUTION P WITH 4 DIMENSIONS
state_preparation = StatePreparation()
state_preparation_parameters = StatePreparationParameters(
    distribution=Normal(mean=0, variance=1),  # Load the univariate unit Gaussian
    n_dimensions=4,                           # Four identical copies
    n_qubits=6,                               # Same number of qubits for each dimension
    xl=-5,                                    # Same xl for each dimension
    delta=10/63,                              # Same delta for each dimension
    ansatz="HWE",                             # (or "TN") Specify the ansatz/loader
    n_layers=6,                               # Same number of ansatz layers for each dimension
    )                              
distribution_circuit = state_preparation.get_circuit(state_preparation_parameters)
# Get the distribution circuit for the multi-variate normal (to be used in QMCI)

# USING ENHANCED-P BUILDER TO ENHANCE THE LOADED STATE
operations = [Sum(4, 1), Sum(5, 2), Sum(6, 3)]
# The Brownian motion is prepared in dimensions 4,5,6,7

thresholds = 
  [Threshold(dimension=4, value=log(1.1), type=BoundType.Upper),  # Knock-out condition, 1st time-slice
   Threshold(dimension=5, value=log(1.1), type=BoundType.Upper),  # Knock-out condition, 2nd time-slice   
   Threshold(dimension=6, value=log(1.1), type=BoundType.Upper),  # Knock-out condition, 3rd time-slice
   Threshold(dimension=7, value=log(1.1), type=BoundType.Upper),  # Knock-out condition, 4th time-slice
   Threshold(dimension=7, value=log(1.05), type=BoundType.Lower)] # Check strike price, 4th time-slice 

esop = ESOP(products=[[(0, True), (1, True), (2, True), (3, True), (4, True)]])
# If the knock-out conditions haven't occurred AND the price is above the strike price

# USING MONTE-CARLO INTEGRATION MODULE TO RUN QMCI
amplitude_estimator = AmplitudeEstimator(type="MLQAE") # Specify type of amplitude estimator

function_applied = Exponential()   # In return space -- use exponential
function_applied.flag_shift = 1.05 # Re-base final time-slice support by subtracting strike price

monte_carlo_integration = MonteCarloIntegrationBuilder.build(
            distribution_circuit=distribution_circuit,
            dimension=7, # Integrate over the last dimension (i.e. the final time slice)
            amplitude_estimator=amplitude_estimator,
            function_applied=function_applied,
            operations=operations,
            thresholds=thresholds,
            esop = esop,
            condition=5, # Conditional expectation
            resource_only=True, # Run in resource mode 
        ) # Build Monte Carlo Integrator

target_rmse = 1.04E-2 # Set desired RMSE for estimate
monte_carlo_integration_result = monte_carlo_integration.estimate_result(target_rmse=target_rmse)
resource_estimates = monte_carlo_integration_result.resource_estimates # Get resource estimates
\end{minted}
\end{center}
\caption{Pseudocode for estimating the payoff of a four time-slice knock-out call option. Note that the indicator qubits are implicitly indexed (starting at 0) in order of creation, hence indicator qubits 0 -- 4 are the threshold conditions in L20 -- 24, and indicator qubit 5  is the esop in L26 -- and it is upon this that the expectation is conditioned in L43.}
\label{fig:barrier-code}
\end{figure}

For each instrument, we quantify the resources required to compute the expected payoff when the payoff is each of the value of the asset, and a fixed \textit{binary} payoff (when the option is exercised)\footnote{As autocallables are already somewhat sophisticated instruments, combining different payoffs under different scenarios, it is doubtful that one would further define the payoff to be binary -- however as it is possible to construct such calculations we report the binary payoff for completeness.}. For the former we set the target RMSE as $10^{-2}(e^{0.5} - e^{-0.5}) = 1.04 \times 10^{-2}$ and for the latter the target RMSE as $10^{-2}$. In each case, these RMSE values correspond to $1\%$ of the range of the support and have been chosen to approximately coincide with the cross-over to quantum advantage in sample complexity. As we use MLQAE to generate the results, Table~\ref{tab:QMCI-convergence} shows that RMSE of 3.12\% of the support and 1.2\% of the support is the cross-over for value and binary payoffs respectively, and so we have indeed entered into the regime of quantum advantage in sample complexity. Fig.~\ref{fig:barrier-code} gives pseudocode for the four time-slice barrier option.\\

It is also worth mentioning that the specified error of $10^{-2}$ also motivates the choice of using six qubits for the circuits sampling \iid~Gaussian random variables. Recalling that the measurement of the state prepared by the six-qubit circuit amounts to sampling a discrete random variable supported on $2^6 = 64$ points, and that the PMF is a discrete approximation of a Gaussian PDF out to $\pm 5 \sigma$, simple statistics tells us that after summing four such random variables (as is the final time-slice of the shorter of the two time-series that we consider) the corresponding $\pm 5 \sigma$ will be supported by $\sqrt{4} \times 64 = 128$ points. Hence we use the ability to apply a function only to part of the support (i.e., the range of the support corresponding to $\pm 5 \sigma$), as detailed in Section~\ref{subsect:func-part-support}, and so it follows that, with 128 supported points (and hence a resolution of $1/128 = 0.0078$ of the support), it is reasonable to calculate the RMSE to $1\% = 0.01$ of the total support.

\subsection{Benchmark results}

\begin{table}[!t]
    \centering
    \begin{tabular}{c c  c  c c  c c }
    \hline
\hline\\
         \textbf{Instrument} & \textbf{Payoff} & \textbf{Qubits} & \textbf{Gate count}  & \textbf{CNOT count} &  \textbf{Gate depth} & \textbf{CNOT depth} \\[1.5ex]  

        \hline&&&&&&\\[-0.8ex]
         Barrier & Value  & $119$ & $4.95\times 10^6$  & $2.44\times 10^6$ & $2.42\times 10^6$ & $1.30\times 10^6$ \\
         (4 slices) & Binary  & $117$ & $1.84\times 10^6$  & $8.98\times 10^5$ & $8.63\times 10^5$ & $4.56\times 10^5$\\
          &&&&&&\\[-0.8ex]
         Barrier & Value  & 
         $307$ & $1.30\times 10^7$  & $6.44\times 10^6$ & $5.15\times 10^6$ & $2.71\times 10^6$ \\
         (8 slices) & Binary  &  $305$ & $4.98\times 10^6$  & $2.46\times 10^6$ & $1.96\times 10^6$ & $1.02\times 10^6$ \\
          &&&&&&\\[-0.8ex]
         Look-back & Value  & $167$ & $9.26\times 10^6$  & $4.62\times 10^6$ & $4.91\times 10^6$ & $2.72\times 10^6$ \\
         (4 slices) & Binary  & $165$ & $4.20\times 10^6$  & $2.09\times 10^6$ & $2.24\times 10^6$ & $1.23\times 10^6$\\
          &&&&&&\\[-0.8ex]
         Look-back & Value  & $439$ & $2.45\times 10^7$  & $1.22\times 10^7$ & $1.25\times 10^7$ & $6.90\times 10^6$ \\
         (8 slices) & Binary  & $437$ & $1.12\times 10^7$  & $5.57\times 10^6$ & $5.76\times 10^6$ & $3.15\times 10^6$\\
          &&&&&&\\[-0.8ex]
         Autocallable & Value  &  $123$ & $1.07\times 10^7$  & $5.27\times 10^6$ & $5.41\times 10^6$ & $2.89\times 10^6$ \\
         (4 slices) & Binary  & $121$ & $5.93\times 10^6$  & $2.89\times 10^6$ & $3.00\times 10^6$ & $1.59\times 10^6$\\
          &&&&&&\\[-0.8ex]
         Autocallable & Value  & $315$ & $7.76\times 10^7$  & $3.83\times 10^7$ & $3.22\times 10^7$ & $1.68\times 10^7$\\
         (8 slices) & Binary  &  $313$ & $2.87\times 10^7$  & $1.41\times 10^7$ & $1.21\times 10^7$ & $6.33\times 10^6$\\
         \hline 
         \hline  \\[-0.5ex]
        \multicolumn{7}{c}{(A)} \\
    \end{tabular}
    \vskip 1cm
    \begin{tabular}{c c  c  c c  c c }
        \hline \hline \\
         \textbf{Instrument} & \textbf{Payoff} & \textbf{Qubits} & \textbf{Gate count}  & \textbf{CNOT count} &  \textbf{Gate depth} & \textbf{CNOT depth} \\[1.5ex]  
         \hline&&&&&&\\[-0.8ex]
         Barrier & Value  & $119$ & $8.37\times 10^4$  & $4.13\times 10^4$ & $4.14\times 10^4$ & $2.22\times 10^4$ \\
         (4 slices) & Binary  &$117$ & $3.10\times 10^4$  & $1.52\times 10^4$ & $1.48\times 10^4$ & $7.80\times 10^3$ \\
         &&&&&&\\[-0.8ex]
         Barrier & Value  & $307$ & $2.20\times 10^5$  & $1.09\times 10^5$ & $8.87\times 10^4$ & $4.67\times 10^4$ \\
         (8 slices) & Binary  &$305$ & $8.41\times 10^4$  & $4.16\times 10^4$ & $3.37\times 10^4$ & $1.76\times 10^4$\\
         &&&&&&\\[-0.8ex]
         Look-back & Value  & $167$ & $9.50\times 10^4$  & $4.75\times 10^4$ & $5.10\times 10^4$ & $2.82\times 10^4$ \\
         (4 slices) & Binary  &$165$ & $7.07\times 10^4$  & $3.52\times 10^4$ & $3.80\times 10^4$ & $2.08\times 10^4$ \\
         &&&&&&\\[-0.8ex]
         Look-back & Value  & $439$ & $2.49\times 10^5$  & $1.25\times 10^5$ & $1.29\times 10^5$ & $7.11\times 10^4$ \\
         (8 slices) & Binary  &$437$ & $1.88\times 10^5$  & $9.40\times 10^4$ & $9.76\times 10^4$ & $5.34\times 10^4$ \\
         &&&&&&\\[-0.8ex]
         Autocallable & Value  & $123$ & $1.14\times 10^5$  & $5.64\times 10^4$ & $5.77\times 10^4$  & $3.08\times 10^4$ \\
         (4 slices) & Binary  & $121$ & $5.60\times 10^4$  & $2.75\times 10^4$ & $2.83\times 10^4$  & $1.49\times 10^4$  \\
         &&&&&&\\[-0.8ex]
         Autocallable & Value  & $315$ & $5.29\times 10^5$  & $2.62\times 10^5$ & $2.35\times 10^5$  & $1.22\times 10^5$\\
         (8 slices) & Binary  & $313$ & $2.63\times 10^5$  & $1.30\times 10^5$ & $1.19\times 10^5$  & $6.17\times 10^4$ \\
         \hline 
         \hline\\[-0.5ex]
         \multicolumn{7}{c}{(B)} \\
    \end{tabular}
   \caption[NISQ resource quantification for the benchmark problems]{NISQ resource quantification for the benchmark problems: (A) total; (B) largest circuit.}
   \label{tab:benchmarks-NISQ}
\end{table}
\begin{table}[!t]
    \centering
    \begin{tabular}{c c  c   c c  c c }
      \hline
\hline&&&&&&\\[-0.3ex]
    & & & \multicolumn{2}{c}{\textbf{Total} }  & \multicolumn{2}{c}{\textbf{Largest circuit} } \\
         \textbf{Instrument} & \textbf{Payoff} & \textbf{Qubits} & \textbf{T count}  & \textbf{T depth } & \textbf{T count}  & \textbf{T depth }  \\[1.5ex]  
         \hline&&&&&&\\[-0.8ex]
         Barrier & Value  &$119$ & $7.27\times 10^7$  & $3.03\times 10^7$  &$1.18\times 10^6$  & $4.91\times 10^5$ \\ 
         (4 slices) & Binary  & $117$ & $1.46\times 10^7$  & $1.04\times 10^6$ & $2.72\times 10^5$  & $1.96\times 10^4$    \\
         &&&&&&\\[-0.8ex]
         Barrier & Value  & 
         $307$ & $1.19\times 10^8$  & $3.55\times 10^7$ &  $1.93\times 10^6$  & $5.76\times 10^5$    \\
         (8 slices) & Binary  & $305$ & $3.05\times 10^7$  & $1.55\times 10^6$  & $5.68\times 10^5$  & $2.93\times 10^4$    \\
         &&&&&&\\[-0.8ex]
         Look-back & Value  &  $167$ &  
 $8.72\times 10^7$  & $3.52\times 10^7$ &   $8.22\times 10^5$  & $3.32\times 10^5$  \\
         (4 slices) & Binary  & $165$ &  $2.10\times 10^7$  & $2.60\times 10^6$ &  $3.91\times 10^5$  & $4.86\times 10^4$ \\
         &&&&&&\\[-0.8ex]
         Look-back & Value  & $439$ & $1.64\times 10^8$  & $5.01\times 10^7$ &  $1.63\times 10^6$  & $5.00\times 10^5$ \\
         (8 slices) & Binary  &$437$ & $4.80\times 10^7$  & $5.72\times 10^6$   &  $8.95\times 10^5$  & $1.07\times 10^5$  \\
         &&&&&&\\[-0.8ex]
         Autocallable & Value  & $123$ & $1.38\times 10^8$  & $4.79\times 10^7$ & $1.53\times 10^6$  & $6.40\times 10^5$  \\
         (4 slices) & Binary  & $121$ & $5.39\times 10^7$  & $4.66\times 10^6$ & $4.10\times 10^5$  & $3.08\times 10^4$   \\
         &&&&&&\\[-0.8ex]
         Autocallable & Value  & $315$ & $7.54\times 10^8$  & $1.10\times 10^8$  & $4.71\times 10^6$  & $1.26\times 10^6$   \\
         (8 slices) & Binary  &$313$ & $2.27\times 10^8$  & $1.32\times 10^7$  & $1.39\times 10^6$  & $7.84\times 10^4$   \\
         \hline 
         \hline     
    \end{tabular}
    \caption[Fault-tolerant resource quantification for the benchmark problems]{Fault-tolerant resource quantification for the benchmark problems.}
   \label{tab:benchmarks-FT}
\end{table}

We give resource quantifications for both NISQ and fault-tolerant settings, and for each values are given for both the total algorithm and also the largest single circuit. The results are shown in Tables~\ref{tab:benchmarks-NISQ} and \ref{tab:benchmarks-FT}, and a first observation is that the number of qubits needed for these calculations is not excessive. It can be seen that non-trivial QMCI calculations, inspired by financial problems can be performed with total qubit number in the 100 -- 500 range, and whilst these numbers will increase to account for the discrepancy between the actual time-series model and its discrete approximation and to represent a more finely time-sliced discretisation of the financial time-series, this increase will not be exorbitant. Specifically, the qubit number will grow only logarithmically with the reciprocal of the target RMSE and quasi-linearly with the number of time-steps (when working in return space). Furthermore, building the operator $Q$ requires a large multi-controlled-NOT gates, and at present we decompose this in a manner that is extremely ancilla-hungry -- approximately half of the total number of qubits are ancillas -- and there is room to vastly reduce the ancilla count with only a modest increase to the total gate count.\\

Thus our attention turns to the number of operations, and the question of whether physical or error-corrected, logical qubits will be required. We can see from Table~\ref{tab:benchmarks-NISQ} that in all likelihood it will be the latter for the eight time-slice benchmarks. However, for the four time-slice benchmarks it is foreseeable that NISQ execution may be possible. Consider, for instance, the four time-slice barrier option, for which the largest circuit contains $1.44 \times 10^{4}$ CNOT gates -- a simple \textit{rule-of-thumb} calculation suggests that about $1- 1/ (1.52 \times 10^{4}) = 99.99\%$ two-qubit gate fidelity would therefore be required.\\

Note that this estimation, whilst somewhat useful, implies that we should run the circuit to a depth where one error is expected (and so can expect to obtain an error-free output with only a small number of repeats)\footnote{Note the precise conversion of fidelity to error-rate depends on the noise model.}, however this does not really capture how noise or QAE works. In particular, as discussed in Section~\ref{subsec:NoiseQAE}, we can use noise-aware QAE to run much deeper QMCI circuits than one would expect from such a rule-of-thumb. Moreover, we can reasonably expect future circuit optimisation to reduce the CNOT gate count considerably (especially noting that no special optimisation of the CNOT-heavy enhanced $P$-builder circuits has been undertaken for these benchmarks); and it may be possible to define a slightly further simplified version of the problem, without losing its essential nature. Accordingly, it is entirely reasonable to speculate that a future quantum computer with $\sim 100$ qubits and two-qubit gate fidelity $\sim 99.99\%$ should be capable of running some simple, but not trivial financial QMCI calculations. However, whilst such a putative future quantum computer may be able to obtain an advantage in \textit{sample} complexity for a non-trivial financial Monte Carlo integral -- which would itself constitute a valuable outcome -- it is doubtful that it would make practical sense to price such an option on a quantum computer, as we discuss in more detail in Section~\ref{sec:discussion}.\\

Turning then to the resource counts for fault-tolerant circuit execution, we should first note that $T$-gate counts are typically high for fault-tolerant circuit execution, largely owing to relative paucity of research on optimised synthesis into fault-tolerant gate-sets meaning that long-known conservative upper-bounds must be relied on, as was discussed (along with other reasons that the fault-tolerant resource quantification is likely to come down significantly) in Section~\ref{subsec:tightening}. For these reasons, the absolute values in Table~\ref{tab:benchmarks-FT} are of relatively little meaning, apart from in the relative sense -- that they give a target to aim for in the reduction of fault-tolerant resources.\\

\section{Discussion}
\label{sec:discussion}

This paper has set out the full technical details of Quantinuum's quantum Monte Carlo integration engine, including our statistically robust QAE algorithm, and general framework for building financial derivative and risk calculations. We have also proposed a set of simple benchmarks, inspired by (slightly) simplified versions of MCI calculations that are ubiquitous in mathematical finance. Whilst these benchmarks clearly indicate that there is some significant further hardware scaling required for quantum advantage, we believe that the QMCI engine will provide utility well ahead of quantum advantage, namely in providing a \textit{research and development} tool.

\subsection{The QMCI engine as an R\&D tool}

The QMCI engine allows users to try out different means of encoding mathematical finance calculations in quantum circuits. Notably, users will be able to test out design choices about data-loading -- i.e., constructing quantum circuits that sample from financial time-series (potentially for multiple correlated assets) -- with resource mode quantifying the relative merits of each. Related questions about how finely to slice up the time-series and whether to work in return or price space can also be probed in this way. It should also be noted that the modular design of the QMCI engine is a deliberate choice, and serves to \textit{future proof} it, as we can  easily add in ``full-scale'' components like quantum arithmetic as hardware matures. 

\subsection{From theoretical quantum advantage to \textit{useful} quantum advantage}

In the above analysis, we have largely focused comparing classical and quantum \textit{sample} complexity. This is a natural comparison to make in theoretical analysis, and owing to Ref.~\cite{herbert2} we know that a quantum sample is at least as easy to prepare as a classical sample -- so we can assert that a sample complexity advantage automatically translates into a computational complexity advantage. However, simply beating the classical MCI in terms of sample or even computational complexity does not guarantee the automatic adoption of the quantum algorithm, once quantum computers scale to such a size that they can execute the requisite circuits and hence realise the theoretical advantage. Instead there are a whole host of factors that must be taken into consideration. \\

A first thing to note on this topic is that both classical and quantum MCI (using QPE-free QAE) enjoy the possibility of massive parallelisation -- however such a possibility is already a reality in the world of classical computing, but in the case of quantum requires scaling of quantum hardware from a small set of laboratory experiments, to a fully-fledged industry (as well as the more widely-discussed scaling of the number of qubits in each device). We may summarise the factors outside of our (direct) control which will determine when it is beneficial to run QMCI rather than MCI:
\begin{enumerate}
    \item (as above)the classical and quantum parallelisation factors -- which we assume to subsume the economic cost of using multiple quantum / classical cores as well as the basic question of what is available;
    \item the clock speeds for the classical and quantum hardware;
    \item the error-correction overhead (if applicable) for the quantum algorithm. 
\end{enumerate}
These all depend on a host of factors that could radically swing the time to quantum advantage one way or the other, but as expounded in Ref.~\cite{herbertDCE} there is reason to be optimistic (or at least not unduly pessimistic) about the timeline to useful quantum advantage in QMCI. Indeed, Chakrabarti \textit{et al} make a valiant effort to translate the timeline to quantum advantage in QMCI for finance into quantities such as actual clock-rates and the like \cite{chakrabarti2020threshold}. Our view is that at this stage it is too early to do so, and it is more meaningful to give rigorous and objective comparisons between QMCI and MCI in terms of somewhat abstract quantities such as gate and sample counts; rather than giving extremely speculative comparisons between ``real-world'' quantities such as clock-speeds and parallelisation factors.\\

We do, however, know enough about the source of quantum advantage in QMCI to predict which types of calculations will see the greatest benefits. Notably, as QMCI offers a quadratic improvement in precision with the number of samples relative to classical MCI, it follows we must focus on applications where a sufficiently small error (RMSE) is demanded that QMCI indeed outperforms classical MCI, even when factors such as parallelisation and clock speed are taken into account. In finance, such a setting comes up on (at least) two important occasions. Firstly, when pricing derivatives, whilst estimation of the expected payoff itself (as exemplified in the benchmarks in Section~\ref{sec:benchmarks}) may in many instances not be too onerous to accomplish classically, it is often the case that estimating the (mathematical) derivative of the price with respect to various factors (quantities known colloquially as the \textit{Greeks} in quantitative finance -- as they are typically denoted by Greek letters) is something that would benefit from computational enhancement, as quantum computing promises to deliver. This is because, in simple terms, the Greeks can be computed by taking some finite difference approximation of the derivative, and the accuracy that can be obtained in the point estimates of the payoff therefore directly affects the minimum separation possible to compute the finite difference estimate. (Note that there are tricks to partially mitigate some of the computational difficulties in computing derivatives by finite differences, such as using the same random seed for each point estimate -- and it has been shown that QMCI can benefit from these too \cite{goldmangradients}; in time we will supplement the QMCI engine with methods of this type for efficient calculation of the Greeks.) Secondly, precision is always critical when estimating financial risk quantities such as VaR and CVaR. As these calculations concern whole portfolios (and so the financial time-series for many assets therefore need to be encoded into qubit registers) it follows that the quantum resource requirements will be correspondingly higher than for (for example) derivative instruments on a single or small number of underlying. For this reason, examples risk calculations have been used sparingly throughout this paper, however, in the fullness of time, it is perfectly foreseeable that the most dramatic impact of quantum computing on mathematical finance will be in portfolio risk quantification.\\

Another reason to be optimistic about the outlook for QMCI in finance is that, whilst it has been shown that the computational complexity of sampling is never worse quantumly than classically \cite{herbert2}, the converse does not hold. In contrived instances, should the random process in question be sampling an IQP circuit then, up to widely held complexity conjectures, the computational complexity of classical sampling is exponentially larger than its quantum counterpart \cite{IQP}. For real world data, such as financial time-series, it is not necessarily the case that there will be such a spectacular asymptotic advantage in quantum over classical sampling, however it is reasonable to suppose that the additional functionality that quantum operations offer when navigating through Hilbert space to prepare samples may offer significant reductions in circuit depth in practice. That is, the construction in Ref.~\cite{herbert2} uses a reversible circuit with a single bank of Hadamards followed by gates from the set $\{X, \text{ CNOT}, \text{ Toffoli}\}$ to construct $P$, however it is highly plausible that more ``natively quantum'' sampling circuits, exploiting a quantum-universal gateset, may prepare the desired samples in fewer operations, hence representing an extremely welcome complementary (to the quadratic speed-up in QAE) advantage.\\

With these observations in mind, we can draw up a check-list for applications that are likely to enjoy a useful quantum advantage by deploying QMCI:
\begin{enumerate}
    \item MCI is the best technique classically;
    \item precision is important: specifically, more precision that the current state-of-the-art would be valuable, but is hard to obtain by classical computational means;
    \item there is an additional, complementary advantage in loading the random process as a quantum state.\\
\end{enumerate}

\subsection{Further applications of the QMCI engine}

Whilst we have focused on mathematical finance in this paper, it is of course the case that the potential applications of QMCI are more widespread than this. Myriad problems across business, supply chain / logistics, energy, operations and data-intensive science have computational workflows that rely heavily on MCI, and in principle each of these can be enhanced by QMCI. Broadly speaking, such applications can be broken down into (i) estimation and forecasting, to which the QMCI engine can be applied in its present form, with only minor modifications; and (ii) numerical integration -- where the randomness is merely a device for computing high dimensional integrals. In the latter case, the integrand must be decomposed into the product of a first term that itself integrates to value one (and hence may be thought of as a probability distribution) and can be encoded in a quantum circuit; and a second term that depends on only one variable, and can be expressed as a periodic piecewise function (This decomposition essentially amounts to importance sampling in the classical setting.). A prominent example of computational-intensive numerical integration that has garnered some attention as a potential application of QMCI are the high-dimensional integrals in high-energy physics \cite{QMCI-HEP}.

\newpage
\part{End matter and appendices}
\label{part:end-matter}

\section*{Acknowledgments}
The authors thank Sam Duffield, Konstantinos Meichanetzidis, and Matthias Rosenkranz for kindly reviewing an earlier version of this article. We also thank Edwin Agnew and Jose Gefaell for their work that helped form the background to this paper, and for all of the fruitful discussions we had along the way.

\newpage

\normalem

\bibliography{main}
\bibliographystyle{IEEEtran}
\newpage

\appendix

\section{QAE statistical robustness benchmarks}

\begin{figure}[h!]
    \centering
    \includegraphics[width=\textwidth]{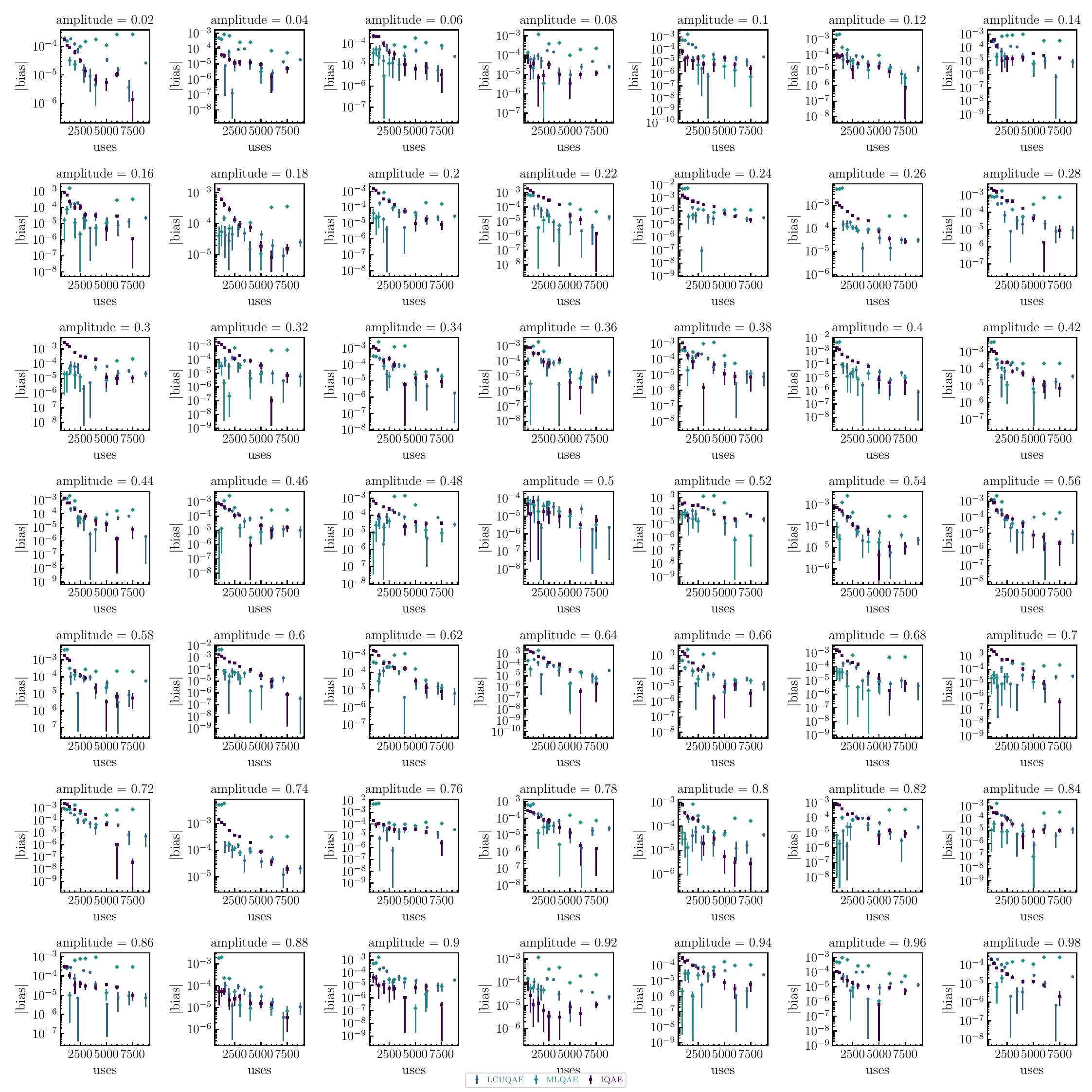}
    \caption{The bias, as found by numerical simulations for the various forms of QAE, for each of the 49 amplitudes. It can be seen that the bias is generally small for all amplitudes and all forms of QAE.}
    \label{fig:bias-by-amp}
\end{figure}

\begin{figure}[t!]
    \centering
    \includegraphics[width=\textwidth]{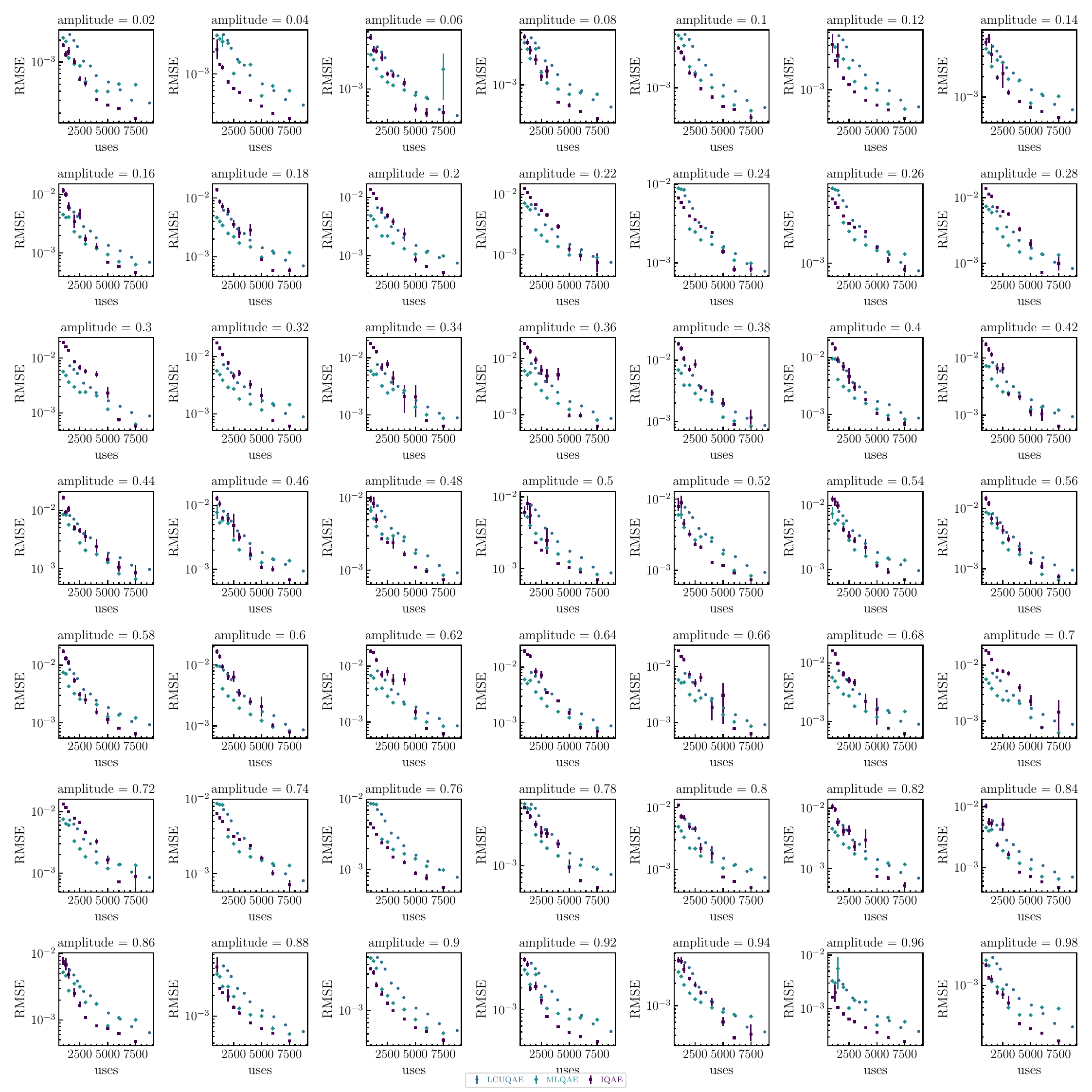}
    \caption{The RMSE, as found by numerical simulations for the various forms of QAE, for each of the 49 amplitudes. This is addressed extensively elsewhere, and included here merely for completeness.}
    \label{fig:rmse-by-amp}
\end{figure}

\begin{figure}[t!]
    \centering
    \includegraphics[width=\textwidth]{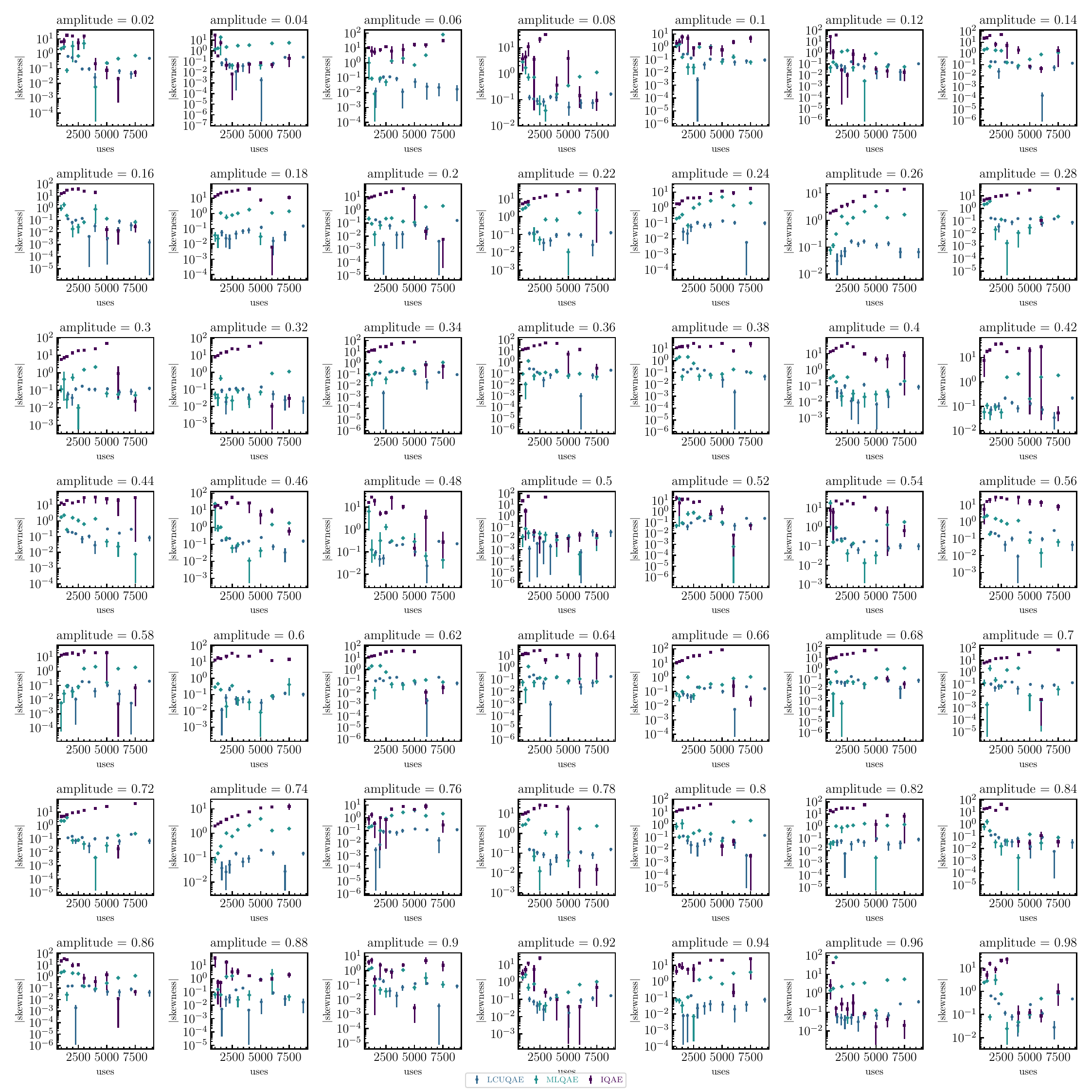}
    \caption{The skewness, as found by numerical simulations for the various forms of QAE, for each of the 49 amplitudes. It can be seen that LCU QAE generally has good skewness, whereas there are certain amplitudes for which MLQAE and IQAE have poor skewness.}
    \label{fig:skew-by-amp}
\end{figure}

\begin{figure}[t!]
    \centering
    \includegraphics[width=\textwidth]{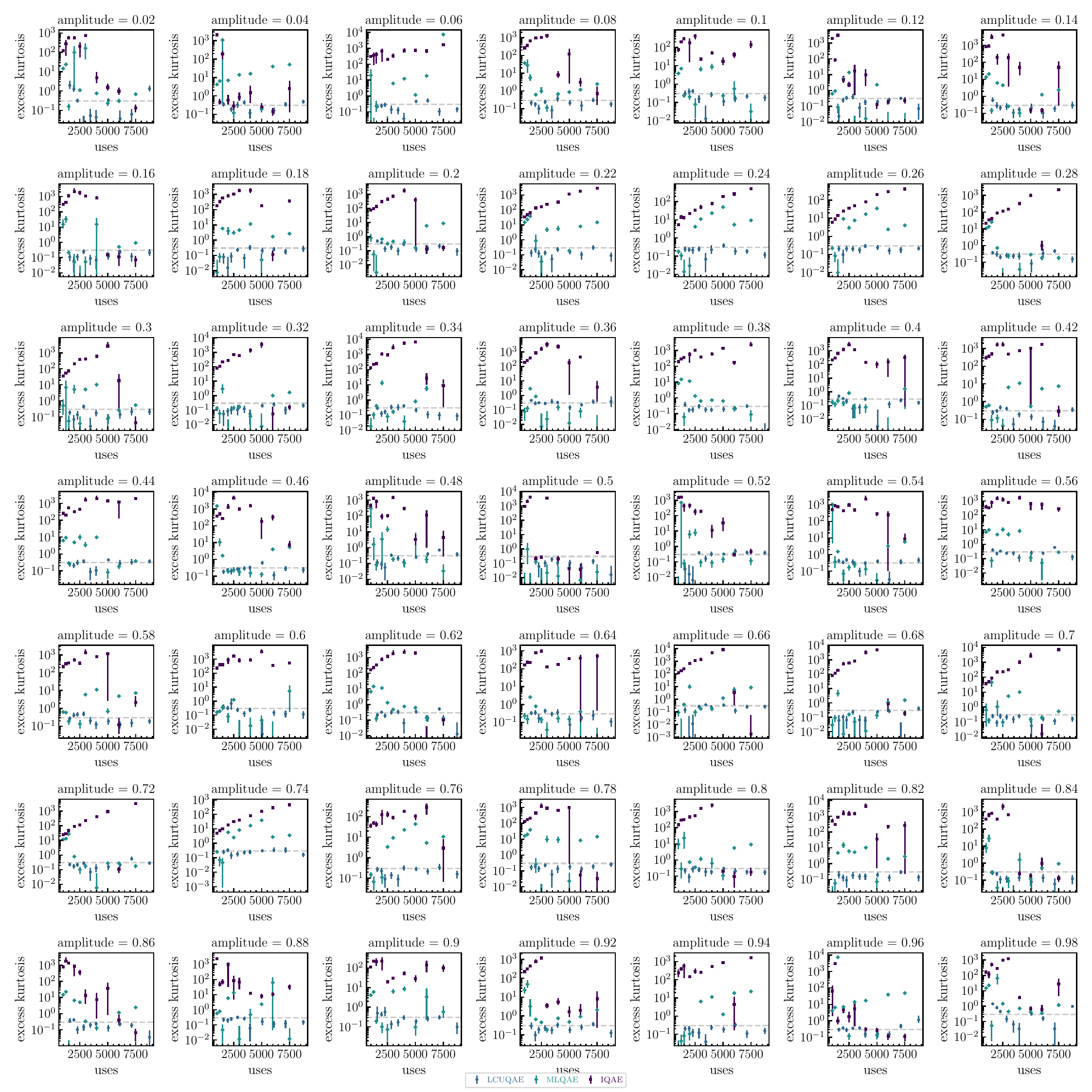}
    \caption{Excess kurtosis for the various forms of QAE, as found by numerical simulations for each of the 49 amplitudes. It can be seen that, aside from the odd outlier, LCU QAE has very low excess kurtosis, whereas neither of the other forms of QAE have good performance for all amplitudes.}
    \label{fig:kurt-by-amp}
\end{figure}

This section includes the (absolute) bias (Fig.~\ref{fig:bias-by-amp}), RMSE (Fig.~\ref{fig:rmse-by-amp}), (absolute) skewness (Fig.~\ref{fig:skew-by-amp}) and kurtosis (Fig.~\ref{fig:kurt-by-amp}) of each form of QAE for each amplitude (over a finely-spaced grid covering the entire range).

\clearpage

\clearpage

\clearpage

\clearpage

\section{RMSE convergence plots}

This Appendix includes values from numerical simulation for RMSE at various values of number of uses, for each of the 49 amplitudes simulated. Lines of best fit are used to find numerical values of $\cqae$. 

\begin{figure}[h!]
    \centering
    \includegraphics[width=\textwidth]{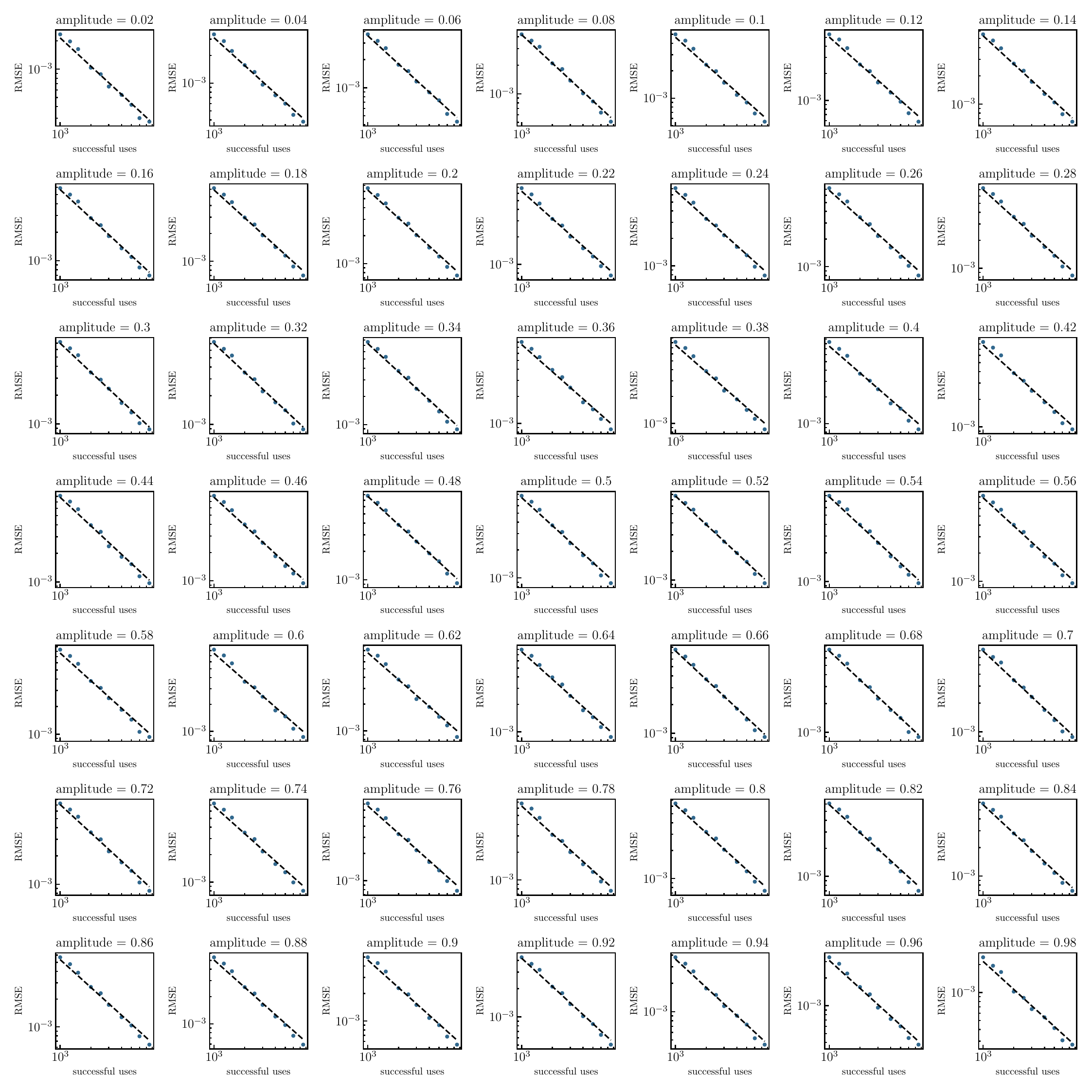}
    \caption{RMSE convergence for LCU QAE versus successful uses. In this case, the line of best fit gives the asymptotic value of $\cqae$, which is equal to 7.83 -- as is the worst case which occurs at amplitude = 0.56.}
    \label{fig:lcu-successful-rmse-fit}
\end{figure}

\clearpage

\begin{figure}[t!]
    \centering
    \includegraphics[width=\textwidth]{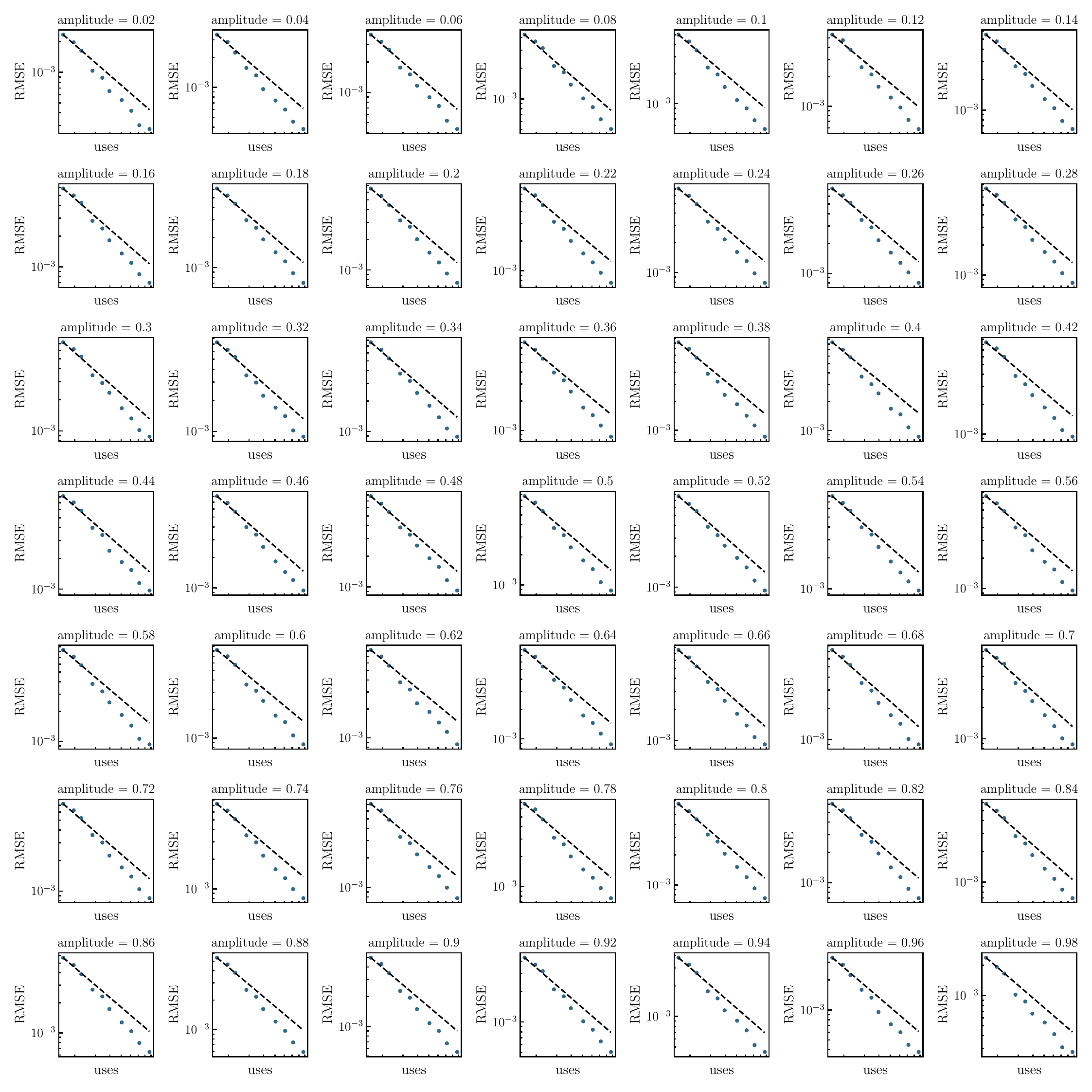}
    \caption{RMSE convergence for LCU QAE versus expected total uses. In this case, the upper-bounding line gives a conservative value of $\cqae$ that is appropriate for finite numbers of uses, and is equal to 13.3 -- as is the worst case which occurs at amplitude = 0.58.}
    \label{fig:lcu-expected-rmse-fit}
\end{figure}

\clearpage

\begin{figure}[t!]
    \centering
    \includegraphics[width=\textwidth]{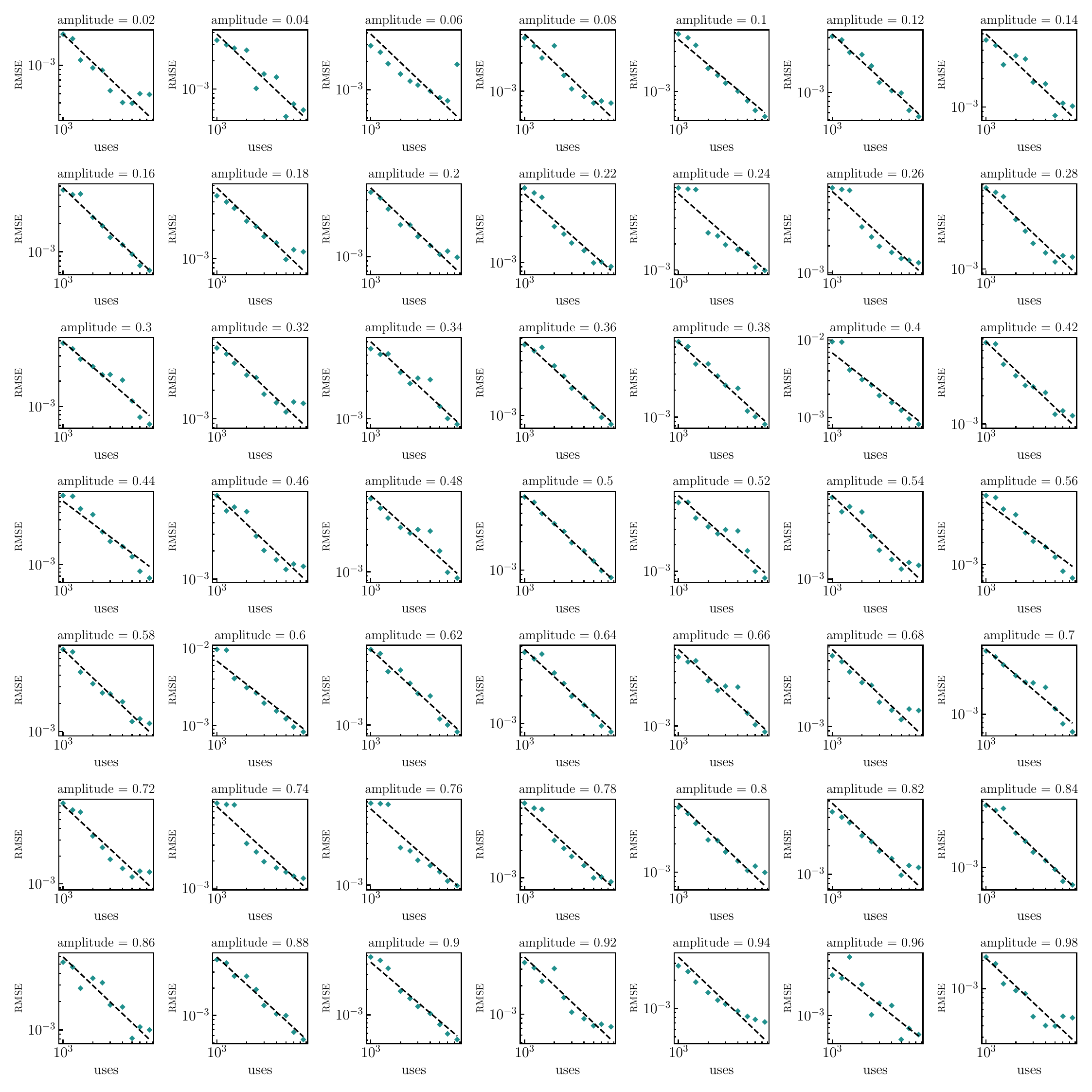}
    \caption{RMSE convergence for MLQAE versus total uses. In this case, the line of best fit gives a value of $\cqae$ which is equal to 8.02 -- as is the worst case which occurs at amplitude = 0.26.}
    \label{fig:mlqae-rmse-fit}
\end{figure}

\clearpage

\begin{figure}[t!]
    \centering
    \includegraphics[width=\textwidth]{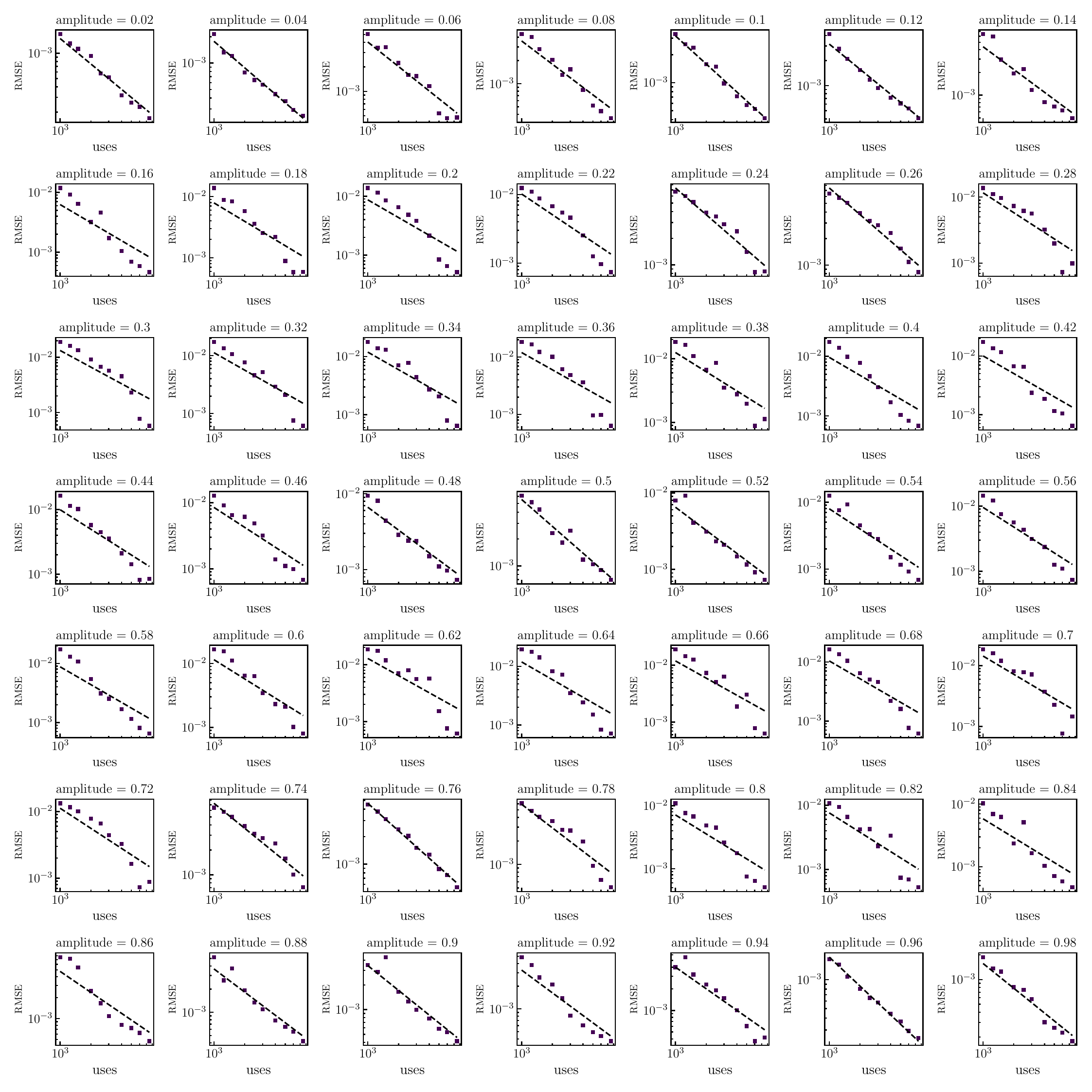}
    \caption{RMSE convergence for IQAE versus total uses. In this case, the line of best fit gives a value of $\cqae$ which is equal to 14.4 -- as is the worst case which occurs at amplitude = 0.7.}
    \label{fig:iqae-rmse-fit}
\end{figure}

\section{Circuits for benchmarks}
\label{sec:benchmark-lognormals}

Below we give the circuits used in the benchmarks in Section~\ref{sec:benchmarks}.
\begin{landscape}
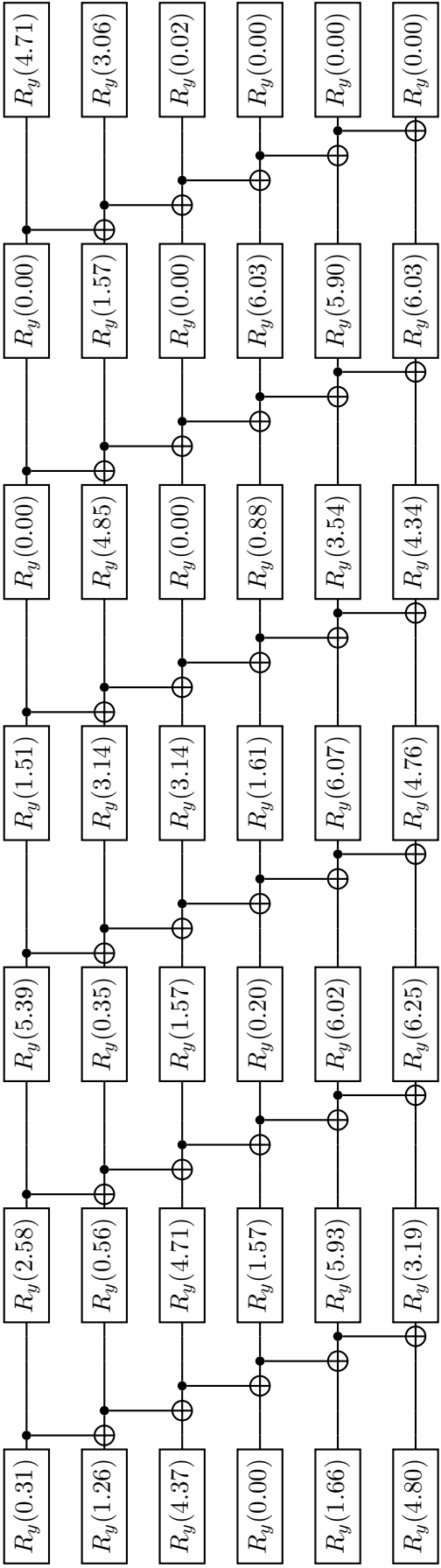
\begin{figure}
\centering
\begin{quantikz}[column sep=0.05cm]
    \gate{R_y(0.31)} & \ctrl{1} & \qw & \qw & \qw & \qw & \gate{R_y(2.58)} & \ctrl{1} & \qw & \qw & \qw & \qw & \gate{R_y(5.39)} & \ctrl{1} & \qw & \qw & \qw & \qw & \gate{R_y(1.51)} & \ctrl{1} & \qw & \qw & \qw & \qw & \gate{R_y(0.00)} & \ctrl{1} & \qw & \qw & \qw & \qw & \gate{R_y(0.00)} & \ctrl{1} & \qw & \qw & \qw & \qw & \gate{R_y(4.71)}\\
    \gate{R_y(1.26)} & \targ{} & \ctrl{1} & \qw & \qw & \qw & \gate{R_y(0.56)} & \targ{} & \ctrl{1} & \qw & \qw & \qw & \gate{R_y(0.35)} & \targ{} & \ctrl{1} & \qw & \qw & \qw & \gate{R_y(3.14)} & \targ{} & \ctrl{1} & \qw & \qw & \qw & \gate{R_y(4.85)} & \targ{} & \ctrl{1} & \qw & \qw & \qw & \gate{R_y(1.57)} & \targ{} & \ctrl{1} & \qw & \qw & \qw & \gate{R_y(3.06)}\\
    \gate{R_y(4.37)} & \qw & \targ{} & \ctrl{1} & \qw & \qw& \gate{R_y(4.71)} & \qw & \targ{} & \ctrl{1} & \qw & \qw& \gate{R_y(1.57)}  & \qw & \targ{} & \ctrl{1} & \qw & \qw& \gate{R_y(3.14)}  & \qw & \targ{} & \ctrl{1} & \qw & \qw& \gate{R_y(0.00)}  & \qw & \targ{} & \ctrl{1} & \qw & \qw& \gate{R_y(0.00)}  & \qw & \targ{} & \ctrl{1} & \qw & \qw& \gate{R_y(0.02)}\\
    \gate{R_y(0.00)} & \qw & \qw & \targ{} & \ctrl{1}& \qw & \gate{R_y(1.57)} & \qw & \qw & \targ{} & \ctrl{1}& \qw & \gate{R_y(0.20)} & \qw & \qw & \targ{} & \ctrl{1}& \qw & \gate{R_y(1.61)} & \qw & \qw & \targ{} & \ctrl{1}& \qw & \gate{R_y(0.88)} & \qw & \qw & \targ{} & \ctrl{1}& \qw & \gate{R_y(6.03)} & \qw & \qw & \targ{} & \ctrl{1}& \qw & \gate{R_y(0.00)}\\
    \gate{R_y(1.66)} & \qw & \qw & \qw & \targ{}  & \ctrl{1} & \gate{R_y(5.93)} & \qw & \qw & \qw & \targ{}  & \ctrl{1} & \gate{R_y(6.02)} & \qw & \qw & \qw & \targ{}  & \ctrl{1} & \gate{R_y(6.07)} & \qw & \qw & \qw & \targ{}  & \ctrl{1} & \gate{R_y(3.54)} & \qw & \qw & \qw & \targ{}  & \ctrl{1} & \gate{R_y(5.90)} & \qw & \qw & \qw & \targ{}  & \ctrl{1} & \gate{R_y(0.00)}\\
    \gate{R_y(4.80)} & \qw & \qw & \qw & \qw & \targ{} & \gate{R_y(3.19)} & \qw & \qw & \qw & \qw & \targ{} & \gate{R_y(6.25)} & \qw & \qw & \qw & \qw & \targ{} & \gate{R_y(4.76)} & \qw & \qw & \qw & \qw & \targ{} & \gate{R_y(4.34)} & \qw & \qw & \qw & \qw & \targ{} & \gate{R_y(6.03)} & \qw & \qw & \qw & \qw & \targ{} & \gate{R_y(0.00)}
\end{quantikz}
\caption{The 6-qubit circuit that prepares the 64-point PMF of a discrete approximation of a random variable distributed as $\mathcal{N}(0,1)$, where all rotations are given in radians (based on the standard definition of a rotation operator) and are specified up to global phases; $x_\ell = -5$ and $\Delta = 10/63$, such that the support out to $\pm 5 \sigma$ is covered. Here the angles are ``absolute'' values -- not given as multiples of $\pi$.}
\label{fig:normal-discrete}
\end{figure}
\end{landscape}

\clearpage

\begin{landscape}
\begin{figure}[t!]
\centering
\begin{quantikz}[column sep=0.05cm]
    \gate{R_y(3.09)} & \ctrl{1} & \qw & \qw & \qw & \qw & \gate{R_y(1.47)} & \ctrl{1} & \qw & \qw & \qw & \qw & \gate{R_y(3.09)} & \ctrl{1} & \qw & \qw & \qw & \qw & \gate{R_y(0.93)} & \ctrl{1} & \qw & \qw & \qw & \qw & \gate{R_y(3.21)} & \ctrl{1} & \qw & \qw & \qw & \qw & \gate{R_y(5.05)} & \ctrl{1} & \qw & \qw & \qw & \qw & \gate{R_y(0.10)}\\
    \gate{R_y(3.00)} & \targ{} & \ctrl{1} & \qw & \qw & \qw & \gate{R_y(3.67)} & \targ{} & \ctrl{1} & \qw & \qw & \qw & \gate{R_y(6.19)} & \targ{} & \ctrl{1} & \qw & \qw & \qw & \gate{R_y(0.20)} & \targ{} & \ctrl{1} & \qw & \qw & \qw & \gate{R_y(6.06)} & \targ{} & \ctrl{1} & \qw & \qw & \qw & \gate{R_y(0.18)} & \targ{} & \ctrl{1} & \qw & \qw & \qw & \gate{R_y(0.43)}\\
    \gate{R_y(3.65)} & \qw & \targ{} & \ctrl{1} & \qw & \qw& \gate{R_y(3.26)} & \qw & \targ{} & \ctrl{1} & \qw & \qw& \gate{R_y(0.27)}  & \qw & \targ{} & \ctrl{1} & \qw & \qw& \gate{R_y(3.97)}  & \qw & \targ{} & \ctrl{1} & \qw & \qw& \gate{R_y(0.43)}  & \qw & \targ{} & \ctrl{1} & \qw & \qw& \gate{R_y(1.63)}  & \qw & \targ{} & \ctrl{1} & \qw & \qw& \gate{R_y(2.14)}\\
    \gate{R_y(2.48)} & \qw & \qw & \targ{} & \ctrl{1}& \qw & \gate{R_y(2.41)} & \qw & \qw & \targ{} & \ctrl{1}& \qw & \gate{R_y(6.28)} & \qw & \qw & \targ{} & \ctrl{1}& \qw & \gate{R_y(0.80)} & \qw & \qw & \targ{} & \ctrl{1}& \qw & \gate{R_y(5.60)} & \qw & \qw & \targ{} & \ctrl{1}& \qw & \gate{R_y(5.80)} & \qw & \qw & \targ{} & \ctrl{1}& \qw & \gate{R_y(4.28)}\\
    \gate{R_y(3.92)} & \qw & \qw & \qw & \targ{}  & \ctrl{1} & \gate{R_y(1.51)} & \qw & \qw & \qw & \targ{}  & \ctrl{1} & \gate{R_y(0.10)} & \qw & \qw & \qw & \targ{}  & \ctrl{1} & \gate{R_y(4.42)} & \qw & \qw & \qw & \targ{}  & \ctrl{1} & \gate{R_y(5.77)} & \qw & \qw & \qw & \targ{}  & \ctrl{1} & \gate{R_y(4.98)} & \qw & \qw & \qw & \targ{}  & \ctrl{1} & \gate{R_y(5.18)}\\
    \gate{R_y(0.74)} & \qw & \qw & \qw & \qw & \targ{} & \gate{R_y(5.01)} & \qw & \qw & \qw & \qw & \targ{} & \gate{R_y(0.56)} & \qw & \qw & \qw & \qw & \targ{} & \gate{R_y(0.30)} & \qw & \qw & \qw & \qw & \targ{} & \gate{R_y(5.62)} & \qw & \qw & \qw & \qw & \targ{} & \gate{R_y(0.70)} & \qw & \qw & \qw & \qw & \targ{} & \gate{R_y(0.26)}
\end{quantikz}
\caption{The 6-qubit circuit that prepares the 64-point PMF of a discrete approximation of a random variable distributed as $\mathcal{LN}(0,1/800)$, where all rotations are given in radians (based on the standard definition of a rotation operator) and are specified up to global phases; $x_\ell = 0.83$ and $\Delta = 0.01$, such that the support out to $\pm 5 \sigma$ is covered. Here the angles are ``absolute'' values -- not given as multiples of $\pi$.}
\label{fig:lognormal-discrete-1}
\end{figure}
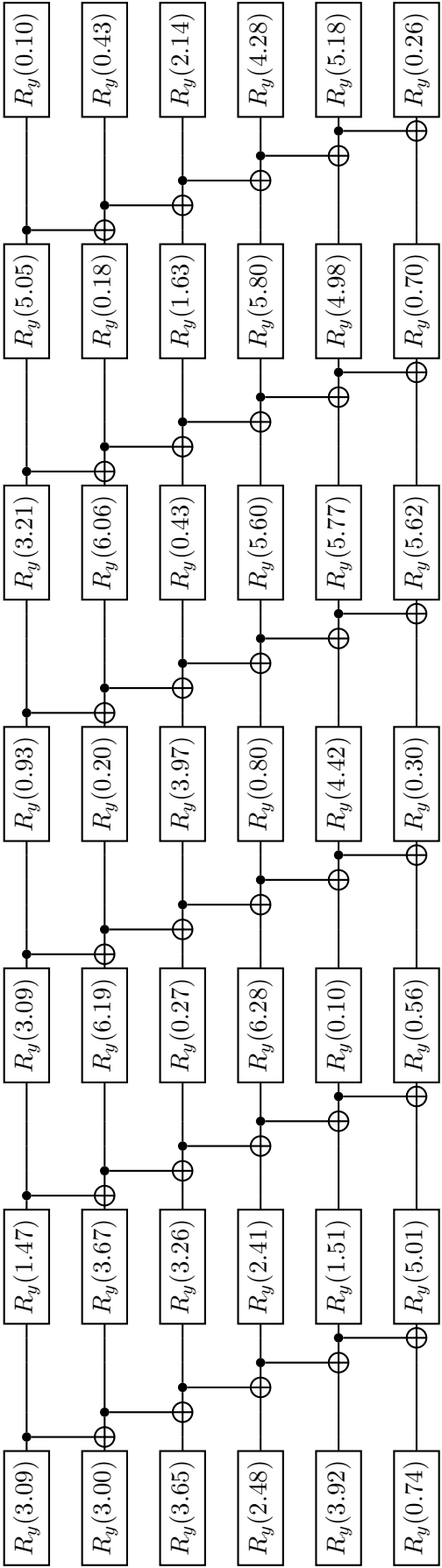
\end{landscape}

\clearpage

\begin{landscape}
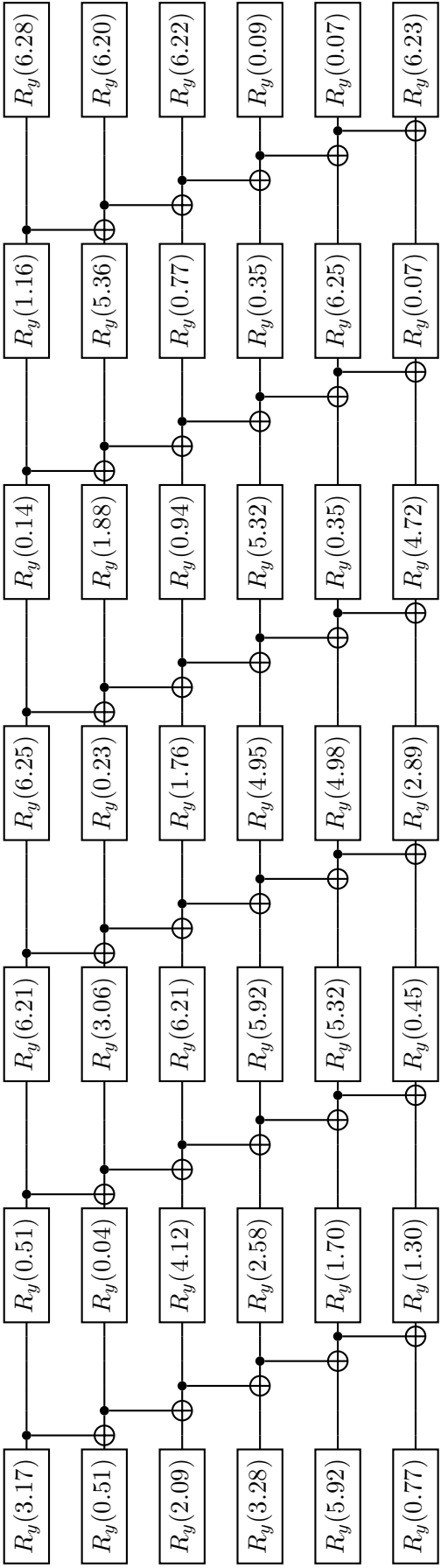
\begin{figure}[t!]
\centering
\begin{quantikz}[column sep=0.05cm]
    \gate{R_y(3.17)} & \ctrl{1} & \qw & \qw & \qw & \qw & \gate{R_y(0.51)} & \ctrl{1} & \qw & \qw & \qw & \qw & \gate{R_y(6.21)} & \ctrl{1} & \qw & \qw & \qw & \qw & \gate{R_y(6.25)} & \ctrl{1} & \qw & \qw & \qw & \qw & \gate{R_y(0.14)} & \ctrl{1} & \qw & \qw & \qw & \qw & \gate{R_y(1.16)} & \ctrl{1} & \qw & \qw & \qw & \qw & \gate{R_y(6.28)}\\
    \gate{R_y(0.51)} & \targ{} & \ctrl{1} & \qw & \qw & \qw & \gate{R_y(0.04)} & \targ{} & \ctrl{1} & \qw & \qw & \qw & \gate{R_y(3.06)} & \targ{} & \ctrl{1} & \qw & \qw & \qw & \gate{R_y(0.23)} & \targ{} & \ctrl{1} & \qw & \qw & \qw & \gate{R_y(1.88)} & \targ{} & \ctrl{1} & \qw & \qw & \qw & \gate{R_y(5.36)} & \targ{} & \ctrl{1} & \qw & \qw & \qw & \gate{R_y(6.20)}\\
    \gate{R_y(2.09)} & \qw & \targ{} & \ctrl{1} & \qw & \qw& \gate{R_y(4.12)} & \qw & \targ{} & \ctrl{1} & \qw & \qw& \gate{R_y(6.21)}  & \qw & \targ{} & \ctrl{1} & \qw & \qw& \gate{R_y(1.76)}  & \qw & \targ{} & \ctrl{1} & \qw & \qw& \gate{R_y(0.94)}  & \qw & \targ{} & \ctrl{1} & \qw & \qw& \gate{R_y(0.77)}  & \qw & \targ{} & \ctrl{1} & \qw & \qw& \gate{R_y(6.22)}\\
    \gate{R_y(3.28)} & \qw & \qw & \targ{} & \ctrl{1}& \qw & \gate{R_y(2.58)} & \qw & \qw & \targ{} & \ctrl{1}& \qw & \gate{R_y(5.92)} & \qw & \qw & \targ{} & \ctrl{1}& \qw & \gate{R_y(4.95)} & \qw & \qw & \targ{} & \ctrl{1}& \qw & \gate{R_y(5.32)} & \qw & \qw & \targ{} & \ctrl{1}& \qw & \gate{R_y(0.35)} & \qw & \qw & \targ{} & \ctrl{1}& \qw & \gate{R_y(0.09)}\\
    \gate{R_y(5.92)} & \qw & \qw & \qw & \targ{}  & \ctrl{1} & \gate{R_y(1.70)} & \qw & \qw & \qw & \targ{}  & \ctrl{1} & \gate{R_y(5.32)} & \qw & \qw & \qw & \targ{}  & \ctrl{1} & \gate{R_y(4.98)} & \qw & \qw & \qw & \targ{}  & \ctrl{1} & \gate{R_y(0.35)} & \qw & \qw & \qw & \targ{}  & \ctrl{1} & \gate{R_y(6.25)} & \qw & \qw & \qw & \targ{}  & \ctrl{1} & \gate{R_y(0.07)}\\
    \gate{R_y(0.77)} & \qw & \qw & \qw & \qw & \targ{} & \gate{R_y(1.30)} & \qw & \qw & \qw & \qw & \targ{} & \gate{R_y(0.45)} & \qw & \qw & \qw & \qw & \targ{} & \gate{R_y(2.89)} & \qw & \qw & \qw & \qw & \targ{} & \gate{R_y(4.72)} & \qw & \qw & \qw & \qw & \targ{} & \gate{R_y(0.07)} & \qw & \qw & \qw & \qw & \targ{} & \gate{R_y(6.23)}
\end{quantikz}
\caption{The 6-qubit circuit that prepares the 64-point PMF of a discrete approximation of a random variable distributed as $\mathcal{LN}(0,1/400)$, where all rotations are given in radians (based on the standard definition of a rotation operator) and are specified up to global phases; $x_\ell = 0.77$ and $\Delta = 0.01$, such that the support out to $\pm 5 \sigma$ is covered. Here the angles are ``absolute'' values -- not given as multiples of $\pi$.}
\label{fig:lognormal-discrete-2}
\end{figure}
\end{landscape}

\newpage

\end{document}